\documentclass[a4paper,11pt]{article}
\usepackage{graphicx,amssymb,amstext,amsmath}
\usepackage{color}

\definecolor{cbl}{rgb}{0,0,1}                % bleu

\newtheorem{rema}{Remark}[section]

\newcommand{\bc}{\begin{center}}
\newcommand{\ec}{\end{center}}
\def\ba#1{\begin{array}{#1}\displaystyle}
\newcommand{\ea}{\end{array}}

\newcommand{\beq}{\begin{equation}}
\newcommand{\eeq}{\end{equation}}
\newcommand{\beqa}{\begin{eqnarray}}
\newcommand{\eeqa}{\end{eqnarray}}
\newcommand{\no}{\nonumber}
\newcommand{\n}{\nonumber\\}
\newcommand{\bi}{\begin{itemize}}
\newcommand{\ei}{\end{itemize}}
\def\lt#1{\left#1}
\def\rt#1{\right#1}
\def\t#1{\tilde{#1}}

\def\frc#1#2{\frac{#1}{#2}}

\newcommand{\p}{\partial}

\newcommand{\vac}{{\rm vac}}
\newcommand{\bra}{\langle}
\newcommand{\ket}{\rangle}

\newcommand{\Or}{{\cal O}}

\newcommand{\ep}{\epsilon}
\newcommand{\varep}{\varepsilon}

\newcommand{\Tr}{{\rm Tr}}

\newcommand{\ri}{{\rm i}}

\begin{document}

\begin{titlepage}
\vspace{0.2cm}
\begin{center}

{\Large {\bf Thermodynamic Bethe ansatz for non-equilibrium steady\\[0.1cm]  states: exact energy current and fluctuations in integrable QFT}}

\vspace{0.8cm} {\large \text{Olalla Castro-Alvaredo${}^{\clubsuit}$, Yixiong Chen${}^{\spadesuit}$, Benjamin Doyon${}^{\spadesuit}$, Marianne Hoogeveen${}^{\spadesuit}$}}

\vspace{0.2cm}
{{\small ${}^{\clubsuit}$} Department of Mathematics, City University London, Northampton Square EC1V 0HB, UK }\\
{{\small ${}^{\spadesuit}$} Department of Mathematics, King's College London, Strand WC2R 2LS, UK }
\end{center}

\vspace{1cm}
We evaluate the exact energy current and scaled cumulant generating function (related to the large-deviation function) in non-equilibrium steady states with energy flow, in any integrable model of relativistic quantum field theory (IQFT) with diagonal scattering. Our derivations are based on various recent results of D. Bernard and B. Doyon. The steady states are built by connecting homogeneously two infinite halves of the system thermalized at different temperatures $T_l$, $T_r$, and waiting for a long time. We evaluate the current $J(T_l,T_r)$ using the exact QFT density matrix describing these non-equilibrium steady states and using Al.B.~Zamolodchikov's method of the thermodynamic Bethe ansatz (TBA). The scaled cumulant generating function is obtained from the extended fluctuation relations which hold in integrable models.  We verify our formula in particular by showing that the conformal field theory (CFT) result is obtained in the high-temperature limit. We analyze numerically our non-equilibrium steady-state TBA equations for three models: the sinh-Gordon model, the roaming trajectories model, and the sine-Gordon model at a particular reflectionless point. Based on the numerics, we conjecture that an infinite family of non-equilibrium $c$-functions, associated to the scaled cumulants, can be defined, which we interpret physically. We study the full scaled distribution function and find that it can be described by a set of independent Poisson processes. Finally, we show that the ``additivity'' property of the current, which is known to hold in CFT and was proposed to hold more generally, does not hold in general IQFT, that is $J(T_l,T_r)$ is not of the form $f(T_l)-f(T_r)$.
\vfill

{\ }\hfill \today

\end{titlepage}

\tableofcontents

\section{Introduction}

In recent years, there has been a surge of interest in the study of the thermodynamics of quantum systems out of equilibrium (for reviews, see \cite{Harbola, Campisi11RMP83}). In part this is due to the recent advances in experimental techniques, making it possible to drive quantum systems away from equilibrium in a controlled way, and to study their non-equilibrium properties, see for instance
\cite{utsumi2010bidirectional,nakamura2010nonequilibrium,nakamura2011fluctuation,sanchez2012detection,saira2012test,battista2013quantum}.
This is also due to theoretical advances which include the discovery of several types of fluctuation theorems, generalizing the fluctuation-dissipation theorem \cite{nyquist1928} to systems far from equilibrium (see \cite{Harbola,Campisi11RMP83} for extensive discussions). These fluctuation theorems express ``universal'' properties of certain distribution functions. In non-equilibrium steady states, where constant flows of particles, charge or energy exist, one object of interest is the scaled cumulant generating function (SCGF)\footnote{This is related to the large-deviation function, commonly used in large-deviation theory, by a Legendre transform.}. The SCGF fully characterises the statistics of the transferred quantity at large times. Fluctuation theorems are symmetry relations for the SCGF; see for instance \cite{Kurchan,Yukawa,deRoeck,Mukamel,JarzWoj,BD3} concerning the energy-flow SCGF.

In this manuscript we obtain for the first time the exact non-equilibrium current and SCGF for energy transfer in a wide family of integrable models. For this purpose, we propose, develop and test a new approach to the study of quantum integrable models in non-equilibrium steady states, concentrating on the study of non-equilibrium energy flows. We take the setup where two hamiltonian reservoirs at different temperatures are connected homogeneously to each other (local quantum quench), so that at large times an energy current be established. We consider the scaling limit: the quantum system is assumed to be in the universal regime near a quantum critical point (with unit dynamical exponent), with a mass gap and the driving temperatures assumed to be much smaller than any microscopic energy scale. This regime is described by massive quantum field theory (QFT). Our results are the first exact results for the energy SCGF and most general results for the energy current in non-critical models that cannot be described by free particles.

Integrability occurs when enough additional symmetries render the model exactly solvable \cite{das,rajaraman}. Exact solvability means that in principle all energy eigenstates and eigenvalues, and correlation functions of local operators, can be obtained non-perturbatively. Integrability has enjoyed much success over the past four decades. For integrable quantum spin chains several very effective approaches exist such as the algebraic Bethe ansatz (see e.g.~\cite{Faddeev:1996iy}). For integrable massive quantum field theories (IQFTs) the problem of computing scattering amplitudes and vacuum correlation functions of local fields has been solved for many models using factorized scattering theory \cite{Karowski:1978eg,za,abdalla,Mussardo:1992uc,Dorey:1996gd,mussardobook}. Factorized scattering and Bethe ansatz ideas led to a very successful approach for the study of the equilibrium thermodynamic properties of integrable models proposed by Al.B.~Zamolodchikov \cite{tba1}, known as the thermodynamic Bethe ansatz (TBA) approach. Other exact methods are based on free-fermion techniques (Clifford algebras), and the universal regime of exactly gapless quantum models is described by conformal field theory (CFT), for which a very wide variety of exact techniques are available \cite{BPZ,Mathieu}.

Using such techniques, exact results for SCGF in non-equilibrium integrable quantum systems have been obtained in the past, especially for charge flows, in which case the SCGF is referred to as the ``full-counting statistics''. The first exact formula in free fermion models was obtained by Lesovik and Levitov in \cite{LL1,LL2} and further studied in other works \cite{Klich03,PhysRevB.75.205329,AvronEtAl08,BD0}. Other exact formulas were obtained in  Luttinger liquids, which are particular models of CFT, in \cite{GutmanEtAl10}, and in the low-temperature universal regime of quantum critical models in \cite{BD2} using general CFT. In certain integrable interacting impurity models, the SCGF was obtained in \cite{PhysRevB.63.201302,PhysRevLett.107.100601} using TBA-like ideas. Exact charge current and shot noise (zero-temperature second cumulant) studies have also been done from similar ideas before in a variety of integrable models and with a variety of techniques, see for instance \cite{FLS1,FLS2,FLS3,FLS4,KSL1,KSL2,con2,con1,con3,MehtaAndrei,MehtaChaoAndrei,BoulatSaleurPRB,BoulatSaleurPRL}.

However for energy flows, exact results for currents in interacting models and for SCGF in general are more recent. As far as we are aware, the first exact SCGF was obtained for a chain of quantum harmonic oscillators in \cite{SaitoDhar}. Exact results for the energy current and SCGF in the low-temperature universal regime of general quantum critical models were obtained very recently in \cite{BD1,BD2} using CFT, and for the quantum Ising chain in a magnetic field in \cite{DeLuca} using free fermion techniques (results also exist for energy flows in more general ``star-graph'' configurations, see for instance \cite{MintchevNonEquilibrium2011,MintchevSorbaLuttinger2013,DHB}).

There are in fact many ways of theoretically constructing a non-equilibrium steady state (see \cite{BonettoEtAl00} for a discussion); the aforementioned results involve a variety of setups. The hamiltonian-reservoir setup considered here is an old idea going back, in the quantum realm, to Keldysh \cite{Keldysh} and Caroli et. al. \cite{CaroliDirect1971}, and studied in detailed classically in \cite{SpohnLebo}. For energy flows, it was rigorously analyzed in the XY model in \cite{Aschbacher03non-equilibriumsteady} based on \cite{ruelle2000natural}, where the resulting non-equilibrium density matrix was constructed\footnote{The idea of a density matrix for representing a steady state has arisen many times in the past, see for instance \cite{PhysRevLett.70.2134,ruelle2000natural,aschbook,DoyonAndrei}.}, and likewise in CFT in \cite{BD1,BD2,DHB}. The energy-flow non-equilibrium density matrix of the hamiltonian setup in general massive QFT was proposed and derived in \cite{BD1,Dhouches}.

We present here a generalization of the TBA approach to non-equilibrium steady states energy flows with hamiltonian reservoirs: the NESSTBA approach. We base our derivation on the exact massive QFT density matrix derived in \cite{Dhouches}. We deduce a set of coupled integral equations which are analog to the standard TBA equations up to some crucial modifications. Solutions to these equations give the exact energy current. All the exact cumulants are then obtained from the extended fluctuation relation that was derived in \cite{BD3}. We illustrate the working of our approach by numerically evaluating the current and some of the cumulants in three different models: the sinh-Gordon model, the roaming trajectories model, and the sine-Gordon model at the simplest nontrivial reflectionless point. Our work can be seen as a generalization of the CFT results of \cite{BD1} to IQFTs. In particular, as a consistency check we show that our formulation indeed leads to the known CFT result for the current and SCGF when both right and left temperatures are very high as compared to the gap.

The analysis of our results bring out three interesting observations. First, we find a family of non-equilibrium analogs of Zamolodchikov's $c$-functions \cite{tba1}, in terms of the current and all higher cumulants. We numerically verify that our non-equilibrium $c$-functions are monotonic along RG trajectories, and equal to the central charge at fixed points. Hence, following the usual wisdom, they are (non-equilibrium) measures of the number of degrees of freedom as functions of the energy scale. Our $c$-functions in turn give rise to exact upper bounds on the current and all higher cumulants. Second, we show numerically that the process of energy transfer can be understood, in its large-time regime, as a family of independent Poisson processes, one for each energy value and direction of transfer. This generalizes nontrivially what was observed in CFT in \cite{BD1}. It indicates that the generalization of the usual Landauer form of the conductance to the full out of equilibrium statistics is not straightfoward: the weights of the Poisson processes are {\em not} simply related to the state density and occupation (contrary to the CFT case). Finally, it is known that in CFT the current is {\em additive} \cite{BD1}: it satisfies $J(\beta_1,\beta_2) + J(\beta_2,\beta_3) + J(\beta_3,\beta_1)=0$ (where $\beta_i$ are inverse temperatures), equivalently $J(\beta_l,\beta_r)=f(\beta_l)-f(\beta_r)$ for some function $f$. This was suggested to hold more generally from various arguments \cite{Moore,Dhouches}, most importantly from DMRG numerics \cite{Moore}. We provide a proof that additivity does not hold in a large family of IQFT (including the three models we consider), a fact which is explained via the nontrivial relation between state occupation and state density in interacting models. This result is not in disagreement with the DMRG numerics.

The paper is organized as follows:
In section 2, we describe the physical situation and give the density matrix that describes the non-equilibrium steady state. We also give an overview of some existing results that we will use. In section 3  we derive the NESSTBA equations. In section 4  we introduce the integrable models that we will consider. In section 5 we present numerical results for the current and the $L$-functions of these models. In section 6, we introduce a family of $c$-functions that can be obtained by normalizing higher order cumulants, and present numerical evidence of their monotonicity. We discuss the physical implications of this property. In section 7 we provide evidence of the Poissonian nature of the energy transfer process. In section 8 we test the additivity property of the current and show that it is non-additive for three different integrable models. We provide a general argument as well as a mathematical proof as to why this must be the case for IQFTs. Finally, we present our conclusions in section 9. Various derivations, proofs, particular limits and consistency checks of our formulation are presented in appendices A-E.

\section{Physical situation and overview of relevant previous results}

\subsection{Physical description}

Consider a homogeneous open quantum chain of length $L$ with local interactions. Suppose it is initially broken into two contiguous halves of lengths $L/2$ with commuting hamiltonians $H_l^L$ and $H_r^L$ (for instance, some links are broken near some site, so that the two halves do not interact with each other). The two halves are independently thermalized at temperatures\footnote{In this paper we set $k_B=1$ and $\hbar=1$.} $T_l=\beta_l^{-1}$ and $T_r = \beta_r^{-1}$ respectively (left and right), so that the system is described by the density matrix $\rho_0 = e^{-\beta_l H_l^L-\beta_r H_r^L}$. At time $0$ the two halves are connected in order to form the original homogeneous quantum chain of length $L$, with hamiltonian $H^L = H_l^L+H_r^L + \delta H$. Here $\delta H$ represents the connection energy between the two halves, which must be local, i.e. involving only few sites around the connection point, and which we assume to be independent of $L$. The system is then allowed to $H$-evolve unitarily until time $t$. One can view this situation as starting with a profile of temperature on the full quantum chain that has a step change around one site, with $T_l$ on the left and $T_r$ on the right of it, and then evolving unitarily from this profile -- this point of view requires the notion of a local temperature, which we do not discuss here.

\begin{figure}[h!]
 \includegraphics[width=16cm]{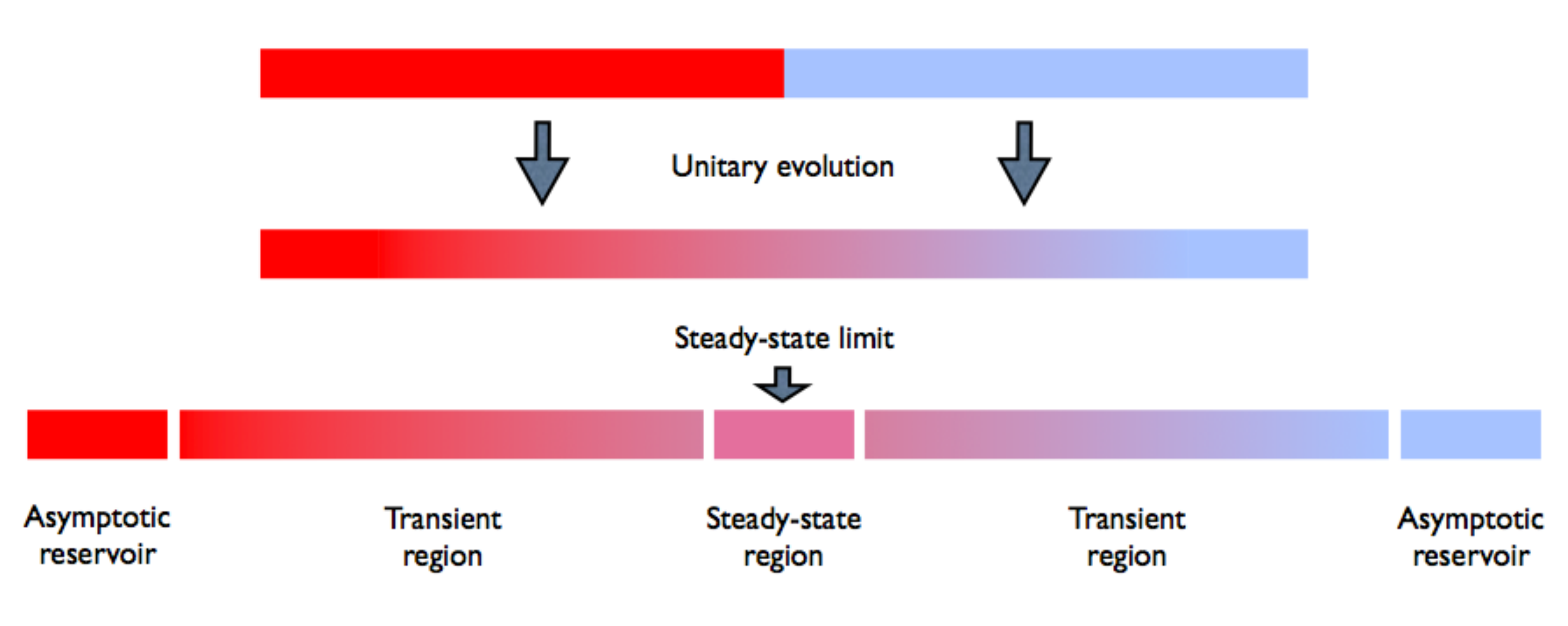}
\caption{The steady state is established by connecting two halves of the system that have been independently thermalized, and letting them evolve unitarily for a time $t$. After a long time and in a very large system, there will be asymptotically far hot and cold reservoirs which have not yet received any excitation from the connection point, two transient regions, and the steady-state region (the region described by the steady-state limit). In integrable massive QFT, excitations scatter elastically, and different points in the transient regions will receive excitations at different times depending on the excitation's velocity. Hence the length of the transient regions should scale proportionally to $t$ at large $t$. There is no prediction for the length of the steady-state region, which may be observable-dependent and which may scale more slowly than $Ct$ at large $t$ for every $C>0$. The only requirement is that every point a finite distance away from the connection point will converge to the steady state.}
\label{steady-state-fig}
\end{figure}

Averages of observables $\Or$ at time $t$ are described by
\beq
	\bra \Or\ket(L;t) = \Tr\lt( e^{-\ri tH^L}\,\frak{n}[\rho_0]\, e^{\ri tH^L}
	\,\Or\rt),
	\quad \rho_0 = e^{-\beta_l H_l^L-\beta_r H_r^L}
\eeq
where here and below we use $\frak{n}[\rho] = \rho / \Tr(\rho)$. The {\em steady-state limit} is the limit $L\to\infty$ followed by $t\to\infty$ of such an average, with the observation lenght $\ell$ kept fixed, see Figure \ref{steady-state-fig} (below we will denote by $H$, $H_l$ and $H_r$ the $L\to\infty$ limits, when it makes sense, of the operators $H^L$, $H_l^L$ and $H_r^L$ respectively). The observation length is the length of the region over which lie both the support of $\Or$ and the contact point $x=0$.

If this limit exists, then we say that the system reaches a steady state with respect to the observable $\Or$:
\beq\label{sslimit}
	\bra\Or\ket_{\rm stat}:= \lim_{t\to\infty}\lim_{L\to\infty}\bra \Or\ket(L;t).
\eeq
For local observables, we expect the steady-state limit to exist thanks to the locality of the hamiltonians. Physically, this is because we expect the two infinite ($L\to\infty$) halves of the system to play the role, at large enough times, of far thermal reservoirs, able to absorb and emit independent thermalized excitations unboundedly for all times $t\ll L/v$ where $v$ is a propagation velocity. Indeed for such times, excitations do not have time to bounce off the far endpoints and thus create non-thermal correlations or be re-emitted. Then, observing only locally, the detailed reservoir information is lost and we only see the steady state. In particular, the energy current observable
\beq
	{\cal J} := \frc \ri2 [H^L,H_r^L-H_l^L] = \frc \ri2 [\delta H,H_r^L-H_l^L]
\eeq
is local (and independent of $L$) thanks to the locality of $\delta H$, wherefore the system should reach a steady state with respect to the energy current.

The first detailed study of this steady-state limit appeared for classical harmonic oscillators in \cite{SpohnLebo}. In quantum systems, similar steady-state limits were discussed and shown to exist in quantum models under certain conditions in \cite{ruelle2000natural} and more particularly in the XY model in \cite{Aschbacher03non-equilibriumsteady}; in the context of charge transfer in the Kondo model in \cite{DoyonAndrei} and in the resonant-level model in \cite{DoyonNotes,BD0}; and in the context of energy and charge transfer in conformal models in \cite{BD1,BD2}.

As was discussed in many works, the effect of the connection will propagate from the center outwards as time evolves, so that the effective reservoirs, i.e.~the regions which are still thermal, will be situated further and further away from the connection point; specifically, the temperature profile will become flat. This holds true in general local models, not just in CFT (although there are difference in details). Hence, if energy transport is purely diffusive, then the steady-state current should be zero (by Fourier's law). However, if the initial temperatures are different and the energy propagation in the system has a ballistic component, then we expect that the steady-state limit will support an energy flow, $\bra{\cal J}\ket_{\rm stat}\neq0$: it is then a nontrivial non-equilibrium steady state. It is expected that any integrable system will display a ballistic component to the energy transport: the conserved charges give enough constraints to counteract diffusion \cite{castella1995integrability,PhysRevB.46.9325,2009PhRvL.103u6602S}. In particular, the conserved charge associated to the energy current is usually one of the first nontrivial ``higher'' conserved charges in integrable quantum spin chains.

\begin{rema} In CFT and QFT, the momentum is always conserved; hence, as explicitly shown in \cite{BD1,BD2} in CFT, the steady-state limit exists and supports a nonzero energy flow. This means that in the scaling limit (low-temperature, small-gap universal regime) of any quantum chain (with unit dynamical exponent), integrable or not, the steady-state limit should exist (we refer to \cite{BD2,Dhouches} for a discussion of the interplay between the steady-state limit and the scaling limit). This was indeed observed using DMRG numerics for a particular non-integrable quantum chain in \cite{Moore}. Of course, in any quantum chain, the nonzero temperatures and, possibly, nonzero gap mean that the system is never exactly in the scaling limit. Hence, if the microscopic system is not integrable, there may be some large, non-universal time scale where non-integrability  ``kicks-off'' and the current starts to decrease. An interesting open question is the nature of this time scale, or by what other mechanism microscopic non-integrability may appear. In the present paper we discuss integrable QFT, which may be thought of as coming from integrable microscopic systems, where there is true ballistic transport.
\end{rema}

\begin{rema} We may attempt to make the steady-state limit a bit more precise as follows. We require that the limit exist when each of the conditions $L\gg v t$ and $vt \gg \ell$ hold almost surely for all $v$, where $v$ are propagation velocities of excitations, and $\ell$ is the observation length scale. Here we have in mind a distribution of propagation velocities associated to the available excited states, and the corresponding measure (discrete / continuous parts of the Hilbert space, etc.); the phrase ``almost surely'' means that the strong inequalities hold for all velocities except possibly for a set of velocities that is of measure zero. Then, the fact that the steady-state limit exists for any local observable can only be true if there is no stationary excitation with nonzero measure: the Hilbert space shouldn't have a discrete part containing 0-velocity excitations.
\end{rema}

\subsection{Steady state in massive QFT}\label{ssectQFT}

Assume now that the quantum chain has a parameter $g$ (for instance, an external magnetic field) such that there is a quantum critical point at $g=g_c$ with unit dynamical exponent. Then, at $g=g_c$, the energy gap $\Delta$ is zero. At $g=g_c$ and low temperatures, the chain is described by a CFT. Let us now take the scaling limit where $g\to g_c$ and $T_{l,r}\propto \Delta$ (which tends to zero). This is the low-temperature, small-gap region near the quantum critical point. This limit is universal, and described by massive QFT, with the mass representing the infinitesimal gap $\Delta$. The main result of \cite{BD1,Dhouches}, which we take as our starting point, is a proposed exact description of the non-equilibrium density matrix $\rho_{\rm stat}$ in the scaling limit, that represents the steady-state limit \eqref{sslimit},
\beq\label{ness}
	\bra \Or\ket_{\rm stat} = \Tr\lt(\frak{n}[\rho_{\rm stat}]\,\Or\rt).
\eeq
It is expressed in terms of massive QFT data, and was derived using general QFT arguments.

The non-equilibrium steady-state density matrix $\rho_{\rm stat}$  is defined as follows. Let us consider a model of relativistic QFT with a spectrum of $\ell$ particle types, with masses $m_1,\ldots,m_\ell$. The basis of asymptotic states (here we take them implicitly as asymptotic $in$ states), which span the Hilbert space, is described by
\beq
	|\vac\ket,\quad |\theta_1,\ldots,\theta_n\ket_{i_1,\ldots,i_n}\;:\;
	\theta_1>\ldots>\theta_n,\quad i_1,\ldots,i_n \in\{ 1,2,\ldots,\ell\},
	\quad n=1,2,\ldots
\eeq
where $|\vac\ket$ is the vacuum, and $\theta_j$ are the rapidities of the asymptotic particles. The density matrix $\rho_{\rm stat}$ acts diagonally on the basis of asymptotic states. Its eigenvalues are simply given by $\rho_{\rm stat}|\vac\ket=1$ and
\beq\label{rhostat}
	\rho_{\rm stat}|\theta_1,\ldots,\theta_n\ket_{i_1,\ldots,i_n} = e^{-\sum_k W_{i_k}(\theta_k)}|\theta_1,\ldots,\theta_n\ket_{i_1,\ldots,i_n}
\eeq
where
\beq\label{W}
	W_i(\theta) := m_i\cosh\theta\cdot \lt\{\ba{ll} \beta_l & (\theta>0) \\
	\beta_r & (\theta<0).  \ea\rt.
\eeq
If we factorize the Hilbert space into a product of the space formed by positive-rapidity particles, and that formed by negative-rapidity particles, then \eqref{rhostat} indicates that the steady-state density matrix factorizes accordingly. Specifically, on each factor, it is thermal with inverse temperature $\beta_{l,r}$, respectively. Note that this density matrix is stationary and homogeneous (translation invariant).

The density matrix (\ref{rhostat}) agrees with the mathematical description of the energy-flow steady state in the $XY$ model in \cite{Aschbacher03non-equilibriumsteady} and with that obtained in the Ising model from less rigorous arguments in \cite{DeLuca}. Its natural counterpart in CFT, where right-movers and left-movers are thermalized at different temperatures, was proven in \cite{BD1,BD2}.

\begin{rema}
The physical interpretation of the density matrix $\rho_{\rm stat}$ is rather simple. The steady-state density matrix is the long-time evolution $\rho_{\rm stat}=\lim_{t\to\infty} e^{-\ri Ht}\rho_0 e^{\ri Ht}$ of the initial density matrix $\rho_0$. The initial density matrix represents the left- and right-hand sides of the system separately thermalized at inverse temperatures $\beta_l$ and $\beta_r$, respectively, and the time evolution is generated by the full hamiltonian $H$ where both halves are connected. In the scattering language, $\rho_{\rm stat}=\lim_{t\to\infty} e^{-\ri Ht}e^{\ri H_0 t}\rho_0 e^{-\ri H_0 t}e^{\ri Ht} = S\rho_0 S^{-1}$ where $S=\lim_{t\to\infty} e^{-\ri Ht}e^{\ri H_0 t}$ is the scattering operator. Also, asymptotic states are, by definition, states on the line with the property that when evolved back in time, they become well-separated, well-defined wave packets behaving like free particles (the incoming asymptotic particles). In the scattering language, an asymptotic state has the form $|v\ket = S|v\ket_0$ where $|v\ket_0$ is a free-particle state representing the asymptotically free particles. Hence, the action of $\rho_{\rm stat}$ on asymptotic states is isomorphic to the action of $\rho_0$ on the well-separated, well-defined wave packets: if $|v\ket_0$ is an eigenvector of $\rho_0$, then $|v\ket$ is an eigenvector of $\rho_{\rm stat}$ with the same eigenvalue. Since for positive rapidities these wave packets are far on the left, and for negative rapidities they are far on the right, the expression (\ref{W}) follows. A  more complete argument is presented in \cite{Dhouches}. More precisely, this argument shows that for any local operator, the contributions of asympotic states with finite numbers of particles to the steady-state average is given by \eqref{rhostat} and \eqref{W}.
\end{rema}

\subsection{Current and fluctuations}

In QFT, the energy current operator is simply the momentum density ${\cal J}=p(x)$ at the position $x=0$, so that the average current is
\[
	J = \bra  p(0) \ket_{\rm stat}.
\]

Besides the current, its fluctuations are also of great interest in non-equilibrium steady states. In quantum mechanics, one has to define fluctuations of the current via appropriate measurement protocols: see for instance \cite{LL1,LL2} for discussions of an indirect measurement protocol, \cite{Klich03,JarzWoj,Schonhammer,BD0,BD1} for results within a two-time measurement protocol and connection with the former protocol,
and the review \cite{Harbola}. In order to be specific, let us consider the two-time measurement protocol. We perform a first von Neumann  measurement of the quantity
\beq
	Q:=\frc12 (H_r-H_l)
\eeq
at time 0, returning the value $q_0$, and then another von Neumann  measurement at time $t$, returning the value $q_t$. We are interested in the statistics of the difference $q:=q_t-q_0$. By standard quantum mechanics, the probability distribution for $q$ is
\beq
	\Omega_t(q) = \Tr\lt(P_{Q=q_t} e^{-\ri Ht} P_{Q=q_0} \frak{n}[\rho_0]
	P_{Q=q_0} e^{\ri Ht} P_{Q=q_t}\rt)
\eeq
where $P_{Q=a}$ is the projector onto the eigenspace of eigenvalue $a$ of the operator $Q$. One may then evaluate the associated scaled cumulants
\beq\label{defCn}
	C_n := \lim_{t\to\infty}t^{-1} \bra q^n\ket^{\rm cumulant}_{\Omega_t},
\eeq
where $\bra q^n\ket^{\rm cumulant}_{\Omega_t}$ is the $n^{\rm th}$ cumulant of $q$ with respect to $\Omega_t$. These may be gathered into a generating function as usual,
\beq\label{defF}
	F(z) := \sum_{n=1}^\infty C_n z^n/n!,\quad
	F(z) = \lim_{t\to\infty} t^{-1} \log\big(
	\bra e^{qz}\ket_{\Omega_t}\big)
\eeq
(here $z$ is a formal parameter). Note that the first cumulant $C_1 = J$ is the average current.

A result of the discussions cited above is that one expects $F(z)$ to have the following expression in terms of averages of quantum-mechanical operators:
\beq\label{exprF}
	F(z) = \lim_{t\to\infty} t^{-1} \log\lt( \bra e^{zQ(t)} e^{-zQ}\ket_{\rm stat}
	\rt).
\eeq
In fact, we expect that in general, the scaled cumulants can also be expressed in terms of connected correlation functions\footnote{The term ``connected'' has  the same combinatoric meaning as ``cumulant'', but is more common in the context of quantum field theory.} of time-evolved current operators,
\beq\label{Cn}
	C_n = 	\lim_{\ep\to0^+} \int_{-\infty}^\infty du_1\cdots du_{n-1}\,
	\bra {\cal J}(u_{n-1}+(n-1)\ri\ep)\cdots{\cal J}(u_1+\ri\ep){\cal J}(0)\ket_{\rm stat}^{\rm connected}
\eeq
where $\ri\ep$ is an imaginary time shift that regularizes the correlators. This expression is derived, under certain assumptions that hold for integrable models, in Appendix \ref{appcumul}.

Note that an important symmetry relation satisfied by $F(z)$ is the fluctuation relation
\beq\label{FR}
	F(\beta_l-\beta_r-z) = F(z).
\eeq
This energy-flow fluctuation relation was derived in various situations in \cite{Kurchan,Yukawa,deRoeck,Mukamel,JarzWoj} (but see \cite{Akagawa} for limitations in the case of finite systems); \cite{Harbola,Campisi11RMP83} present good reviews. It was derived recently in \cite{BD3} within the present physical setup using a scattering formalism.

A recent result of Bernard and Doyon \cite{BD3} is that in any system where the energy is purely transmitted (that is, $QS = -SQ$ where $S$ is the system's scattering operator, see \cite{BD3}), the scaled cumulant generating function $F(z)$ can be evaluated solely from the current at shifted inverse temperatures\footnote{Here the current describes the energy transfer from the left to the right, in contrast to the convention used in \cite{BD3}.},
\beq\label{EFR}
	\frc{d F(z)}{dz} = J(\beta_l-z,\beta_r+z)\quad\Leftrightarrow\quad
	F(z) = \int_0^z dz'\, J(\beta_l-z',\beta_r+z')
\eeq
(recall that $F(0)=0$).
The upshot is that the scaled cumulants are simply given by derivatives of the current with respect to inverse temperatures,
\beq\label{cumul}
	C_{n+1} = \frc{d^n}{dz^n} J(\beta_l-z,\beta_r+z)\Big|_{z=0}
\eeq
for $n=0,1,2,\ldots$. Equation \eqref{EFR} was referred to as an {\em extended fluctuation relation} (EFR) in \cite{BD3}, as it actualy implies the usual non-equilibrium fluctuation relations \eqref{FR} if one assumes parity symmetry. As explained in \cite{BD3}, pure energy transmission, and therefore the EFR, is valid in any homogeneous integrable model and in CFT. Note that it implies a relation between the non-equilibrium differential conductivity $G:=\beta^2 d J/d\gamma$ (where $\beta = (\beta_r+\beta_l)/2$ and $\gamma = \beta_r-\beta_l$) and the noise (second cumulant), which reads $\beta^2 C_2=2G$.

\subsection{Previous exact results}

Exact results for energy flow and fluctuations in the present physical setup are available in the XY and Ising models \cite{Aschbacher03non-equilibriumsteady,DeLuca,Aschbacher}, and in conformal field theory (critical models) \cite{BD1,BD2,DHB}. To our knowledge, the first exact cumulant generating function for energy transfer in quantum systems was found in \cite{SaitoDhar}, within a slightly different physical construction of the non-equilibrium steady-state using Caldeira-Leggett baths attached to a chain of harmonic oscillators. Here we just recall the CFT results, as these are most relevant for our purposes.

In general unitary CFT, the current was first calculated in \cite{BD1} and found to be
\beq\label{JCFT}
	J_{\rm CFT} = \frc{c\pi}{12} (T_l^2-T_r^2).
\eeq
This has a clear interpretation via the Stefan-Boltzmann law thanks to \cite{CardyStef}. The scaled cumulant generating function $F_{\rm CFT}(z)$ was first evaluated exactly in \cite{BD1}, giving the expression $F_{{\rm CFT}}(z) = (c\pi z/12)(\beta_l^{-1}(\beta_l-z)^{-1} - \beta_r^{-1}(\beta_r+z)^{-1})$, in agreement with \eqref{EFR}. This means that the scaled cumulants are
\beq\label{CnCFT}
	C_{n\,{\rm CFT}} = \frc{c\pi n!}{12}(T_l^{n+1} + (-1)^n T_r^{n+1}).
\eeq
The function $F_{\rm CFT}(z)$ was evaluated in \cite{BD1} assuming the fluctuation relations \eqref{FR}; it was proven without this assumption in the CFT context using a local operator formalism in \cite{BD2}, and using the Virasoro algebra in \cite{DHB} (where a more general star-graph configuration is considered).  The expression \eqref{JCFT} was verified numerically in the XXZ chain by Karrasch, Iland and Moore \cite{Moore}.

\section{The non-equilibrium steady-state TBA equations} \label{sectTBA}

In the present paper, we obtain for the first time the exact energy current and scaled cumulant generating function in interacting (integrable) models.

In integrable models of relativistic QFTs, two crucial properties hold: the scattering of asymptotic particles is purely elastic (the set of momenta is preserved) and the scattering amplitudes always factorize into two-particle scattering processes. Hence, the only dynamical data necessary in order to fully define the model and its ``local structure'' (its full scattering and its set of local observables) are the two-particle scattering amplitudes. In particular, two-particle scattering amplitudes are solutions of the Yang-Baxter equation, which is nontrivial whenever there are internal degrees of freedom and the two-particle scattering processes is not diagonal in the internal space. This is the essence of factorized scattering theory.

We now consider a general  integrable model of relativistic QFT, under the simplifying assumption that the two-particle scattering matrix is diagonal in the internal space (there is no ``back-scattering'': in a two-particle scattering process, only phases may occur).

\subsection{The integral equations}

According to \eqref{ness}, the average energy current $J$ is obtained by evaluating the average momentum density $p(x)$ at an arbitrary point, say $x=0$, using the density matrix (\ref{rhostat}):
\beq\label{nessQFT}
	J=\Tr\lt(\frak{n}[\rho_{\rm stat}]\,p(0)\rt).
\eeq
We have provided the exact density matrix \eqref{rhostat} as an operator acting on asymptotic states with finitely many particles. In order to evaluate this trace, we further need the matrix elements of $p(0)$. But we need a bit more: we need to describe how to perform the trace operation. This implies understanding $\rho_{\rm stat}$ in states with nonzero densities of particles (i.e. infinitely many particles). In such finite-density states, the interaction between particles becomes important, and this affects the way the trace is defined. One could say that the information about the ``local structure'' is encoded both into the matrix elements of $p(0)$ and into the way the trace is performed.

In order to add these two elements of information, we follow Al.B. Zamolodchikov's TBA argument \cite{tba1}. This is an argument that mixes the ideas of factorized scattering and of the Bethe ansatz. For this purpose, we consider the model on a finite periodic space of circumference $R$. On the finite space, according to \cite{tba1}, the description of integrable models of QFT can be done similarly to that of Bethe ansatz integrable systems. Each state is characterized by its contents in quasi-particles, which possess momenta $p_k$ and energies $e_k$ whose sum give the total momentum and energy of the state, with relativistic dispersion relation. Then, we construct a density matrix $\rho_{\rm stat}^R$ that approximates $\rho_{\rm stat}$. The natural choice for $\rho_{\rm stat}^R$ is defined by its action on every state $|v\ket$ as follows:
\beq\label{rhostatL}
	\rho_{\rm stat}^R |v\ket = e^{-\sum_k W_k}|v\ket,\quad
	W_k = e_k\cdot\lt\{\ba{ll} \beta_l & (p_k>0) \\ \beta_r & (p_k<0).
	\ea\rt.
\eeq
The finite-$R$ approximation of the average current is the average of the local momentum density $p(0)$ with respect to $\rho_{\rm stat}^R$:
\beq\label{JL0}
	J_R = \frc{\Tr_{R}\lt[\rho_{\rm stat}^R p(0) \rt]}{\Tr_{R}[\rho_{\rm stat}^R]},\quad \lim_{R\to\infty}J_R = J.
\eeq
Note that $\rho_{\rm stat}^R$ is translation invariant. By translation invariance, we may replace $p(0)$ by $R^{-1}\int_{-R/2}^{R/2} dx\, p(x)$. We then have
\beq\label{JL}
	J_R = R^{-1} \frc{\Tr_{R}\lt[\rho_{\rm stat}^R P_R \rt]}{\Tr_{R}[\rho_{\rm stat}^R]}
\eeq
where $P_R:=\int_{-R/2}^{R/2} dx\, p(x)$ is the total momentum.

The current can then be calculated by conveniently introducing a generating parameter $a$ associated to the momentum:
\beq
	J = -\lim_{R\to\infty} R^{-1}\,
	\lt.\frc{d}{da} \log \Tr_R\lt(\rho_{\rm stat}^R\,e^{-aP_R}\rt)\rt|_{a=0}.
\eeq
We may also define the ``free energy'' $f^a$ and write the current from it:
\beq\label{fa}
	f^a:=-\lim_{R\to\infty} R^{-1} \log \Tr_R\lt(\rho_{\rm stat}^R\,e^{-aP_R}\rt),
	\quad
	J =
	\lt.\frc{d}{da} f^a\rt|_{a=0}.
\eeq
The free energy $f^a$ can be evaluated by extending the TBA arguments of Al. B.~Zamolodchikov presented in \cite{tba1} (a similar extension of TBA arguments beyond Gibb's equilibrium was first done, to our knowledge, in the context of quantum quenches for the Lieb-Liniger model in \cite{Mossel}). This allows us to obtain the following expressions (see Appendix \ref{appTBA}):
\beqa
	J(\beta_l,\beta_r) &=& \sum_{i=1}^\ell \int_{-\infty}^\infty \frc{d\theta}{2\pi}\,
	 \frc{m_i \cosh\theta\;x_i(\theta)}{1+e^{\ep_i(\theta)}}
	 \n
	x_i(\theta)
	&=&  m_i\sinh\theta
	+ \sum_{j=1}^\ell
	\int_{-\infty}^\infty \frc{d\gamma}{2\pi}\,
	\frc{\varphi_{ij}(\theta-\gamma)\,x_j(\gamma)}{
	1+e^{\ep_j(\gamma)}}\n
	\ep_i(\theta) &=& W_i(\theta)-\sum_{j=1}^\ell \int_{-\infty}^\infty \frc{d\gamma}{2\pi}\,
	\,\varphi_{ij}(\theta-\gamma)\,\log(1+e^{-\ep_j(\gamma)})
	\label{exactJ}
\eeqa
where as usual $\varphi_{ij}(\theta) = -\ri\p_\theta S_{ij}(\theta)$ and $S_{ij}(\theta)$ is the two-particle scattering matrix, and where $W_i(\theta)$ is defined in \eqref{W}. This is the exact non-equilibrium steady-state TBA (NESSTBA) expression for the energy current in general diagonal-scattering integrable relativistic QFT. It turns out that there are many equivalent expressions for the current. For instance (see Appendix \ref{appTBA}),
\beqa
	J(\beta_l,\beta_r) &=& \sum_{i=1}^\ell \int_{-\infty}^\infty \frc{d\theta}{2\pi}\,
	 \frc{m_i \sinh\theta\;\mu_i(\theta)}{1+e^{\ep_i(\theta)}}
	 \n
	\mu_i(\theta)
	&=& m_i\cosh\theta
	+ \sum_{j=1}^\ell
	\int_{-\infty}^\infty d\gamma\,
	\frc{\varphi_{ij}(\theta-\gamma)\,
	\mu_j(\gamma)}{1+e^{\ep_j(\gamma)}}.
	\label{exactJ2}
\eeqa

Note that it is customary to define the $L$-functions
\beq
	L_i(\theta) = \log(1+e^{-\ep_i(\theta)}),
\eeq
as these functions possess interesting and perhaps clearer features than the pseudo-energies $\ep_i(\theta)$. Note also that the pseudo-energies, hence the $L$-functions, have jumps at $\theta=0$ contrary to the equilibrium case,
\beq\label{discep}
	\ep_i(+{\bf 0}) - \ep_i(-{\bf 0}) = m(\beta_l-\beta_r).
\eeq

We present various verifications of \eqref{exactJ} in Appendices \ref{apphight} and \ref{applowt}. We verify that the low-temperature expansion of \eqref{exactJ} agrees with an explicit evaluation of the trace \eqref{JL} in the large-$L$ limit, using the finite-volume regularization methods of Pozsgay and Tak\'acs \cite{Pozsgay} (and assuming a single particle spectrum for simplicity).Using and generalizing the analysis performed by Al.B.~Zamolodchikov in \cite{tba1}, we also verify that the high-temperature expansion of \eqref{exactJ} exactly reproduces the CFT prediction \eqref{JCFT} of \cite{BD1,BD2}, with  the correct central charge.

As mentioned, in \cite{BD3} it was shown that for integrable models of relativistic QFT, the energy-flow steady state satisfies the extended fluctuation relation \eqref{EFR}. This relation immediately allows us to calculate the exact scaled cumulant generating function for such models using the NESSTBA expression above. That is, the generating function $F(z)$ defined in \eqref{defF} for the scaled cumulants $C_n$ defined in \eqref{defCn}-\eqref{Cn}, is
\beqa
	F(z) &=& \sum_{i=1}^\ell \int_{-\infty}^\infty
	\frc{d\theta}{2\pi}\,
	 m_i \cosh\theta \int_0^z dz' \frc{x_i(\theta,z')}{1+e^{\ep_i(\theta,z')}} \n
	x_i(\theta,z) &=& m_i\sinh\theta + \sum_{j=1}^\ell
	\int_{-\infty}^\infty \frc{d\gamma}{2\pi}\,
	\frc{\varphi_{ij}(\theta-\gamma)\, x_j(\gamma,z)}{
	1+e^{\ep_j(\gamma,z)}} \n
	\ep_i(\theta,z) &=& W_i(\theta)
	+z\,{\rm sgn}(\theta) \,m_i\cosh\theta
	-\sum_{j=1}^\ell \int_{-\infty}^\infty \frc{d\gamma}{2\pi}\,
	\,\varphi_{ij}(\theta-\gamma)\,\log(1+e^{-\ep_j(\gamma,z)}).
	\label{exactF}
\eeqa

\begin{rema}
Our verification that the low-temperature expansion of the trace itself agrees with the low-temperature expansion of \eqref{exactJ} indicates that Al.B.~Zamolodchikov's TBA arguments indeed can be extended to the non-equilibrium steady-state density matrix \eqref{rhostat} without additional subtleties. This is important, because the derivation presented in \cite{Dhouches} of the form \eqref{rhostat} of the density matrix is, to be correct, only applicable to states with finite numbers of particles, hence to the low-temperature expansion of the trace.
\end{rema}

\begin{rema}
It is important to note that the ratio of traces on the right-hand side of \eqref{JL} is not expected to be related in any way to the average total momentum in the physical non-equilibrium steady state. Indeed, the latter average does not exist because the total momentum is not a local operator (hence does not have a steady-state limit). Expression \eqref{JL} is just a means to obtain a finite-$R$ approximation of the actual steady-state current.
\end{rema}

\begin{rema}
There are many subtleties involved in the derivation of \eqref{exactJ} that we presented. Let us discuss some of them.

First, at equilibrium, the procedure for replacing the infinite-space density matrix by a finite-$R$ approximation is clear, since even on finite periodic space, there is a clear notion of thermal states. Out of equilibrium, the procedure is less clear: the physical setup we use to construct the non-equilibrium steady state fundamentally requires the infinite-length limit to be taken. So, we must rely on a ``non-physical'' density matrix $\rho_{\rm stat}^R$, with the requirement that its infinite-$R$ limit reproduces $\rho_{\rm stat}$ in some ``topology'' (for instance, for every individual matrix element). This is what we have done in \eqref{rhostatL} for integrable models using Al.B. Zamolodchikov's ideas. Of course, the finite-$R$ ``non-physical'' density matrix \eqref{rhostatL} still describes some non-equilibrium steady state in the periodic system: it is a state where there is an imbalance between the populations of particles moving towards the right and of those moving towards the left. Importantly, however, it is a state whose infinite-$R$ limit provides the correct population imbalance of asymptotic states for describing the physical non-equilibrium steady state.

Second, the validity of the statement $\lim_{R\to\infty}J_R = J$ in \eqref{JL0} relies on the fact that thanks to locality, in the infinite-$R$ limit, the average of the local operator $p(0)$ does not depend on the actual boundary conditions taken (in particular, there are no topological effects that connect the energy density to boundary conditions). An argument can be made as follows. One can consider a different finite-$R$ regularization, which by construction correctly gives the physical steady sate. One evolves the initial density matrix $\rho_0$ for a finite time $t$, and observes the system on a region of length $R=vt$ around the contact point, for $v=\tanh\theta_0$ the speed associated to some small rapidity $\theta_0>0$. The reduced density matrix $\t\rho^{R,\theta_0}$ on that region is an IR regularization, and its limit $R\to\infty$ gives by construction the correct steady state with respect to any local observable. But the point is that $\t\rho^{R,\theta_0}$ and $\rho_{\rm stat}^R$ differ from each other only because of two effects: the boundaries are different (in the former, it's the transient region; in the latter, it's a periodic boundary condition), and the presence of excitations at rapidities $|\theta|<\theta_0$. Therefore, by locality and assuming no topological effects, averages of operators supported in the bulk are equal with respect to both density matrices in the infinite-$R$ limit, except possibly for differences of order $O(\theta_0)$. Since $\theta_0$ can be made arbitrarily small for local operators a finite distance away from the contact point (because the boundary at distance $vt$ is very far at large $t$ for any $v> 0$), then the averages with respect with $\rho_{\rm stat}^R$ give the averages in the physical non-equilibrium steady state.
\end{rema}

\subsection{Physical interpretation}

Formula \eqref{exactJ} has a simple interpretation: the function $x_i(\theta)$ is the ``dressed'' momentum at rapidity $\theta$, the factor $L\,d\theta\, m_i\cosh\theta/(2\pi) = L m_i\,d\sinh\theta/(2\pi)$ is the ``bare'' (uninteracting) number of levels around $\theta$, and $1/(1+e^{\ep_i(\theta)})$ is the filling fraction. Formula \eqref{exactJ2} has a similar and perhaps clearer interpretation: the quantity $L\,d\theta\,\mu_i(\theta)/(2\pi)$ is the true, interacting number of levels around $\theta$, and the factor $m_i\sinh(\theta)$ is the momentum at $\theta$. Formula \eqref{exactJ} encodes the steady-state density matrix into the driving term of the pseudo-energy, and the locality information of the QFT, including the state density and the momentum matrix elements, into the dressing operations based on the differential scattering phases $\varphi_{ij}(\theta)$.

The specialization of \eqref{exactJ} to the Ising model, where there is only one particle type and the differential scattering phase is zero, gives a simple formula for the exact energy current,
\beq\label{Ising}
	J_{\rm Ising}(\beta_l,\beta_r) = \frc{m^2}{4\pi} \int_{-\infty}^\infty d\theta \,\frc{\sinh2\theta}{
	1+e^{W(\theta)}} = \frc{1}{2\pi} \int_m^\infty dE\,E\lt(
	\frc1{1+e^{\beta_l E}} - \frc1{1+e^{\beta_rE}}\rt).
\eeq
This can also be obtained in various other ways, and can be extracted for instance from the exact results in the XY and Ising chains in  \cite{Aschbacher03non-equilibriumsteady,DeLuca}.

The second form of $J_{\rm Ising}$ in \eqref{Ising} has the structure of the Landauer formula: with $dE\, E = dp\, p$ where $E$ is the energy and $p$ is the momentum, we see that we have the number of available channels $dp$, times the current through these channels $p$, weighted by the fermionic filling fractions for particles from the right (with temperature $T_l$) and from the left (with temperature $T_r$). Such a form was also observed in CFT in \cite{BD1}: there the filling fraction at energy $E$ can be taken as being proportional to the Boltzmann occupation $e^{-\beta E}$. There is however no immediate Landauer form for the non-equilibrium energy current in the general interacting case, because there is no separation between the state densities and filling fractions of right-movers and left-movers: the presence of particles at some energy affect the state density at other energies.

In fact, the correct interpretation of both the Ising and the interacting energy flow should be obtained following what was done in the CFT context in \cite{BD1}: one must analyze not only the current, but also the fluctuations, and one must look for a formulation where these appear via independent stochastic processes. It turns out that in CFT, the Boltzmann occupation $e^{-\beta E}$ does provide the correct density for independent Poisson processes describing the current and all fluctuations \cite{BD1}. But, as we will see in section \ref{sectPoisson}, the fermionic or interacting filling fractions are not the correct densities; the correct one will be calculated there.

\section{The integrable models considered}

In the coming sections we will study the non-equilibrium steady-state current and cumulants for several models. We have chosen here three models as good representatives of increasingly complex families of IQFTs:  the sinh-Gordon model which may be regarded as the simplest interacting IQFT and therefore provides an ideal starting point for testing our approach; the roaming trajectories model which is still a very simple theory from a particle content and $S$-matrix point of view, but where the dependence of the $S$-matrix on the free parameter $\theta_0$ leads to a number of new interesting features of the current and its cumulants; finally, the sine-Gordon model at the simplest nontrivial reflectionless point which gives us the opportunity to study a theory with a more complicated particle content but still a diagonal $S$-matrix. It is also a theory which has a well-known discrete counterpart (e.g.~the gapped XXZ chain) and therefore results for this model may prove useful when comparing to numerical simulations of the XXZ quantum spin chain.
\subsection{The sinh-Gordon and roaming trajectories model}
The sinh-Gordon model is one of the simplest interacting QFTs and as such it constitutes an ideal testing ground for the ideas above.
It has a single particle spectrum and no bound states. The two-particle $S$-matrix of the model is
\begin{equation}
    S(\theta)=\frac{\tanh\frac{1}{2}\left(\theta-\frac{\ri\pi B}{2}\right)}{\tanh\frac{1}{2}\left(\theta+\frac{\ri\pi B}{2}\right)}.\label{smatrix}
\end{equation}

The scattering matrix, hence the kernel, depend on the parameter $B \in [0,2]$ known as the effective coupling constant. This parameter
 is related to the coupling constant $\beta$ (not related to the inverse temperatures $\beta_l,\beta_r$ !) in the
sinh-Gordon Lagrangian \cite{toda2,toda1}
\begin{equation}\label{lagran}
    \mathcal{L}=\frac{1}{2}(\partial_{\mu}\phi)^2-\frac{m^2}{\beta^2}\cosh(\beta\phi),
\end{equation}
where $m$ is a mass scale and $\phi$ is the sinh-Gordon field, as
\begin{equation}\label{BB}
    B(\beta)=\frac{2\beta^2}{8\pi + \beta^2},
\end{equation}
under CFT normalization \cite{za}. The $S$-matrix is obviously
invariant under the transformation $B\rightarrow 2-B$, a symmetry
which is also referred to as weak-strong coupling duality, as it
corresponds to $B(\beta)\rightarrow B(\beta^{-1})$ in (\ref{BB}).
The point $B=1$ is known as the self-dual point.
In our numerics for simplicity we will concentrate on the $B=1$ case for which
\beq
\varphi(\theta)=2\, \text{sech} \,\theta.\label{sgker}
\eeq

The sinh-Gordon model is intimately related to the roaming trajectories model \cite{roaming}: the latter can be understood as a ``deformation" of the sinh-Gordon model. Its $S$-matrix is obtained by setting $B=1+\frac{2\ri \theta_0}{\pi}$ in (\ref{smatrix}) where $\theta_0 \in \mathbb{{R}}^+$. The resulting $S$-matrix famously satisfies all the required properties (e.g.~unitary and crossing symmetry) thus describing a new integrable model. The new kernel is then
\begin{equation}\label{kernew}
    \varphi(\theta)={\text{sech}(\theta-\theta_0)}+\text{sech}(\theta+\theta_0).
\end{equation}

The presence of the free parameter $\theta_0$ turns out to have a profound effect on the features of all TBA quantities (e.g.~pseudoenergies, $L$-functions, $c$-functions, etc). Indeed the roaming trajectories model's name refers to the special features of
the effective central charge $c_{\text{eff}}(r)$, which is a particular {\em $c$-function}, within the equilibrium TBA approach \cite{tba1} as
calculated in \cite{roaming}. For massive QFTs it is expected
that the function $c_{\text{eff}}(r)$, whose definition we recall in Section \ref{csection}, ``flows" from the value zero
in the infrared (large $r$) to a finite value in the ultraviolet
(small $r$). For many theories, including the sinh-Gordon model,
the constant value reached as $r\rightarrow 0$ is the effective central
charge of the underlying conformal field theory associated to the
model \cite{ceff1,ceff2,zuberceff} (which equals the CFT central charge in unitary models).
In this case, that theory is the free massless boson, a
conformal field theory with central charge $c=1$. Therefore, in
the sinh-Gordon model, the function $c_{\text{eff}}(r)$ flows from
the value zero to the value 1 as $r$ decreases.

Crucially, when the same
function $c_{\text{eff}}(r)$ is computed for the roaming trajectories model it shows a
very different behaviour. It still flows from the value 0 to the
value 1, but it does so by ``visiting" infinitely many
intermediate values of $c$ giving rise to a staircase (or
roaming) pattern. The values of $c$ that are visited
correspond exactly to the central charges of the unitary minimal
models ${\cal M}_p$ of conformal field theory,
\begin{equation}\label{cc}
	1-\frac{6}{p(p-1)},\quad \text{with} \quad p=3,4,5\ldots
\end{equation}
As we will discuss later, interesting staircase patterns for several non-equilibrium thermodynamic quantities can also
be found for this model.

Another observation made in \cite{roaming} is that the size of the
intermediate plateaux that the function $c_{\text{eff}}(r)$
develops at the values (\ref{cc}) is determined by the value of
$\theta_0$.  For
$\theta_0=0$ there is a single plateau at $c=1$, thus the usual
sinh-Gordon behaviour is recovered, whereas the plateaux at
(\ref{cc}) become more prominent as $\theta_0$ is increased. In
the limit $\theta_0\rightarrow \infty $ a single plateau at
$c=\frac{1}{2}$ remains which reflects the fact that the
$S$-matrix (\ref{smatrix}) becomes $-1$ in this limit, wherefore the
model reduces to the Ising field theory. This interesting limiting
behaviour was studied in \cite{Ahn:1993dm} using the form factor
approach.

Generalizations of the roaming trajectories model have been
constructed based on more complex theories in which case the associated functions
$c_{\text{eff}}(r)$ have staircase patterns which visit different families of CFTs \cite{staircase}.
Oher families of theories exist where this staircase patterns also arise naturally, albeit involving a finite number
of steps which are in one-to-one correspondence with the on-set of unstable particles in the spectrum. These theories are the homogeneous sine-Gordon models, whose $S$-matrix was constructed in \cite{smatrix}. TBA analysis of these models has yielded a variety of staircase patterns \cite{CastroAlvaredo:1999em,Julian,Dorey:2004qc} which we would also expect to see replicated in computations of the non-equilibrium normalized current and cumulants.

\subsection{The sine-Gordon model at reflectionless points}

We will now consider the sine-Gordon model. This is a more involved theory including several particle types and bound states. In general, the sine-Gordon model has a non-diagonal scattering matrix, which greatly complicates the TBA equations (in non-diagonal models, there is a different formulation that is more efficient \cite{DdV1,DdV2,DdV3,K1,K2}). These $S$-matrices depend however on a continuous parameter $\nu$ which for particular values leads to a diagonal theory. Those values correspond to so-called reflectionless points. At reflectionless points, the theory consists of two fundamental particles of mass $m$ usually called soliton ($s$) and antisoliton ($\bar{s}$) as well as a ``tower" of bound states of the former and of each other known as breathers. The number and masses of the breathers depend on the values of the coupling constant. The scattering matrices involving the soliton, antisoliton and first breather ($b$) are \cite{Zamolodchikov:1977yy}:
\begin{equation}
    S_{ss}^{ss}(\theta)=S_{\bar{s}\bar{s}}^{\bar{s}\bar{s}}(\theta)= \exp\left(\int_0^\infty \frac{dt}{t} \frac{\sinh\frac{(1-\nu)t}{2}}{\sinh \frac{\nu t}{2}\cosh t} \sinh\frac{t \theta}{\ri\pi}\right),
\end{equation}
\begin{equation}
    S_{s\bar{s}}^{s\bar{s}}(\theta)=S_{\bar{s}{s}}^{\bar{s}s}(\theta)= \frac{\sinh \frac{\theta}{\nu}}{\sinh \frac{\ri\pi-\theta}{\nu}} S_{ss}^{ss}(\theta),
\end{equation}
\begin{equation}
    S_{s\bar{s}}^{\bar{s}s}(\theta)=S_{\bar{s}{s}}^{{s}\bar{s}}(\theta)= \frac{\sinh \frac{\ri\pi}{\nu}}{\sinh \frac{\ri\pi-\theta}{\nu}} S_{ss}^{ss}(\theta),
\end{equation}
\begin{equation}
    S_{sb}^{sb}(\theta)=S_{\bar{s}b}^{\bar{s}b}(\theta)=\frac{\sinh\theta+\ri \sin \frac{\pi(1+\nu)}{2}}{\sinh\theta-\ri \sin\frac{\pi(1+\nu)}{2}},\qquad S_{bb}^{bb}(\theta)=\frac{\sinh\theta+\ri \sin {\pi\nu}}{\sinh\theta-\ri \sin {\pi\nu}}.
\end{equation}
As we can see, the $S$-matrix $ S_{s\bar{s}}^{\bar{s}s}(\theta)$ vanishes whenever $\nu^{-1}$ takes integer values; these are the reflectionless points. In those cases the $S$-matrices become in fact those of a well-known diagonal theory, namely the $D_{\nu^{-1}+1}$-minimal Toda theory (see e.g.~\cite{BCDS}).

Here we will be interested in the simplest reflectionless point, namely when $\nu=\frac{1}{2}$. In this case only the first breather exists and its mass is given by $m_b=\sqrt{2}m$.
The scattering amplitudes above simplify greatly and we obtain:
\begin{equation}
    S_{ss}^{ss}(\theta)=S_{\bar{s}\bar{s}}^{\bar{s}\bar{s}}(\theta)= S_{s\bar{s}}^{s\bar{s}}(\theta)=S_{\bar{s}{s}}^{\bar{s}s}(\theta)=-\frac{\sinh\frac{1}{2}\left(\theta+\frac{\ri\pi}{2}\right)}
{\sinh\frac{1}{2}\left(\theta-\frac{\ri\pi}{2}\right)}.
\end{equation}
As anticipated earlier
$
    S_{s\bar{s}}^{\bar{s}s}(\theta)=S_{\bar{s}{s}}^{{s}\bar{s}}(\theta)=0
$,
and
\begin{equation}
    S_{sb}^{sb}(\theta)=S_{\bar{s}b}^{\bar{s}b}(\theta)=\frac{\sinh\theta+\frac{\ri}{\sqrt{2}} }{\sinh\theta-\frac{\ri}{\sqrt{2}} },\qquad S_{bb}^{bb}(\theta)=\frac{\sinh\theta+\ri }{\sinh\theta-\ri }.
\end{equation}
The corresponding TBA kernels are
\begin{equation}
    \varphi_{ss}(\theta)=\varphi_{\bar{s}\bar{s}}(\theta)=\varphi_{s\bar{s}}(\theta)=\varphi_{\bar{s}{s}}(\theta)=\frac{1}{2}\varphi_{bb}(\theta)=
    -\text{sech}\theta,
\end{equation}
and
\begin{equation}
    \varphi_{sb}(\theta)=\varphi_{\bar{s}b}(\theta)=-2\sqrt{2}\,{\cosh\theta}\;{\text{sech} 2\theta}.
\end{equation}
The sine-Gordon model describes the scaling limit of a well-known spin chain system, namely the XXZ quantum spin chain in the gapped regime. We hope therefore that some of our results for this model may be comparable to spin chain results.

\section{$L$-functions and current in non-equilibrium TBA}\label{Lfunctions}

In this section we present results for the $L$-functions and current of several models obtained numerically by solving equations (\ref{exactJ}). We will start with one of the simplest interacting integrable QFTs: the sinh-Gordon model. Results for this model will provide a benchmark for all the general features of the current and $L$-functions. In addition the subtle differences between the equilibrium and non-equilibrium quantities will become apparent when analysing this model. Throughout this and later secions, we will use the variable
\[
	\sigma = \frc{\beta_r}{\beta_l} = \frc{T_l}{T_r}.
\]

\subsection{The sinh-Gordon model}
\begin{figure}[h!]
 \includegraphics[width=7.5cm]{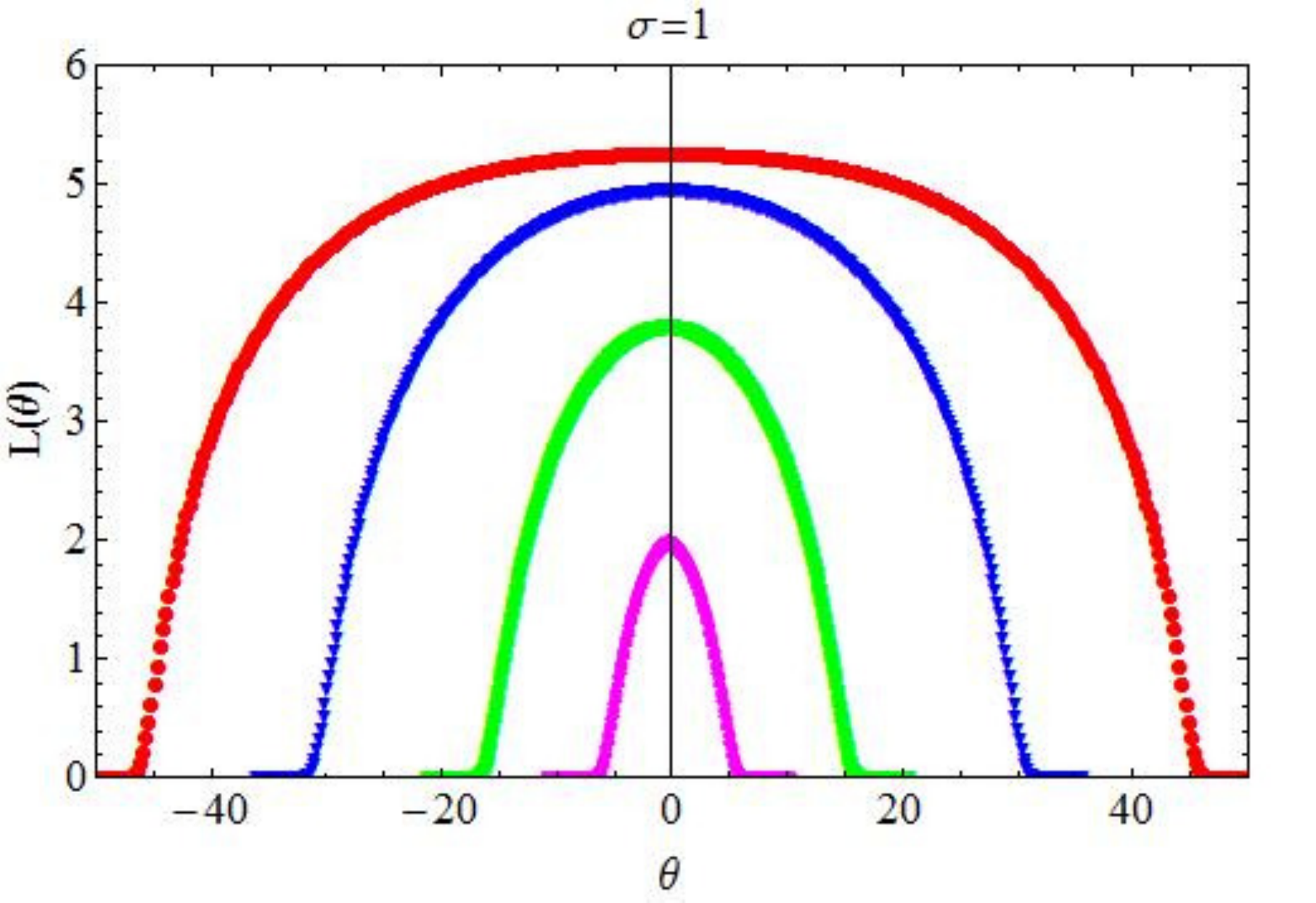}
\includegraphics[width=7.5 cm]{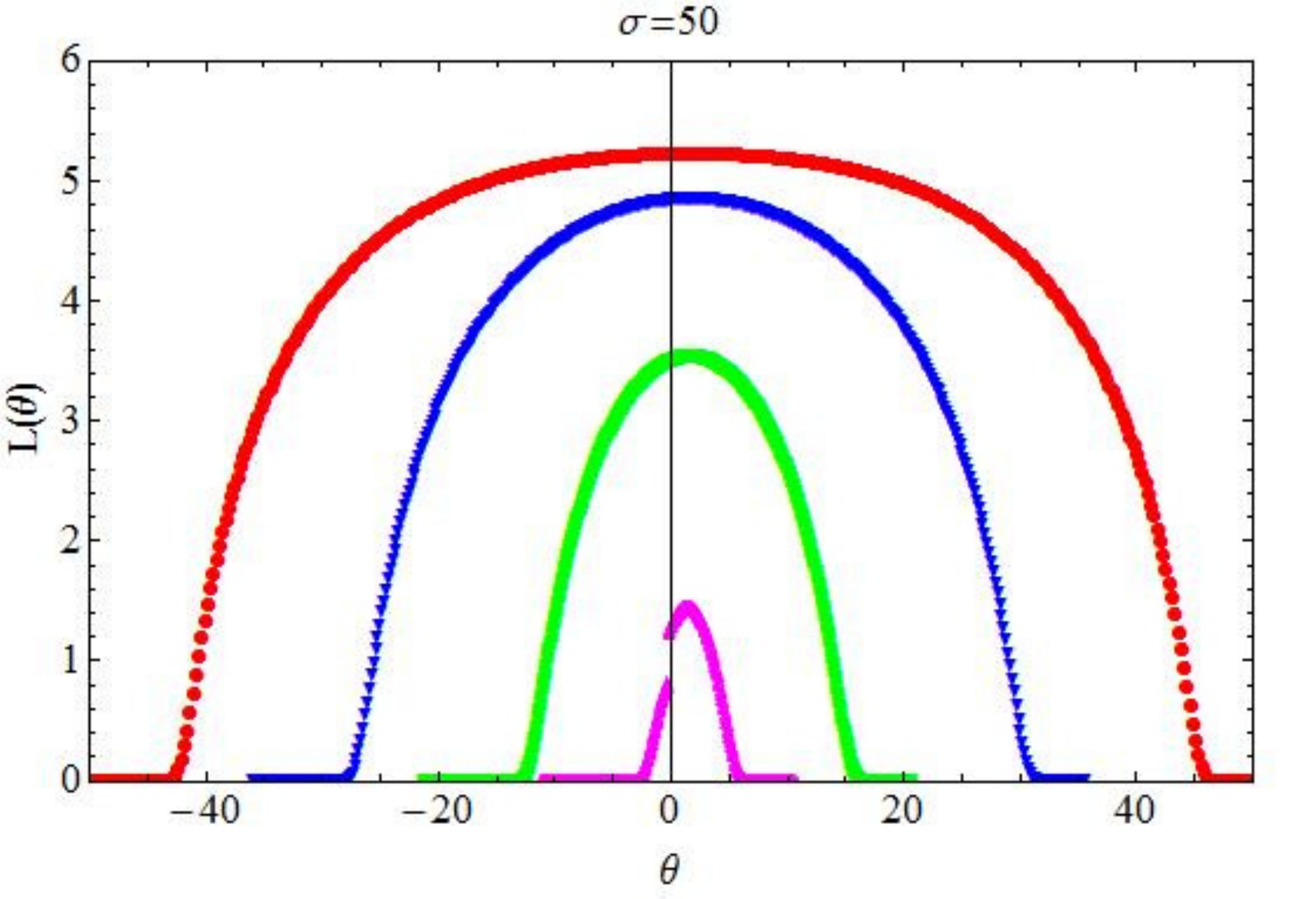}
\caption{The function $L(\theta)$ for $m\beta_l=6e^{-20}$ (red), $m\beta_l=2e^{-13}$ (blue), $m\beta_l=6e^{-7}$ (green) and $m\beta_l=2e^{-5}$ (pink) and two values of  $\sigma=\beta_r/\beta_l$. }\label{ll}
\end{figure}

Since in this model we only have one particle type we will for now drop the particle indices in \eqref{exactJ}. Employing the kernel (\ref{sgker}) in the equations (\ref{exactJ}) we can now solve for $\epsilon(\theta)$  and $L(\theta)$ by applying a numerical recursive algorithm starting with the free solution $\epsilon(\theta)=W(\theta)$. Figure \ref{ll} gives the $L$-functions for different values of $m\beta_l$ and $m\beta_r$, where $m$ is the mass of the particle.\\
The first figure with $\sigma=1$, that is $\beta_l=\beta_r$ gives the equilibrium quantities which are known from solving the standard TBA equations. It is a common observation from TBA studies that both the pseudoenergies and $L$-functions develop a plateau at high energies in the region $-\log\frac{2}{m\beta}\ll \theta \ll \log\frac{2}{m\beta}$ when $\beta=\beta_l=\beta_r$. For most integrable models the values of $L(\theta)$ and $\epsilon(\theta)$ at the plateau can be obtained exactly by solving so-called constant TBA equations which are a high-energy limit of the original equations \cite{tba1,tba2}.

From the figures above it appears that we start to see this plateau in the red curves. However, the sinh-Gordon model is rather an exception in this respect as in the limit $m\beta\rightarrow 0$ the value of $\epsilon(0)$ tends to minus infinity, wherefore $L(0)$ tends to infinity. This fact has been understood more generally in \cite{FKS,FK} where is was noted that it is a feature of all affine Toda field theories (of which the sinh-Gordon model is but the simplest example). We will see later that for other models a clear and finite plateau is reached. Comparing the equilibrium case ($\sigma=1$) to the case when $\sigma=50$ thus $\beta_l\neq \beta_r$ we observe two main changes:
\begin{itemize}
  \item The $L$-functions develop a discontinuity at $\theta=0$. This is because of the discontinuity in the function $W(\theta)$ due to the different left and right temperatures. This discontinuity is present for all temperatures and the size of the jump is linear $\beta_l-\beta_r$ (see \eqref{discep}). Thus when both temperatures are high no discontinuity can be seen. However it is very apparent for relatively large values of $m\beta_l$ as can be seen in the pink curve of the second figure.
  \item The $L$-functions cease to be even functions of $\theta$. Again this can be best appreciated for low energies where it is clear that the maximum of the $L$-functions is not located at the origin. Comparing to the equilibrium case, all functions have experienced a shift towards the right. For high energies we have again a plateau of the same height as in the equilibrium case which now extends in the region $-\log\frac{2}{m\beta_r}\ll \theta \ll \log\frac{2}{m\beta_l}$.
\end{itemize}
In summary, for $\sigma>1$ we observe a discontinuity at the origin and a shift towards the right of the $L$-functions. Due to the symmetry of the TBA equations a shift towards the left will occur for $\sigma<1$.

\begin{figure}[h!]
\begin{center}
  \includegraphics[width=7.7cm]{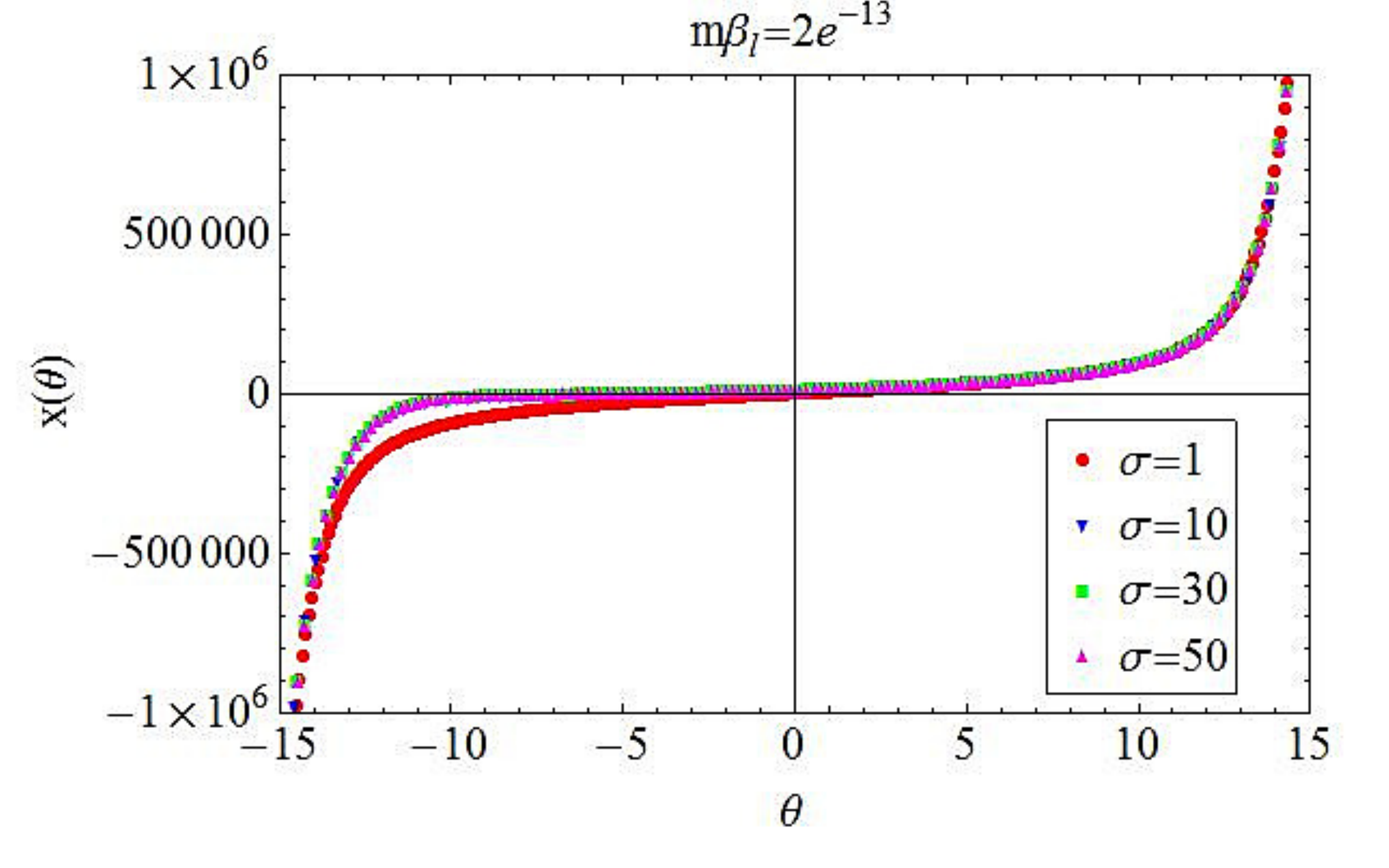}
  \includegraphics[width=7.7cm]{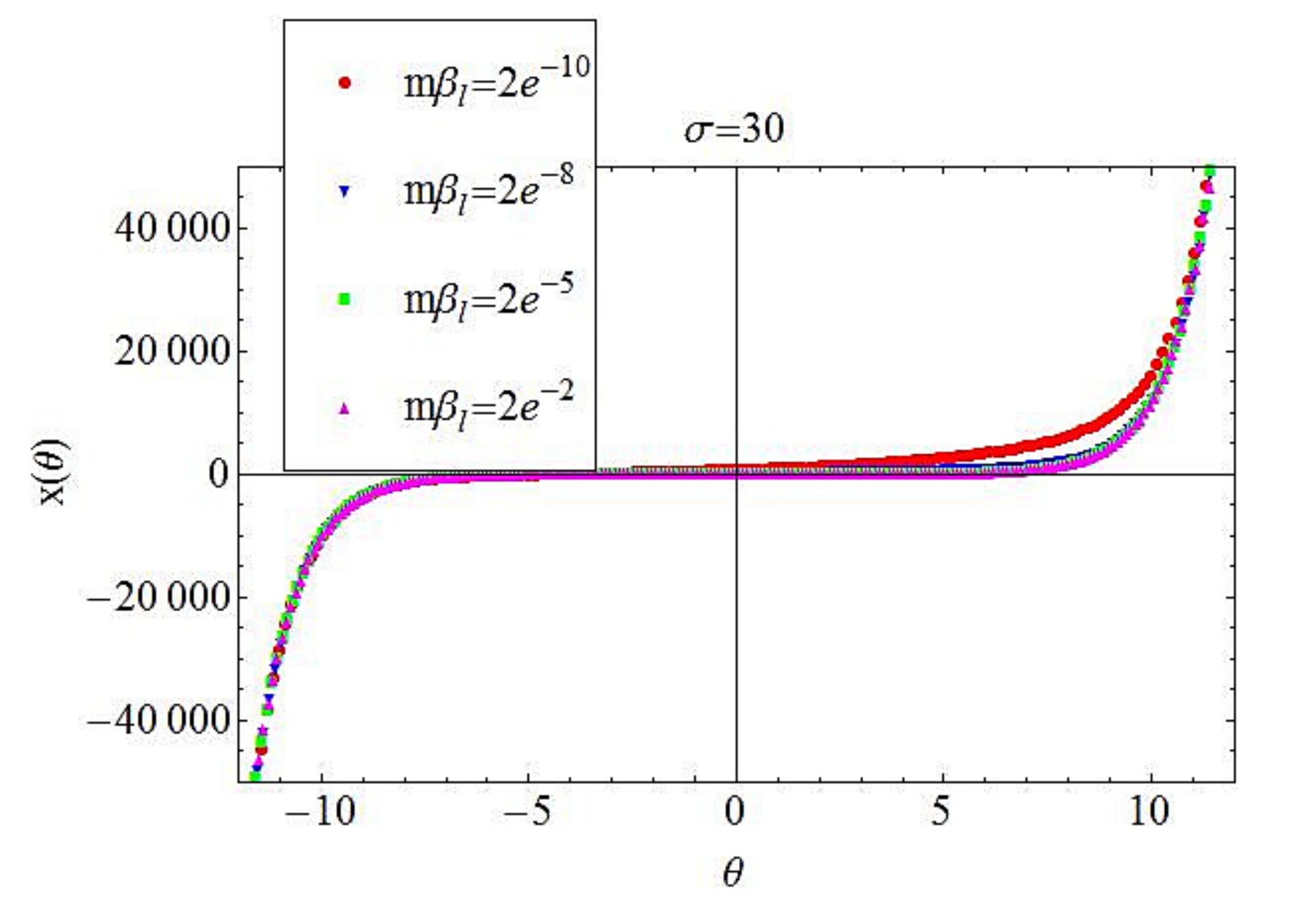}
\caption{The function $x(\theta)$ for fixed $m\beta_l$ and various values of $\sigma$ (first figure) and for fixed $\sigma$ and various values of $m\beta_l$ (second figure).}\label{x0}
\end{center}
\end{figure}

Let us now turn our attention to the functions $x(\theta)$ and $\mu(\theta)$. Numerically solving (\ref{exactJ}) we obtain $x(\theta)$ as shown in figure \ref{x0}.
 As expected we see that the large scale features of all functions are dominated by the $\sinh \theta$ function in the $x$-equation of (\ref{exactJ}).  However, because $ \epsilon(\theta)\neq \epsilon(-\theta)$ we find that in general $x(\theta) \neq - x(-\theta)$. The oddness of the function $x(\theta)$ is only restored in two limiting cases: 1) at equilibrium ($\sigma=1$); 2) when both left and right temperatures tend to zero ($m\beta_l, m\beta_r$ large). For this reason all functions in the second figure (except the red-dotted one) are almost perfectly odd.

\begin{figure}[h!]
\begin{center}
  \includegraphics[width=7.7cm]{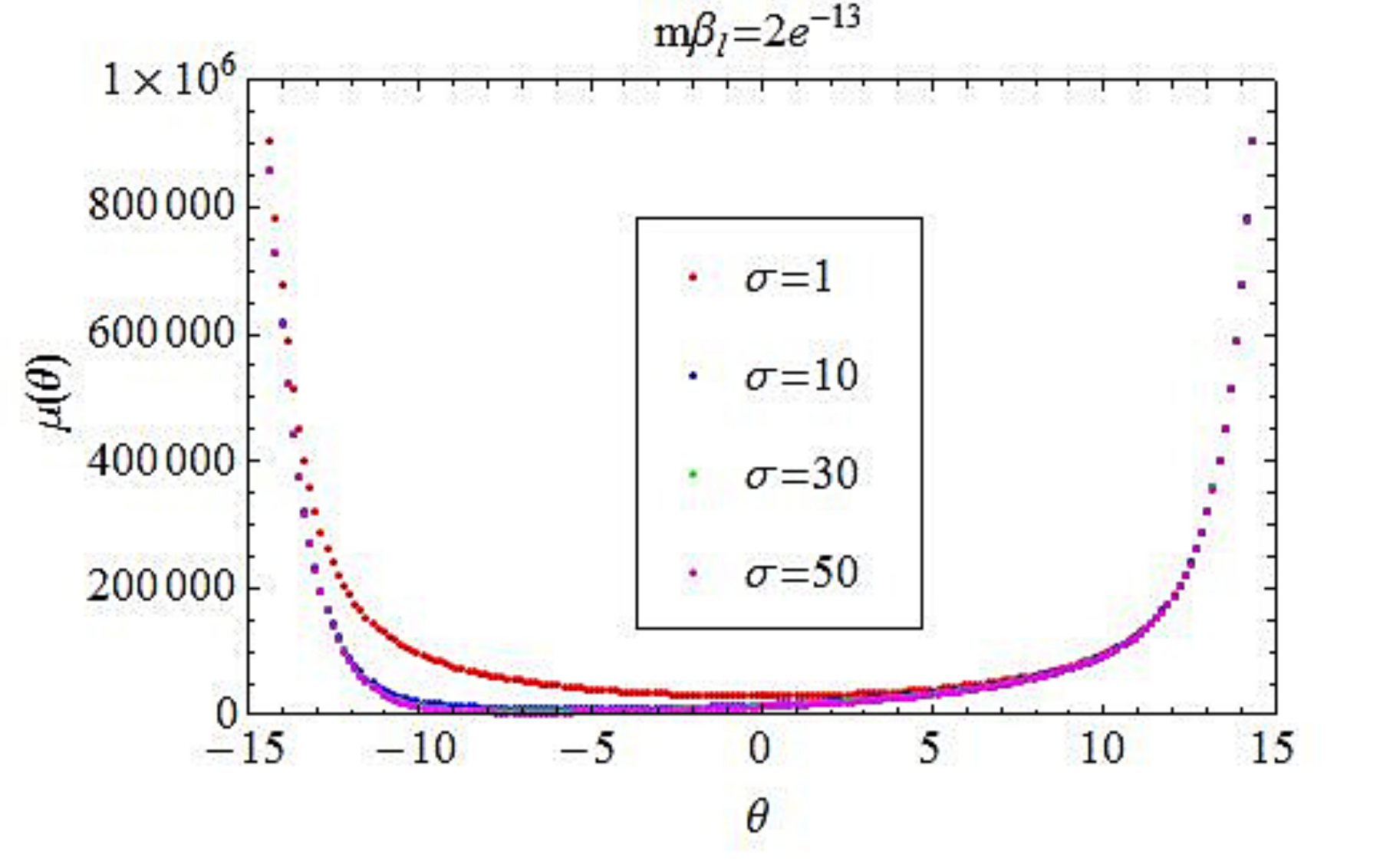}
  \includegraphics[width=7.7cm]{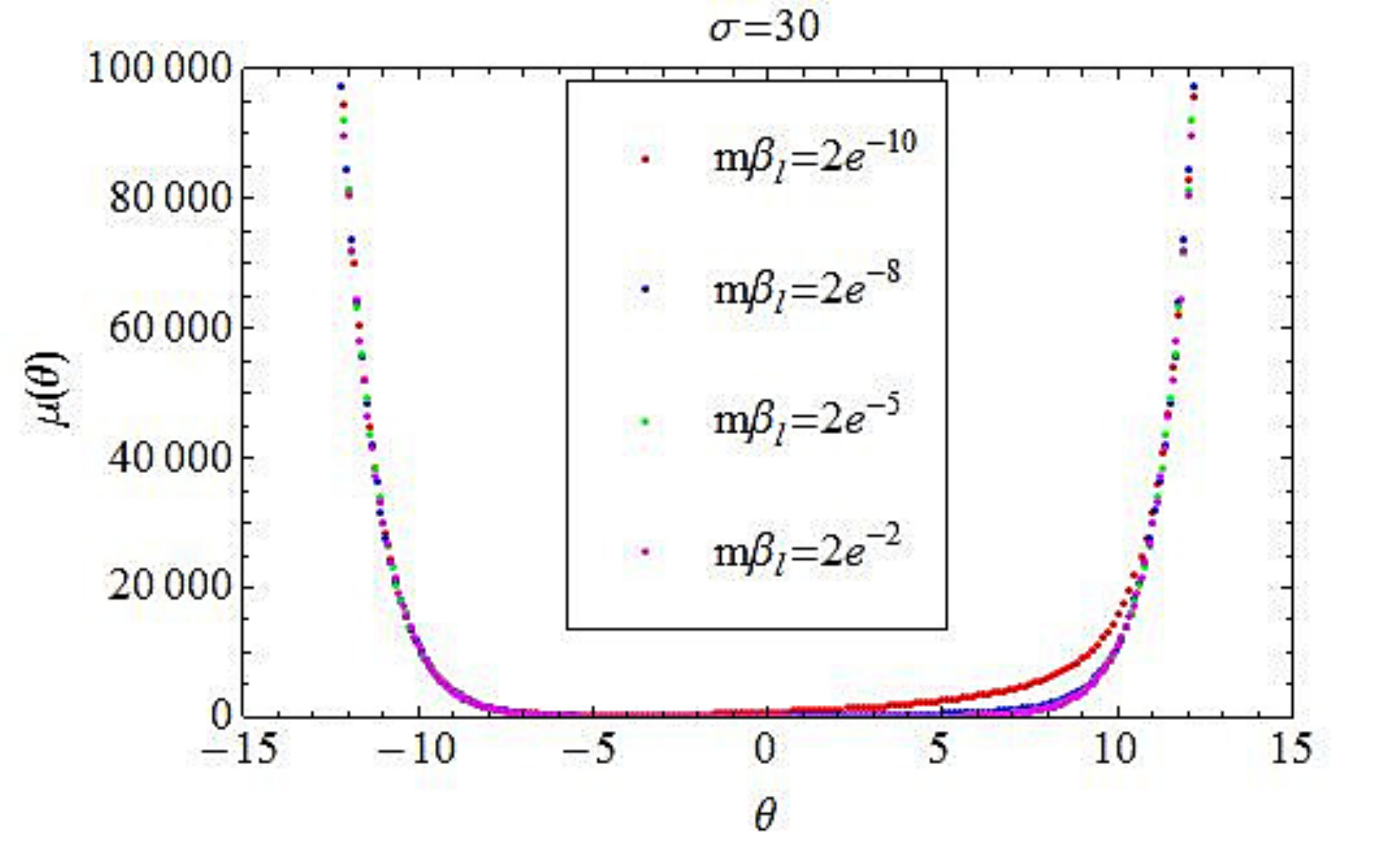}
\caption{The function $\mu(\theta)$ for fixed $m\beta_l$ and various values of $\sigma$ (first figure) and for fixed $\sigma$ and various values of $m\beta_l$ (second figure).}\label{mu}
\end{center}
\end{figure}

 We have seen in the previous section that the sets of equations (\ref{exactJ}) and (\ref{exactJ2}) are in fact equivalent to each other.  Figure~\ref{mu} provides several examples of the function $\mu(\theta)$ for the sinh-Gordon model.
 The features of the level density $\mu(\theta)$ are analogous to those of $x(\theta)$ with the difference that the large scale features are now dictated by the $\cosh\theta$ term in (\ref{exactJ2}). As a consequence all figures look like deformed $\cosh$-functions where the evenness is broken for $\sigma\neq 1$ and large temperatures. Recalling that $\mu(\theta)$ is proportional to the density of energy levels, we see that more energy levels are available at positive rapidities if $\sigma>1$, which agrees with the presence of a nonzero current from the left to the right (more right-moving energy levels available).
\begin{figure}[h!]
\begin{center}
\includegraphics[width=14cm]{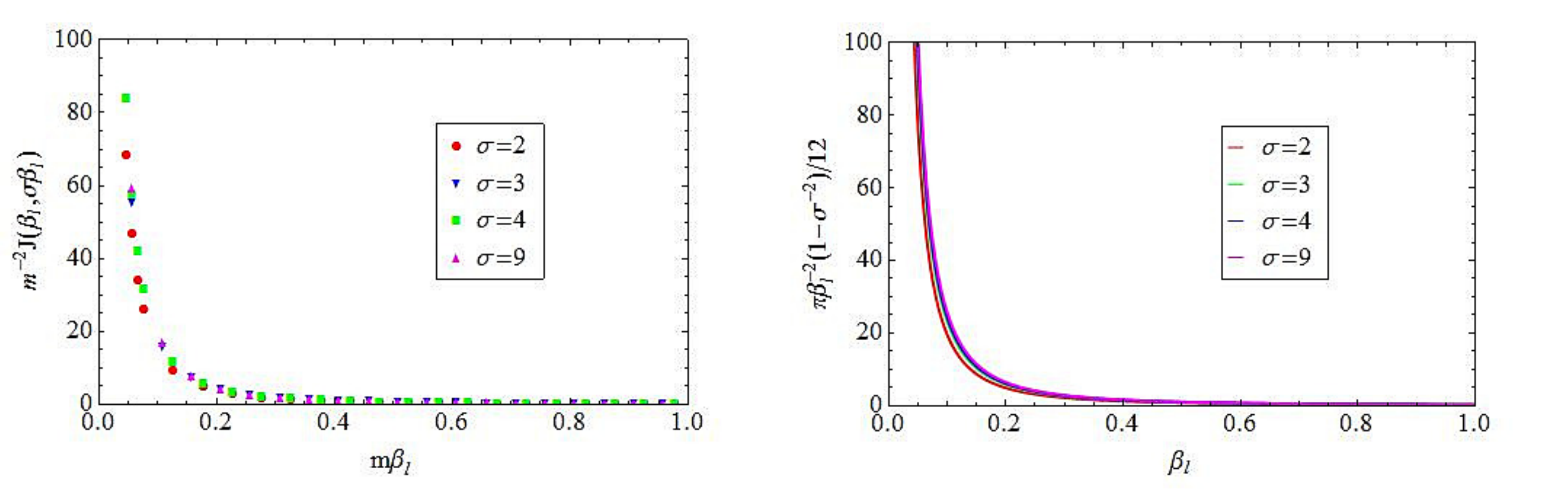}
\caption{The current $J(\beta_l, \sigma\beta_l)$ for the sinh-Gordon model and its CFT counterpart for several values of $\sigma$.  }\label{currentf}
\end{center}
\end{figure}

Finally, we present the results for the current in figure \ref{currentf}. As expected the current in the massive model and its CFT counterpart expression agree very closely for high temperatures, and both tend to zero for low temperatures and the disagreement also appears small in that region.

\subsection{The roaming trajectories model}
\begin{figure}[h!]
\begin{center}
  \includegraphics[width=7.8cm]{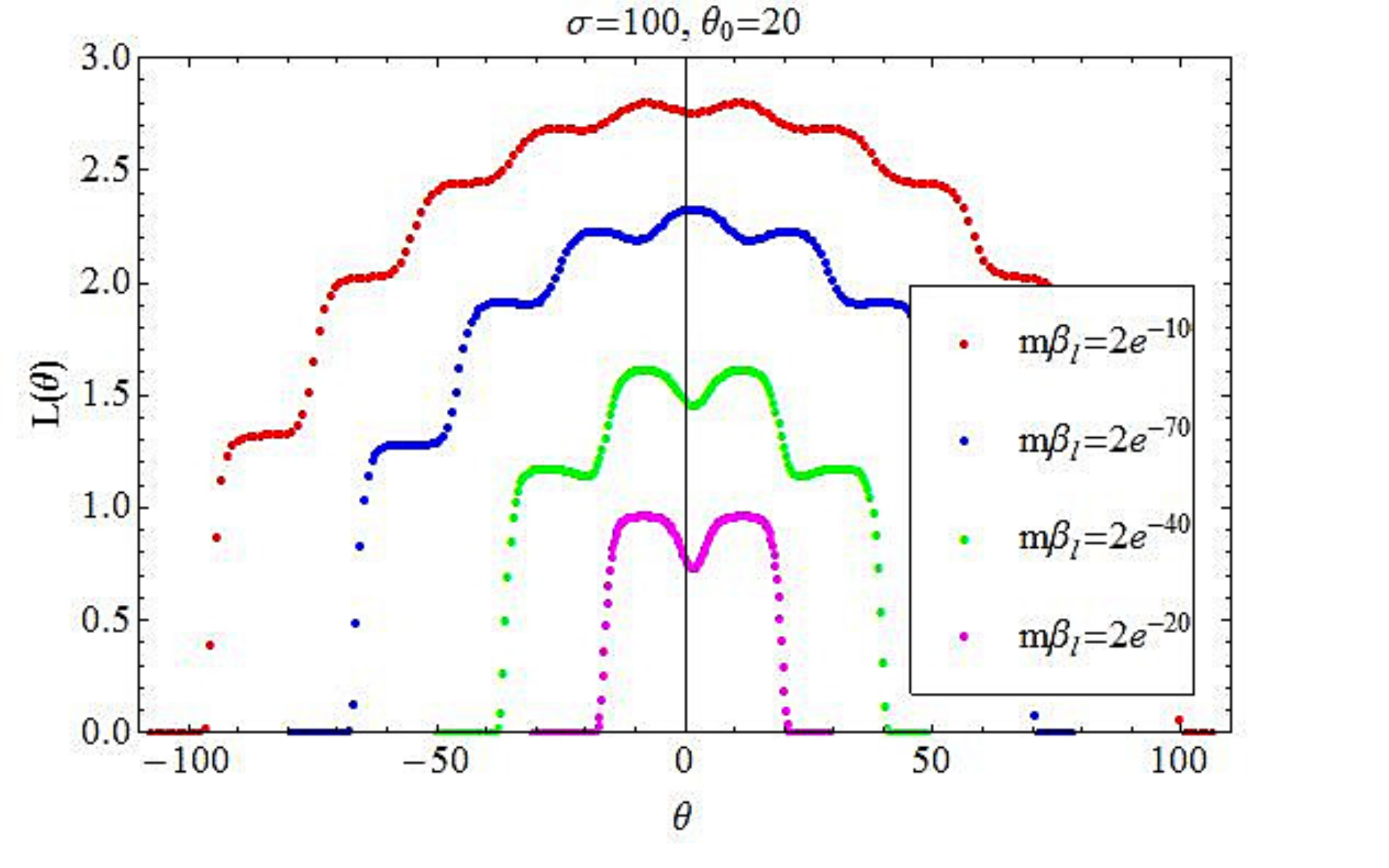}
 \includegraphics[width=7.8cm]{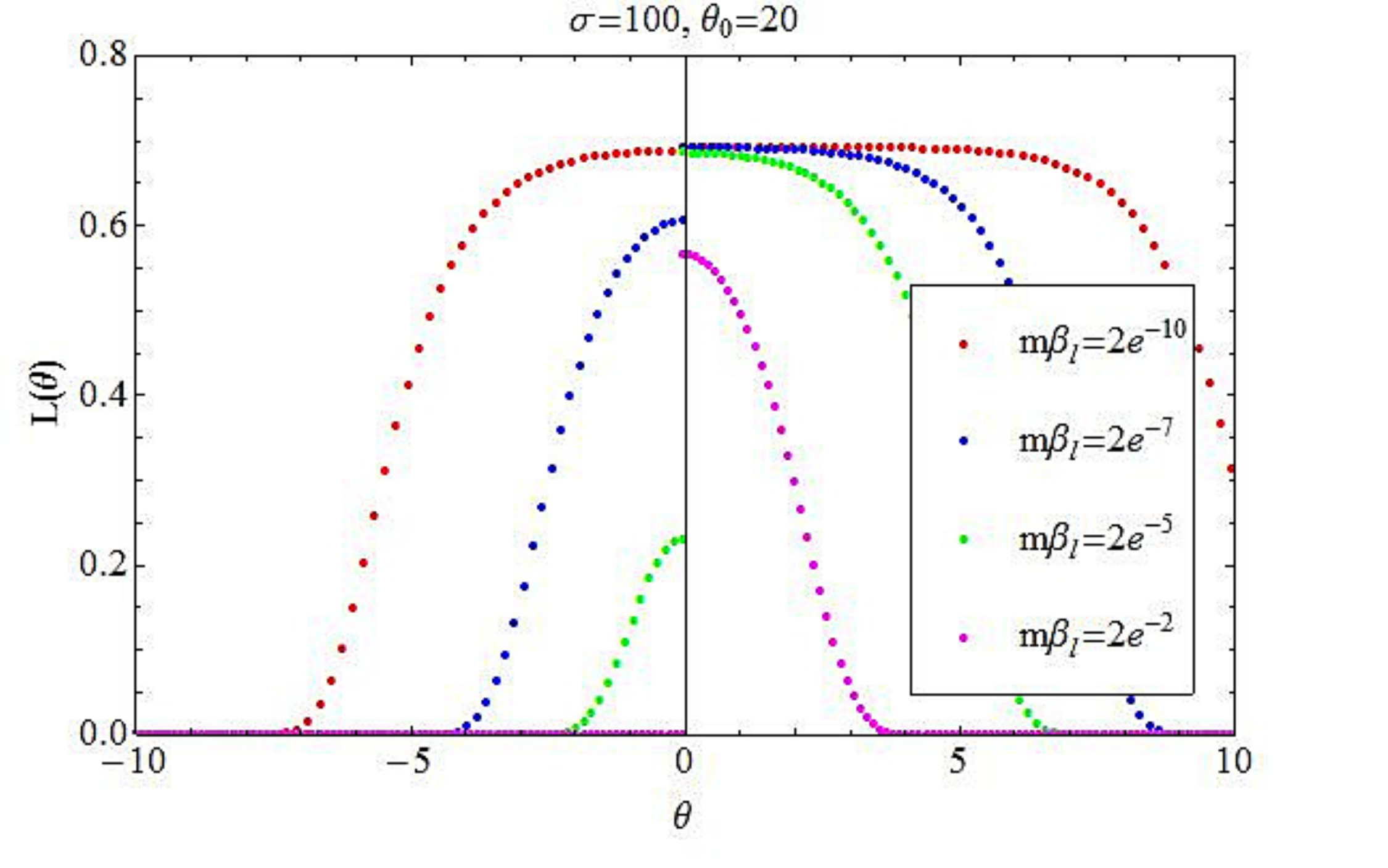}
\caption{$L$-function for various temperatures in the roaming trajectories model with $\theta_0=20$ and $\sigma=100$.}\label{roamingL}
\end{center}
\end{figure}

\begin{figure}[h!]
\begin{center}
  \includegraphics[width=7.5cm]{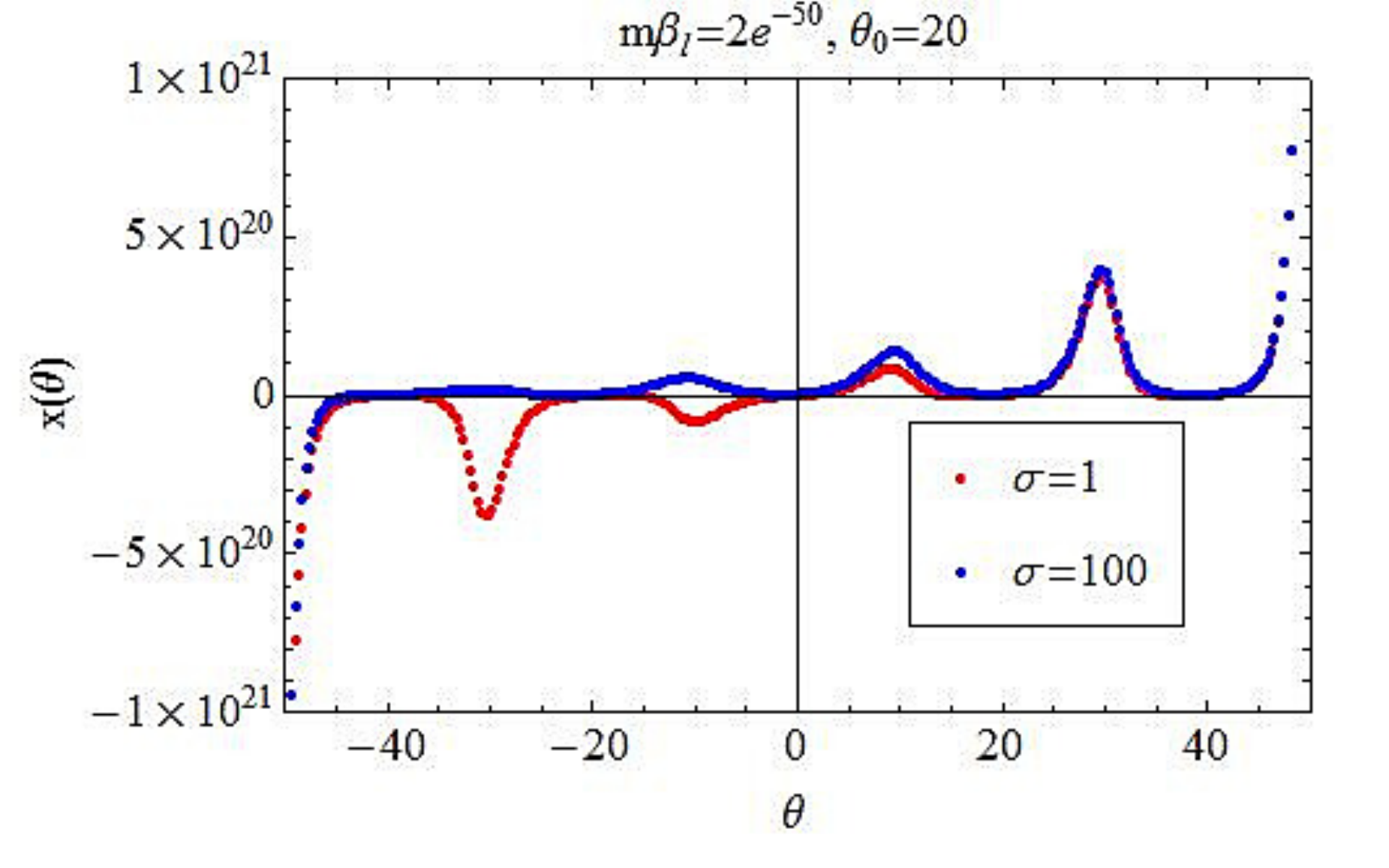}
  \includegraphics[width=7.5cm]{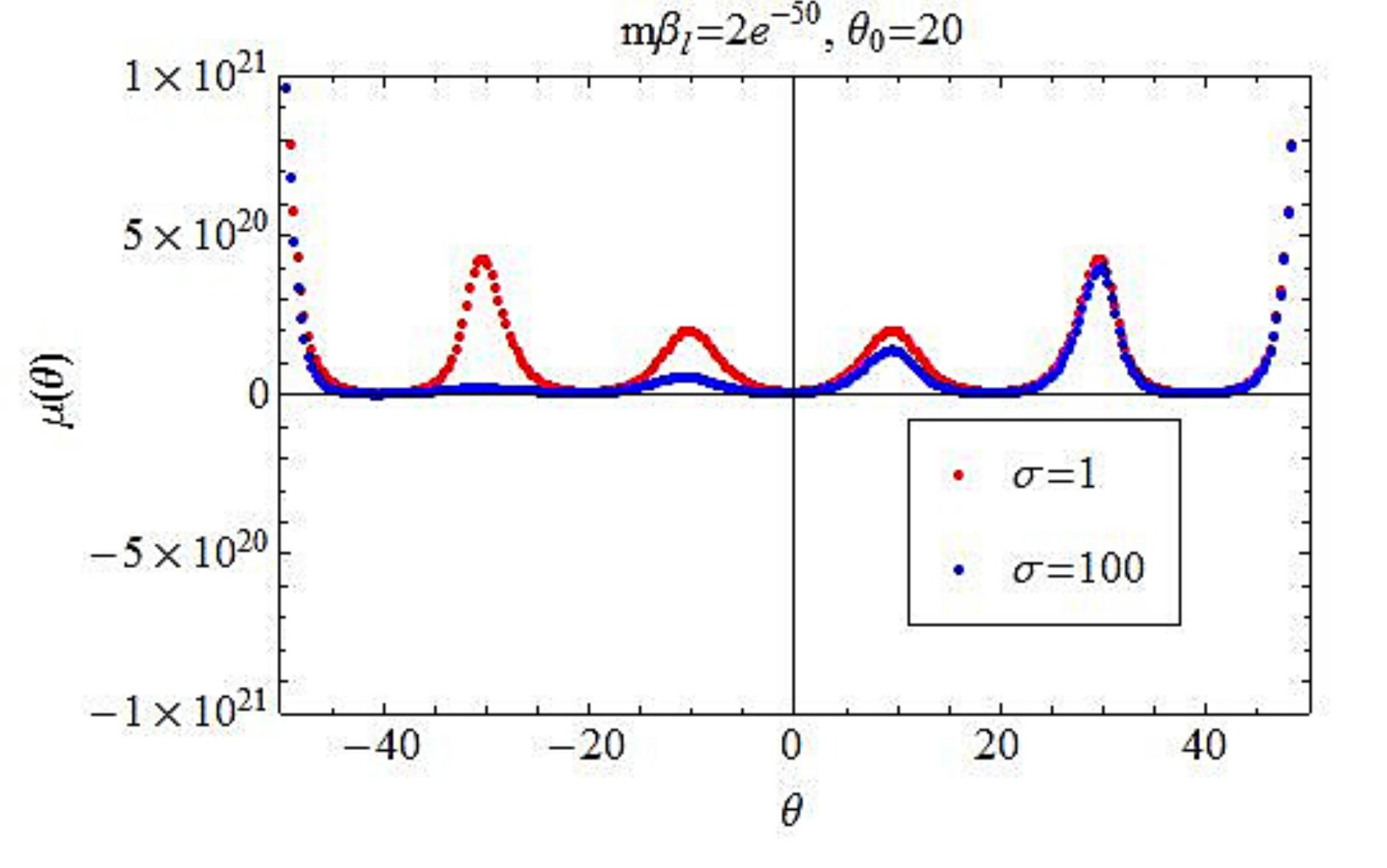}
\caption{The functions $x(\theta)$ and $\mu(\theta)$ at and out of equilibrium for $\theta_0=20$ and $m\beta_l=100$.}\label{roamingx0}
\end{center}
\end{figure}

As indicated above, the roaming trajectories model is an interesting deformation of the sinh-Gordon model (from the $S$-matrix point of view) which displays many new interesting features for all the relevant thermodynamic quantities we study here. Figures \ref{roamingL} and \ref{roamingx0} show the new structure displayed by $L$-functions and $x$-functions.
 The asymmetry of the $L$-functions with respect to the origin is visible in most of the graphs in figure~\ref{roamingL} as all curves appear shifted towards the right with respect to the equilibrium solutions presented in \cite{roaming}. For lower temperatures the discontinuity at the origin becomes apparent once more. The presence of the $\theta_0$ parameter leads to intricate staircase patters reminiscent of the equilibrium results.

The function $x(\theta)$ displays a very interesting new structure with an alternation of maxima and zeroes at values of $\theta$ which appear to be multiples of $\theta_0$. This structure is surprising at first sight as it is rather what we would expect the function $\frac{d\epsilon}{d\theta}$ to look like in view of figure \ref{roamingL}. However, $x(\theta)$ is a derivative w.r.t.~the parameter $a$, not $\theta$, as expressed in  the paragraph after \eqref{faapp} in Appendix \ref{appTBA}. This can be understood as follows. At equilibrium, when $\sigma=1$ the function $x(\theta)$ as defined by the second equation in (\ref{exactJ}) can be identified with $\frac{d\epsilon}{d\theta}$. In other words, if we differentiate the last equation in (\ref{exactJ}) with respect to $\theta$ at $\sigma=1$ we would exactly obtain the second equation in (\ref{exactJ}) up to the identification $x(\theta):=\frac{d\epsilon}{d\theta}$. This ceases to hold when $\sigma \neq 1$, however the equilibrium features of $x(\theta)$ are still preserved to a large extent with certain deformations. In particular for $\theta<0$ the $\sinh$-function now dominates the behaviour of $x(\theta)$ suppressing the negative minima of its equilibrium version. A similar deformation occurs for the function $\mu(\theta)$ where again the out-of-equilibrium features for $\theta<0$ are dominated by the $\cosh$-function.
\begin{figure}[h!]
\begin{center}
  \includegraphics[width=7.2cm]{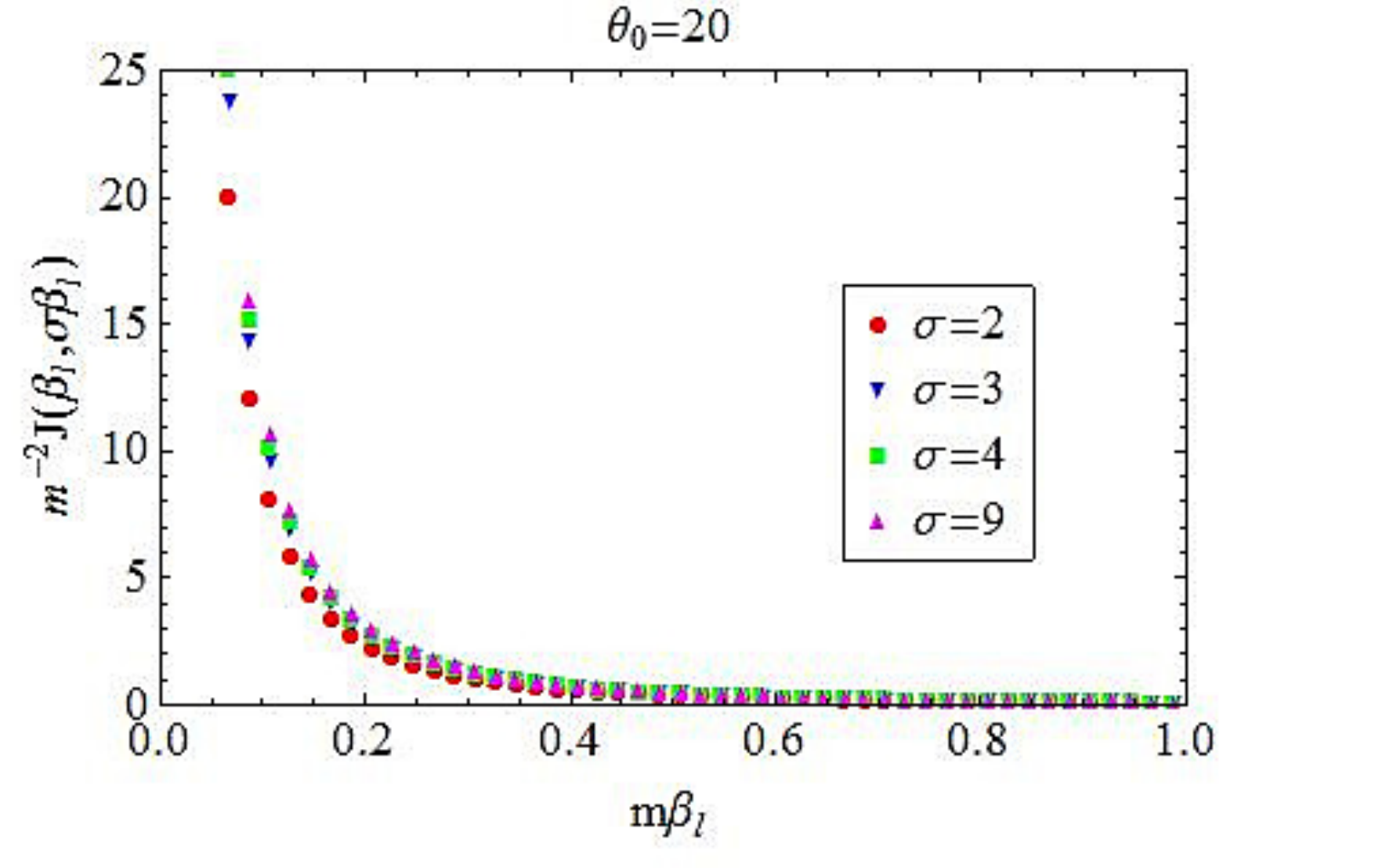}
\caption{The current for several values of $\sigma$.}\label{roamingJ}
\end{center}
\end{figure}

Finally, we take a look at the current in figure \ref{roamingJ}. Comparting to figure~\ref{currentf} we see that the functions seem hardly to have changed. There are however important differences which only become apparent when studying the normalized current (e.g.~the current divided by $T_l^2-T_r^2$) and associated cumulants. We will study these in the next section.

\subsection{The sine-Gordon model at reflectionless points}
\begin{figure}[h!]
\begin{center}
\includegraphics[width=7.2cm]{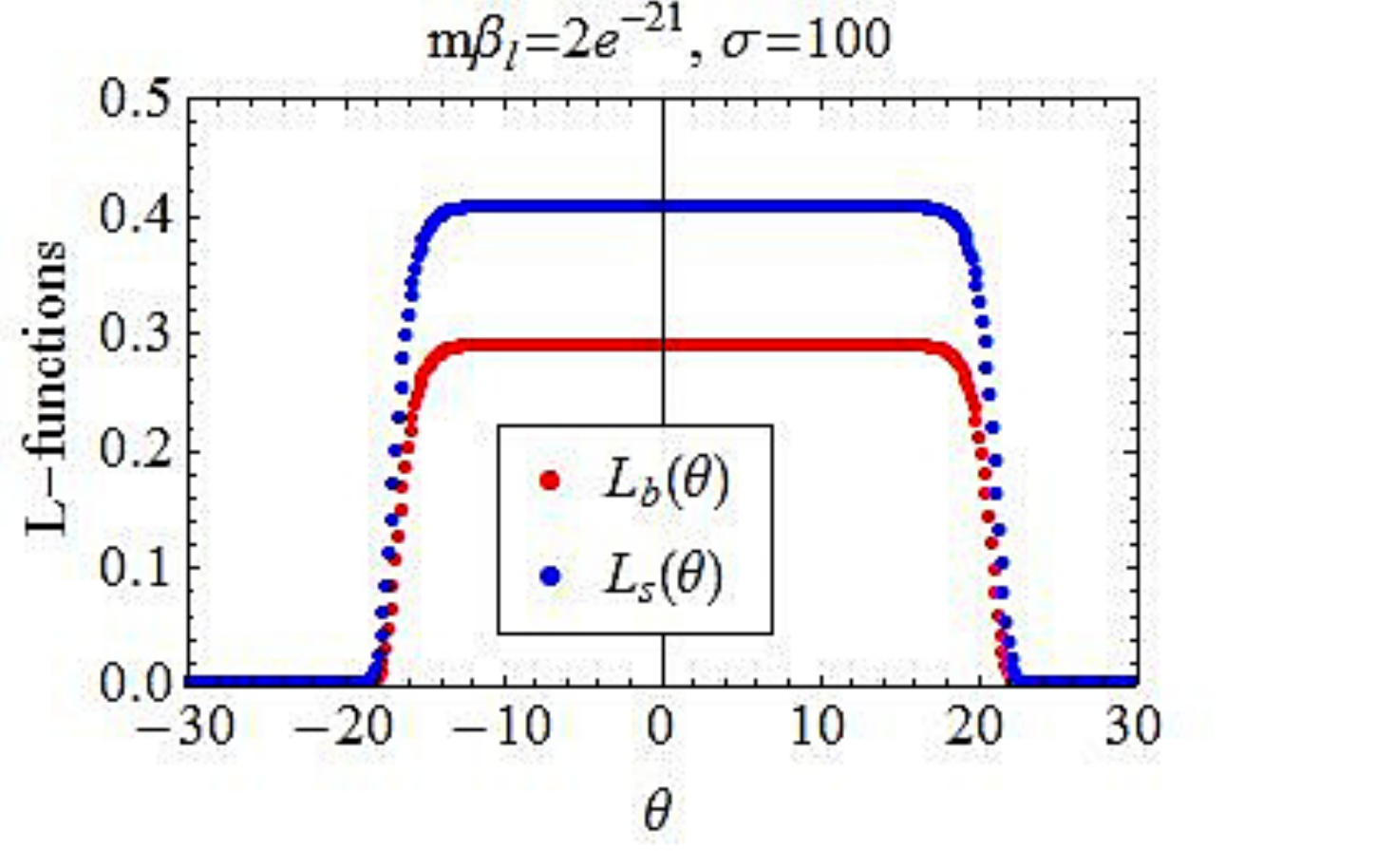}
  \includegraphics[width=7.2cm]{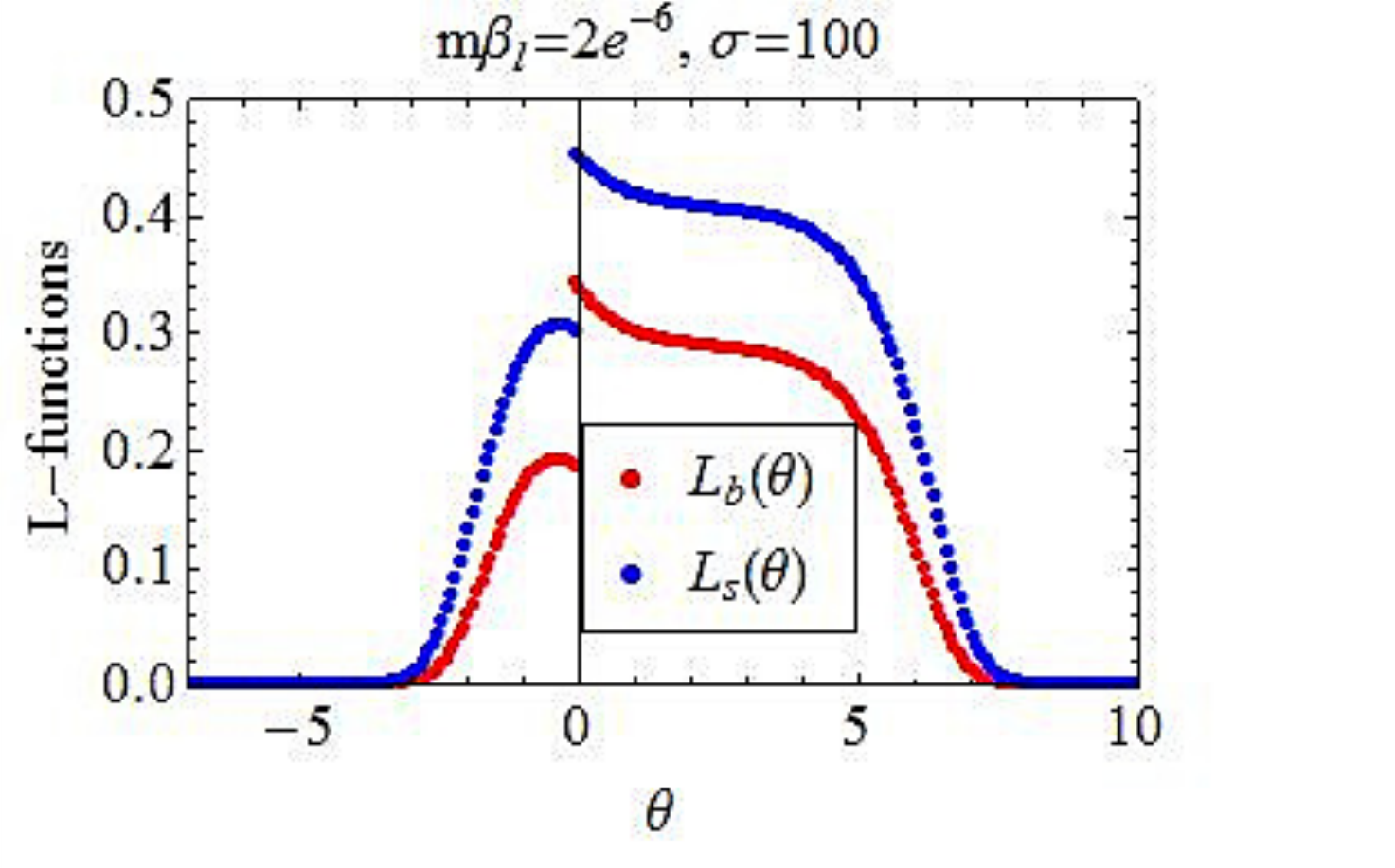}
\caption{Soliton and breather $L$-functions at high and medium temperatures.}\label{ld3}
\end{center}
\end{figure}

As explained earlier we will consider here only a very special case of the sine-Gordon model, namely the situation when the whole spectrum consists of soliton, antisolition and one breather and there is no back-scattering. Once more we will start by examining the general behaviour of the $L$-functions for several values of the energy. These are presented in figure \ref{ld3}. In this case the $L$-functions clearly develop a plateau for high energies in the region $-\log\frac{2}{m\beta_r}\ll \theta \ll \log\frac{2}{m\beta_l}$, which in the first figure corresponds approximately to $-16 \ll \theta \ll 21$. The exact height of the plateu is the same as in the equilibrium case which has been studied in \cite{elastic} by solving the constant TBA equations. The plateaux are located at $L_b(0)=0.287682$ and $L_s(0)=0.405465$ which correspond to the solutions $e^{\epsilon_s(0)}=2$ and $e^{\epsilon_b(0)}=3$.

\begin{figure}[h!]
\begin{center}
\includegraphics[width=7.2cm]{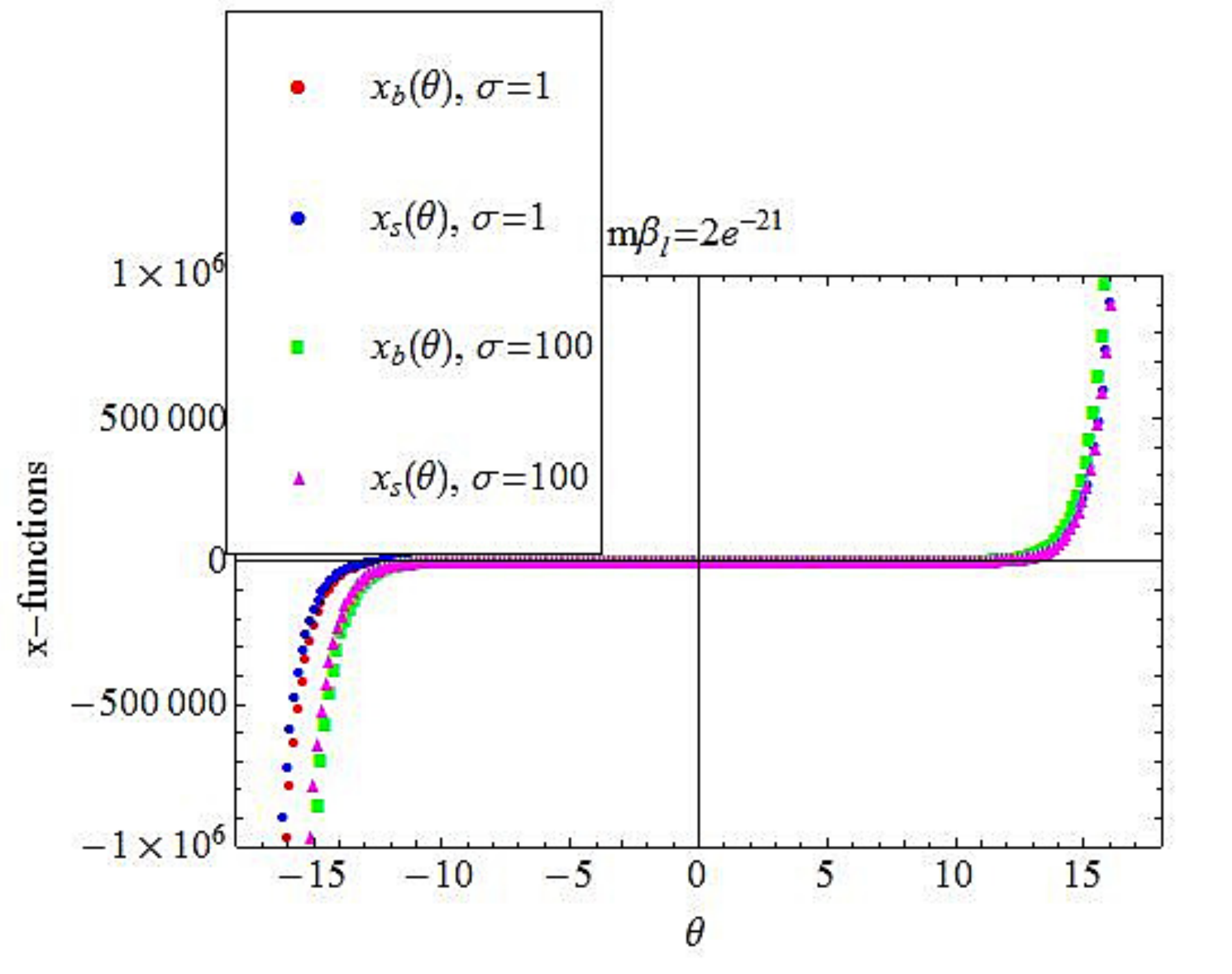}
\includegraphics[width=7.2cm]{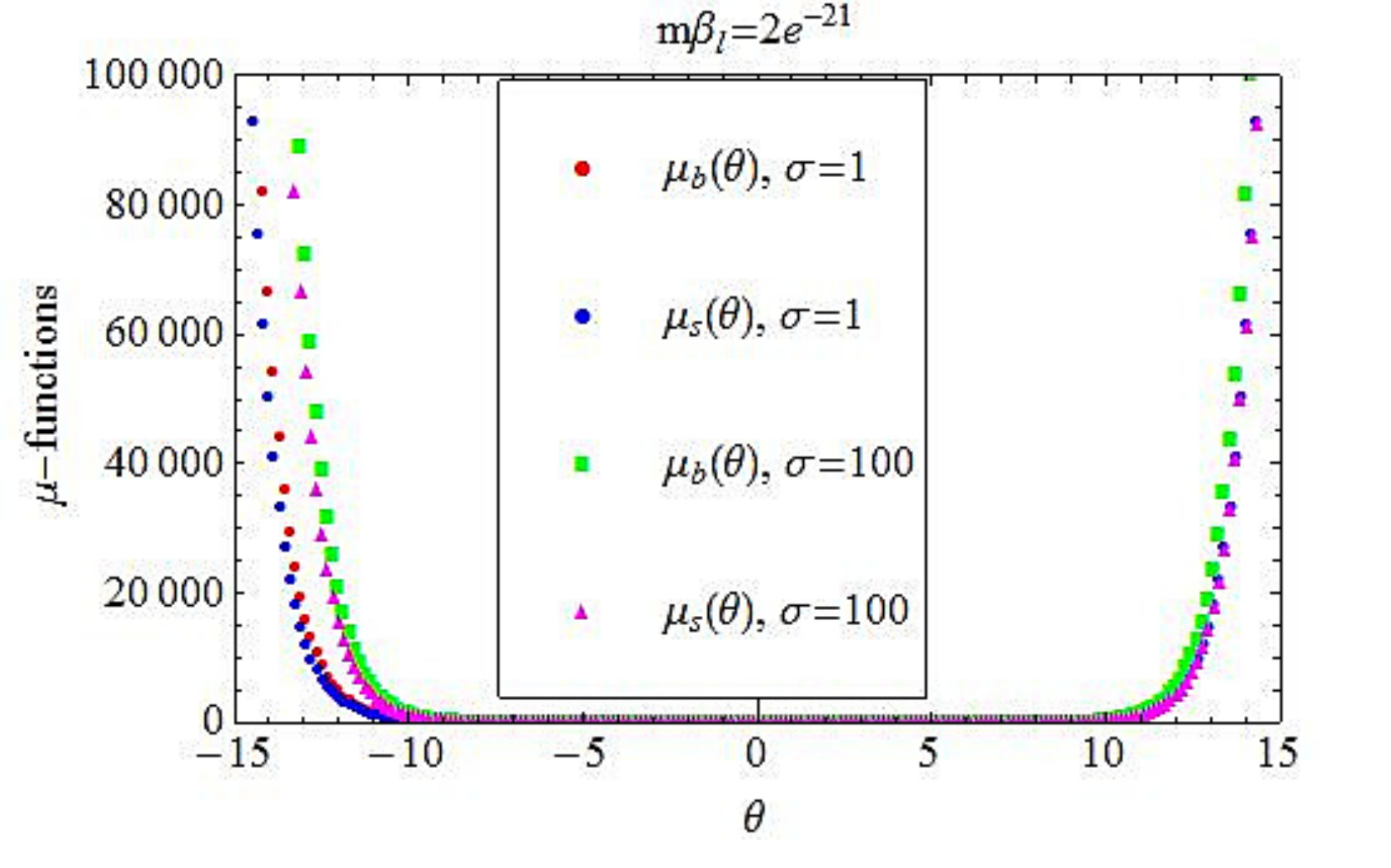}
\caption{The functions $x_{s,b}(\theta)$ and $\mu_{s,b}(\theta)$ for high temperatures.}\label{xmudn}
\end{center}
\end{figure}

The functions $x_{s,b}(\theta)$ and $\mu_{s,b}(\theta)$ in figure~\ref{xmudn} display the same main features already observed for other models. The same applies to the current which for this reason we will not present here.

\section{Non-equilibrium $c$-functions}\label{csection}

A $c$-function is a function of the energy scale (some physical quantity like the temperature or the distance in a correlation function), or of the position on a renormalization group (RG) trajectory, which represents well the effective number of degrees of freedom at this scale or at that point on the trajectory. In unitary models, it should satisfy 2 properties: it should be constant and equal to the CFT central charge at fixed points of the renormalization group (critical points), and it should monotonically increase with the energy scale. The original $c$-function was defined by A.B.~Zamolodchikov in \cite{Zamc} in terms of two-point correlation functions of the stress-tensor, where he used it to prove the inequality $c_{\rm UV}>c_{\rm IR}$ between central charges at UV and IR fixed points of an RG trajectory. Al.B.~Zamolodchikov defined a new $c$-function based on (equilibrium) TBA \cite{tba1}, sometimes referred to as the effective central charge or scaling function,
 \begin{equation}\label{oldc}
    c_{\text{eff}}(r):=\frac{3r}{\pi^2}\int_{-\infty}^\infty d\theta \cosh\theta \,
    \log(1+e^{-\ep(\theta)}),
\end{equation}
where $r$ is the distance scale or inverse temperature scale entering the TBA equations.

Given the behaviour of the current in CFT (\ref{JCFT}), we expect that the normalized current
 \begin{equation}\label{newc}
    c_1(T_l,T_r):=\frac{12 J(T_l^{-1},T_r^{-1})}{\pi(T_l^2-T_r^2)}=\frac{6}{\pi^2 (T_l^2-T_r^2)}\sum_{i=1}^\ell \int_{-\infty}^\infty d\theta\, m_i\cosh\theta \frac{x_i(\theta)}{1+e^{\epsilon_i(\theta)}}
\end{equation}
gives a function which approaches the value of the UV central charge at high temperatures or low masses, i.e.~near quantum critical points. Further, at very low temperatures, the current decays exponentially fast, wherefore $c_1(T_l,T_r)$ tends to zero. This means that $c_1(T_l,T_r)$ satisfies the first property of a $c$-function. It is interesting to enquiry if this function is actually a good $c$-function: that is, if it is also {\em monotonic with temperature scale}. Thanks to the CFT result \eqref{CnCFT} for the higher cumulants, one can similarly enquiry if the functions
\beq
	c_n(T_l,T_r) =  \frc{12\,C_n(T_l,T_r)}{\pi n! \,(T_l^{n+1}+(-1)^n T_r^{n+1})} \label{cfun}
\eeq
are good $c$-functions.
%\subsubsection{The sinh-Gordon and roaming trajectories model}
\begin{figure}[h!]
\begin{center}
 \includegraphics[width=7.2cm]{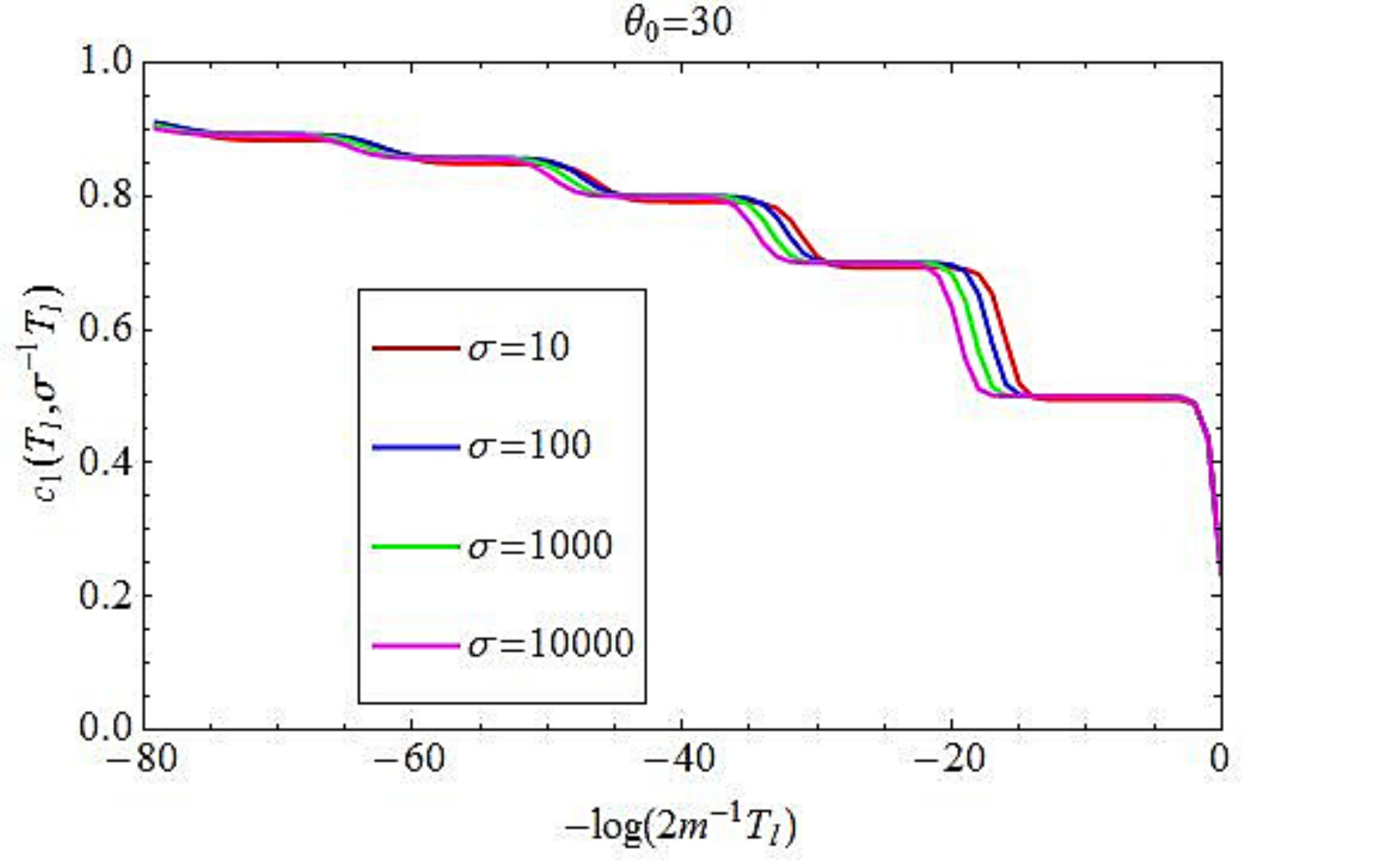}
\includegraphics[width=7.2cm]{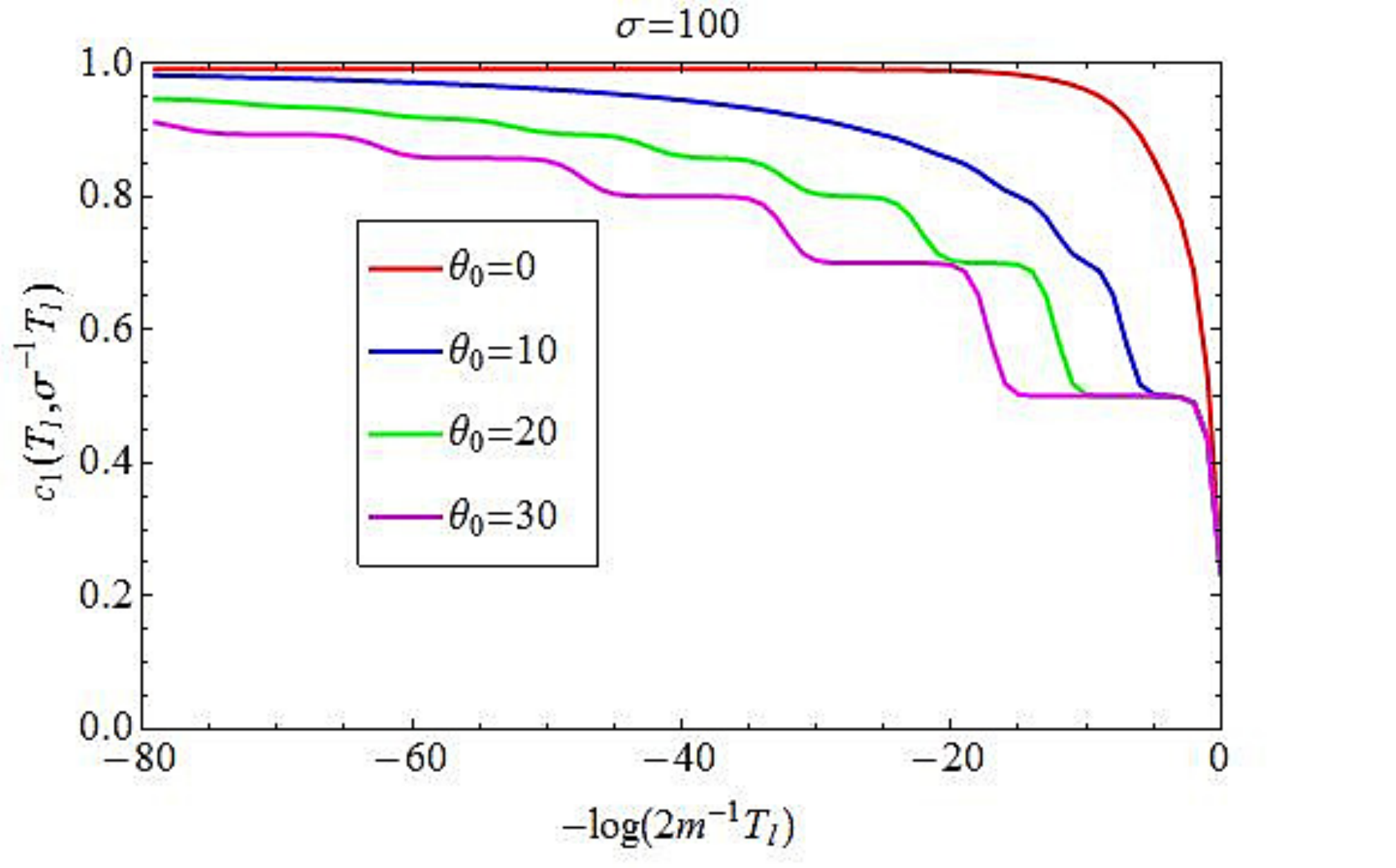}
 \includegraphics[width=7.2cm]{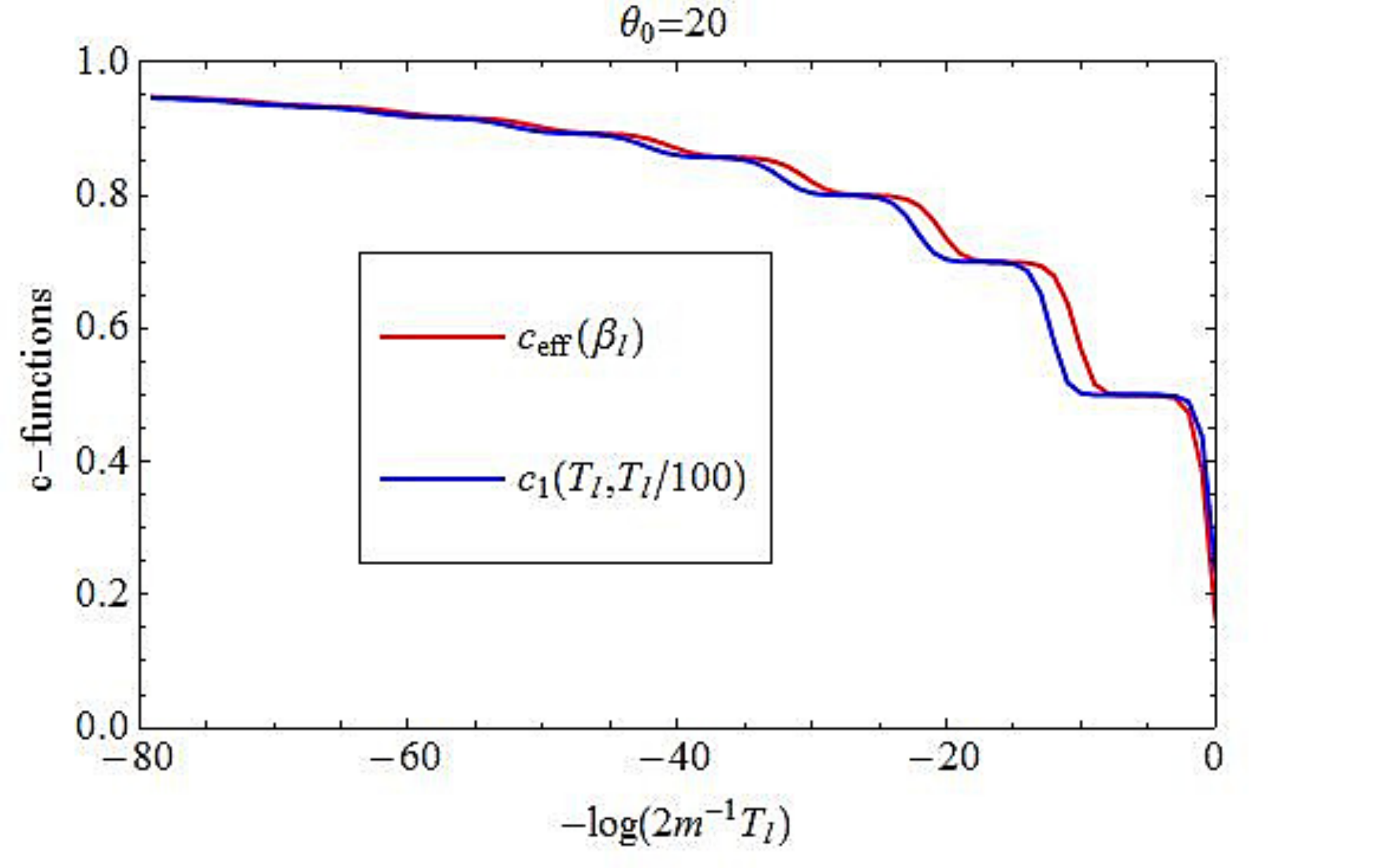}
 \includegraphics[width=7.2cm]{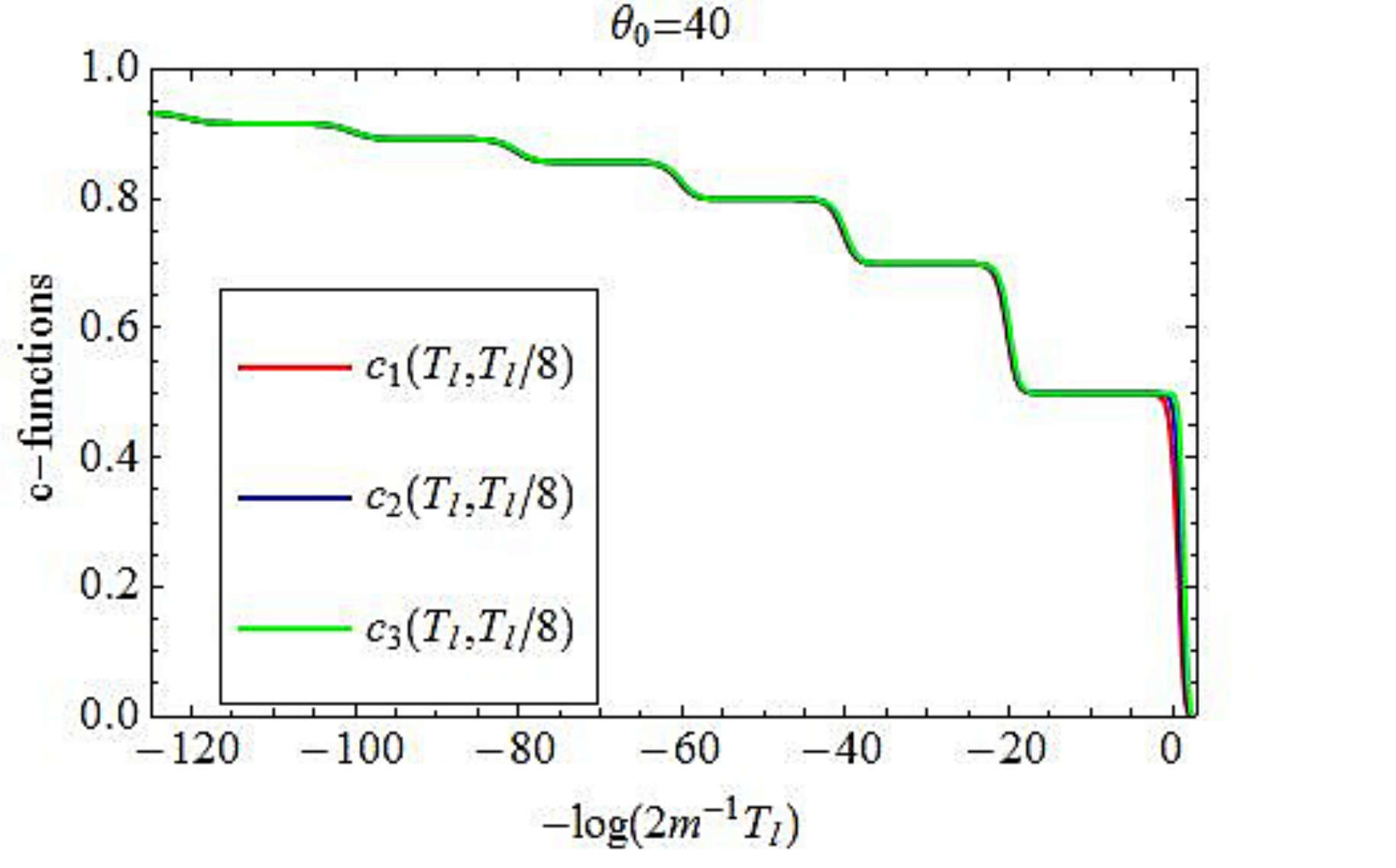}
 \includegraphics[width=7.2cm]{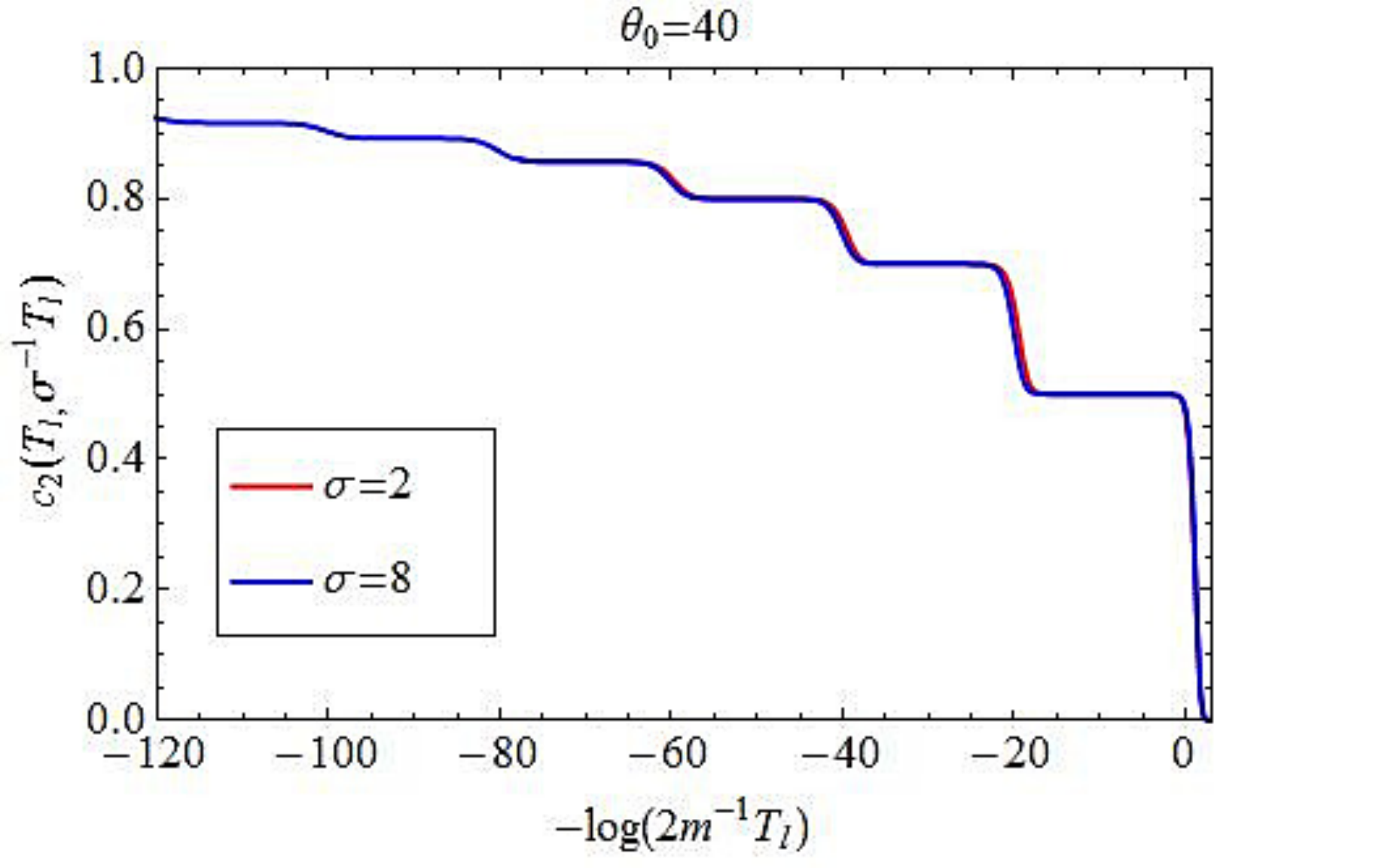}
 \includegraphics[width=7.5cm]{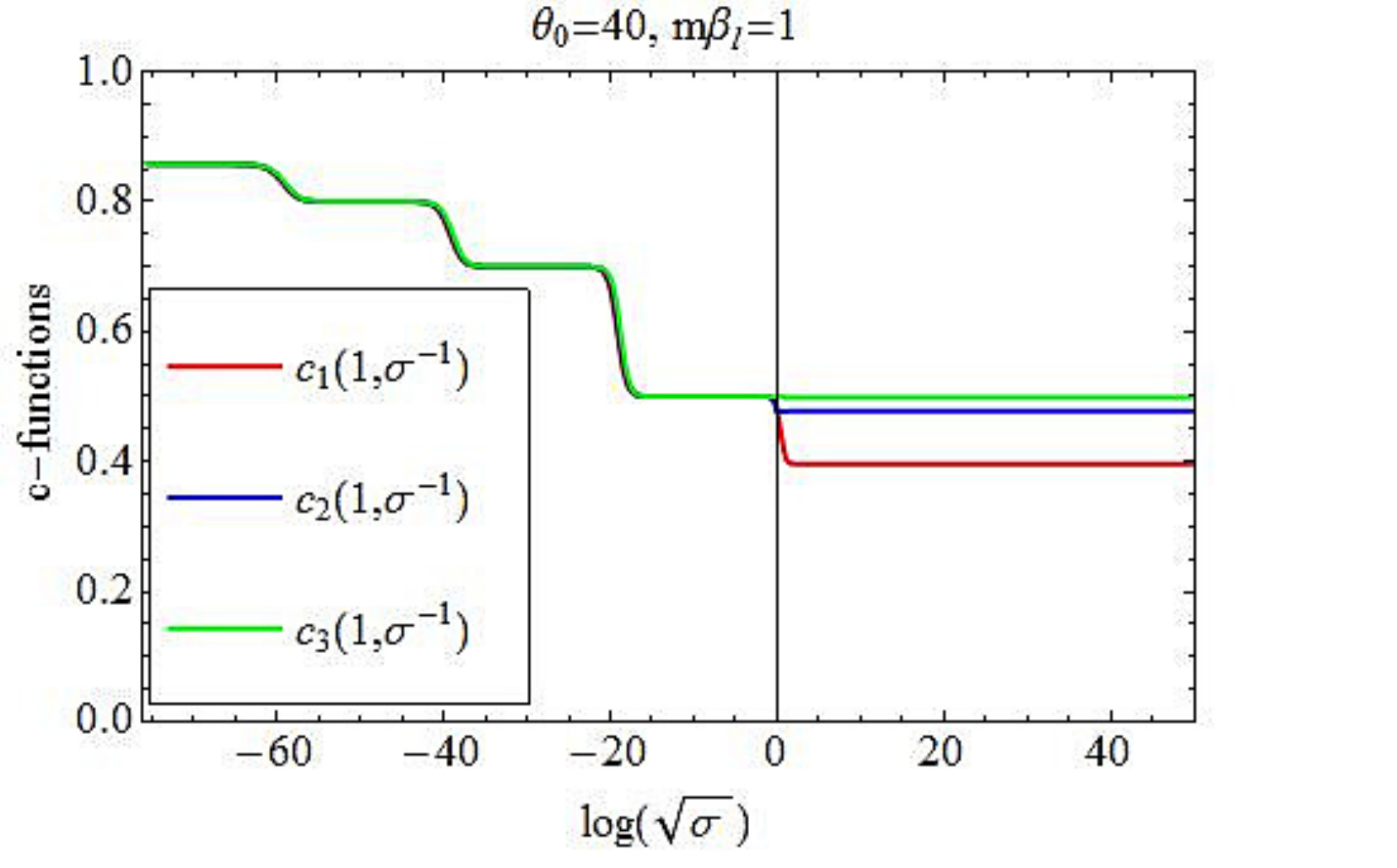}
\caption{$c$-functions for the roaming trajectories model.}\label{roamingc}
\end{center}
\end{figure}

\subsection{Numerical observations}

Our numerics indeed confirm this: we observe that for every order $n$ of the cumulant and every temperature ratio $\sigma = T_l/T_r$, the function \eqref{cfun} is a $c$-function. Note that these are non-equilibrium $c$-functions, defined intrinsically with respect to physical quantities characterizing a non-equilibrium steady state. It is particularly interesting to test this for the roaming trajectories model where the functions above are expected to display a non-trivial plateau structure. A variety of such functions are presented and compared to each other in Figure~\ref{roamingc}.

All insets in figure \ref{roamingc} confirm the assumption that we may regard the functions defined in (\ref{cfun}) as $c$-functions. In fact, in the first figure of the second row we compare to (\ref{oldc}), and we see that the new functions essentially carry the same information as the standard TBA scaling function. Each plateau is indeed located at a height corresponding to the central charge of a minimal model and there is a staircase pattern running from $c_n=0$ (low temperatures) to $c_n=1$ (high temperatures) which can be seen for $n=1,2$ and 3 in the figures. There are however also some subtle and interesting differences with respect to the equilibrium scaling function.

The size of the plateaux  of the function (\ref{oldc}) is exactly $\frac{\theta_0}{2}$ as can be seen in the original work \cite{roaming}. This is due to the structure of the TBA equations  which for high energies can be conveniently mapped into ``massless" TBA equations as explained in \cite{massless}. These equations naturally describe the $c$-function's flow between two different minimal models. In practise, these massless equations are obtained by performing the shifts $\theta\rightarrow \theta \pm \frac{\theta_0}{2}$ and these shifts give rise to a pair of equations which involve the new energy scales $m r e^{\pm {\theta_0}/{2}}$ where $\frac{\theta_0}{2}$ now becomes a natural scale of the problem and $r$ is the inverse temperature.

Observing the first figure in row one carefully we can easily see that the size of the plateaux of $c_1(T_l, T_r)$ depends on $\sigma$ and in fact increases with $\sigma$. The reason for this is very similar to the massless TBA argument given above. At high energies (temperatures) we can once more map our TBA equation into a pair of ``massless" TBA equations. However, we have now two parameters in the problem, that is, $\theta_0$ and $\sigma$. By performing the shifts $\theta \rightarrow \theta \pm \frac{\theta_0+\log\sigma}{2}$ we obtain two massless equations involving the energy scales $m\beta_l e^{\pm \frac{\theta_0+\log\sigma}{2}}$ so that the quantity $\frac{1}{2}(\theta_0+\log{\sigma})$ becomes a new natural scale of the problem. For this reason the size of the plateaux of the new $c$-functions $c_n(T_l,T_r)$ is precisely given by $\frac{1}{2}(\theta_0+\log{\sigma})$. This is confirmed by the first figure where the size of the plateaux is approximately $16, 17, 19$ and 20 for $\sigma=10, 100, 1000$ and 10000, respectively, and the same holds for other figures. For instance, it is noticeable how the plateaux of $c_1(T_l, T_l/100)$ in the first figure of the second row are larger than those of $c_{\text{eff}}(T_l)$. This plateau size argument applies to all the new $c$-functions (\ref{cfun}) so that plotting $c_n(T_l,\sigma^{-1} T_l)$ for various values of $n$ with the same $\sigma$ gives a set of extremely alike functions, as can be seen in the last figure of row two and the first figure of row three.

Finally, the last diagram of Figure~\ref{roamingc} deserves particular attention. At first sight it looks much as other diagrams in the same page. However, it is immediately striking that it has an additional plateau which appears to extend indefinitely on the right hand side, and that the variable in the horizontal axis is now $\frac{1}{2}\log\sigma$. Here we present the functions $c_n(T_l, T_r)$ with $n=1,2,3$ for a fixed value of $m\beta_l=1$. When $\log\sigma<0$, that is $m\beta_r <1$, we observe the standard high temperature behaviour of the $c$-functions, that is plateaux for all values of $c$ corresponding to the minimal models. This is quite natural as we are looking at a region where $m\beta_r \ll 0$. This behaviour changes at $\log \sigma=0$, which corresponds to $m\beta_l=m\beta_r=1$. For $\log \sigma>0$ all $c$-functions reach a plateau (different for each) which can not be matched with the central charge of any minimal model. For $\log \sigma>0$ and $m\beta_l=1$ we are looking at a region where both temperatures are low, thus we do not expect to recover CFT results. In fact it is rather easy to predict, at least approximately, the position of the plateaux that we observe for $\log\sigma>0$. Since we are looking at the low temperature or large $m\beta_{l,r}$ regime this means that at first order the solutions to the TBA equations can be approximated by the free solutions $\epsilon(\theta)=W(\theta)$. Thus the current in this region should be well approximated by the Ising solution given in (\ref{Ising}). In terms of this solution, the functions $c_{1,2,3}(T_l,T_r)$ admit the following expressions:
\begin{eqnarray}
  c_1^{\text{Ising}}(T_l,T_r) &=& \frac{3}{\pi^2(T_l^2-T_r^2)}\int_{0}^\infty d\theta \sinh(2\theta)
  \left(\frac{1}{1+e^{m\beta_l \cosh\theta}}- \frac{1}{1+e^{m\beta_r \cosh\theta}}\right), \nonumber\\
  c_2^{\text{Ising}}(T_l,T_r) &=& \frac{3}{2\pi^2(T_l^3+T_r^3)}\int_{0}^\infty d\theta \cosh\theta\sinh(2\theta)
  \left(\frac{e^{m\beta_l \cosh\theta}}{(1+e^{m\beta_l \cosh\theta})^2}+ \frac{e^{m\beta_r \cosh\theta}}{(1+e^{m\beta_r \cosh\theta})^2}\right), \nonumber \\
  c_3^{\text{Ising}}(T_l,T_r) &=&  \frac{1}{2\pi^2(T_l^4-T_r^4)}\int_{0}^\infty d\theta \cosh^2\theta\sinh(2\theta)\nonumber\\
  && \qquad\qquad\times
  \left(\frac{e^{m\beta_l \cosh\theta}(-1+e^{m\beta_l \cosh\theta})}{(1+e^{m\beta_l \cosh\theta})^3}- \frac{e^{m\beta_r \cosh\theta}(-1+e^{m\beta_r \cosh\theta})}{(1+e^{m\beta_r \cosh\theta})^3}\right).
\end{eqnarray}
Evaluating these functions numerically for $T_l/m=1$ and $T_r/m \rightarrow 0$ we obtain
\begin{equation}\label{values}
    c_1^{\text{Ising}}(1,0)=0.396314,\quad  c_2^{\text{Ising}}(1,0)=0.478062 \quad \text{and} \quad c_3^{\text{Ising}}(1,0)=0.497983,
\end{equation}
which reproduce extremely well (there is agreement up to 5 decimal places) the height of the last plateau in the graph. It is even possible to get a very accurate value for the same functions at $\log\sigma=0$ by taking the limit when $T_l/m\rightarrow T_r/m=1$. This gives
\begin{equation}\label{crossing}
    c_1^{\text{Ising}}(1,1)=0.478062,\quad  c_2^{\text{Ising}}(1,1)=0.478062 \quad \text{and} \quad c_3^{\text{Ising}}(1,1)=0.500284.
\end{equation}
The good agreement of (\ref{values}) and (\ref{crossing}) is remarkable as $m\beta_l=1$ does not really correspond to extremely high temperatures where the Ising approximation should hold best. It appears from this computation that this approximation is in fact extremely accurate already for temperatures around $T_{l,r}/m=1$.

It is also interesting to observe that the size of the plateaux for $\log\sigma<0$ is constant and appears to equal $\frac{\theta_0}{2}$. This can be explained with a similar argument as that employed earlier in predicting the size of the plateaux of other functions. As discussed earlier the quantities $m\beta_le^{\pm\frac{\theta_0+\log\sigma}{2}}$ are natural energy scales of the problem (for high temperatures). We are now fixing $m\beta_l$ and plotting against $\frac{1}{2}\log\sigma$ so that we recover the usual plateau size $\frac{\theta_0}{2}$ at high temperatures.

\subsection{Physical interpretation}

Since $c_n(T_l,T_r)$ are dimensionless, they are functions of the dimensionless variables $T_l/m$ and $T_r/m$ only, where $m$ is the mass scale (say the mass of the lightest particle). Hence, the observation that the $c$-functions $c_n(T_l,T_r)$ expressed in \eqref{cfun} are non-decreasing with the temperature scale, implies that upon a {\em reduction of the mass scale}, all {\em cumulants are non-decreasing}:
\beq
	\frc{d}{dm} C_n \leq 0.
\eeq
This means that as the mass scale is decreased, the average energy transferred increases, and the fluctuations are stronger, the probability distribution becoming overall ``flatter''. This has an immediate interpretation in terms of available energy carriers. By lowering the mass scale, i.e. the gap, while keeping the temperature scale the same, we make more energy carriers available as more quasi-particles can be created. This obviously implies that the energy current should indeed be higher, as the situation is similar to that of lowering the resistance while keeping the voltage the same in an electric circuit. But also, the availability of energy carriers at lower energies mean that the energy carriers cover a wider energy spectrum, provides a wider scope for fluctuations of energy transfer. Naturally, as $m\to0$, one should recover the CFT result \eqref{CnCFT}. This means that we have {\em exact upper bounds} on the cumulants, given by the cumulants of the UV fixed point:
\beq
	C_n \leq \frc{c\pi n!}{12}(T_l^{n+1} + (-1)^n T_r^{n+1}).
\eeq
We expect these results to hold in any integrable model with diagonal scattering, and possibly much more generaly.

\section{The probability distribution and its Poisson process interpretation}
\label{sectPoisson}

In this subsection, for simplicity let us restrict ourselves to a single-particle spectrum, with mass $m$.

As was shown in \cite{BD1}, in CFT, the large-time scaled energy transfer statistics may be described by Poisson processes. Specifically, the scaled cumulant generating function in CFT is that of a combination of independent Poisson processes at every energy $E$ representing jumps in both directions, weighted by the Maxwell-Boltzman factor $e^{-\beta_{l,r} E}$ (for jumps towards the right and left respectively):
\beq\label{FCFTomega}
	F_{\rm CFT}(z) = \int dq\,\omega_{\rm CFT}(q)\,(e^{zq}-1),
	\quad \omega_{\rm CFT} = \frc{c\pi}{12}\cdot
	\lt\{\ba{ll} e^{-\beta_l q} & (q>0) \\ e^{\beta_r q} & (q<0) \ea\rt.
\eeq
where $e^{zq}-1$ is the cumulant generating function for a single Poisson process.

Is the scaled cumulant generating function for energy transfer in IQFT also described by Poisson processes? That is, is it true that
\beq\label{Poisson}
	F(z) = \int dq\,\omega(q)\,(e^{zq}-1)
\eeq
for some nonnegative weights $\omega(q)$? In order to answer this question, we recall that our exact form for $F(z)$ comes from the exact current and the extended fluctuation relations \eqref{EFR}, which relate $F(z)$ to the current at shifted temperatures. This means that if \eqref{Poisson} holds, then the weights $\omega(q)$ are simply related to the Fourier transform of the current as a function of inverse temperature differences:
\beq\label{omega}
	\omega(q) = \frc1q\,\int \frc{d\lambda}{2\pi}\,
	J(\beta_l-i\lambda,\beta_r+i\lambda)\,e^{-i\lambda q}.
\eeq
Note that this is real if $J(\beta_l,\beta_r)$ is a real-analytic function. We may interpret the large-time scale energy transfer as being composed of Poisson processes if and only if this $\omega(q)$ is also nonnegative.

For the Ising model, using \eqref{Ising} in the form
\[
	J_{\rm Ising}(\beta_l,\beta_r) = \frc1{2\pi} \int_m^\infty dE\,E\sum_{n=1}^{\infty}
	(-1)^{n-1} (e^{-n\beta_l E} - e^{-n\beta_r E})
\]
we find
\beq\label{omegaIsing}
	\omega_{\rm Ising}(q) = \sum_{n=1}^\infty \frc{(-1)^{n-1}}{2\pi n^2}
	\cdot \lt\{\ba{ll} e^{-\beta_l q} & \text{for} \quad q>nm \\ e^{\beta_r q} & \text{for} \quad q<-nm \\
    0 & \text{otherwise}
	\ea\rt.
\eeq
For every finite $q$, this is a finite sum. Since $\sum_{n=1}^N \frc{(-1)^{n-1}}{2\pi n^2}$ is positive for every $N$, then indeed $\omega_{\rm Ising}(q)$ is a positive weight. Hence in the Ising model, there is a Poisson process for every transfer $|q|>m$ with finite positive weight proportional to the Maxwell-Boltzmann factor, and as an integer multiple of $m$ is passed, the proportionality factor changes abruptly (increasing or decreasing). In the limit $m=0$ the proportionality factor is constant with $q$, and equals $\sum_{n=1}^\infty \frc{(-1)^{n-1}}{2\pi n^2} = \pi/24$ as it should (CFT with central charge $c=1/2$).

In the interacting case, the question is more complicated, because we cannot analytically Fourier transform the NESSTBA expression \eqref{exactJ} for the current. However, we can analytically study the large-$m$ limit (or low-temperature limit) using the second-order expansion \eqref{Jlowtexp}, derived in Appendix \ref{applowt}. This is divided into two contributions, \eqref{order1} and \eqref{order2}. The contributions \eqref{order1} are exactly of the Ising form (to second order), so that they contribute terms of the form \eqref{omegaIsing} to the weight $\omega(q)$. In fact, to every order there will be terms of that form, so that we can write $\omega(q) = \omega_{\rm Ising}(q) + \omega_{\rm inter}(q)$, where $\omega_{\rm inter}(q)$ incorporates all effects of the particle interactions. The contributions \eqref{order2}, with interaction terms, are purely of second order. This implies that for low enough energies, the scaled energy transfer statistics can indeed be described by a continuum of Poisson processes; more precisely, for all $|q|<2m$, the weight is exactly the Ising weight $\omega_{\rm Ising}(q)$.
\begin{figure}[h!]
\begin{center}
 \includegraphics[width=8cm]{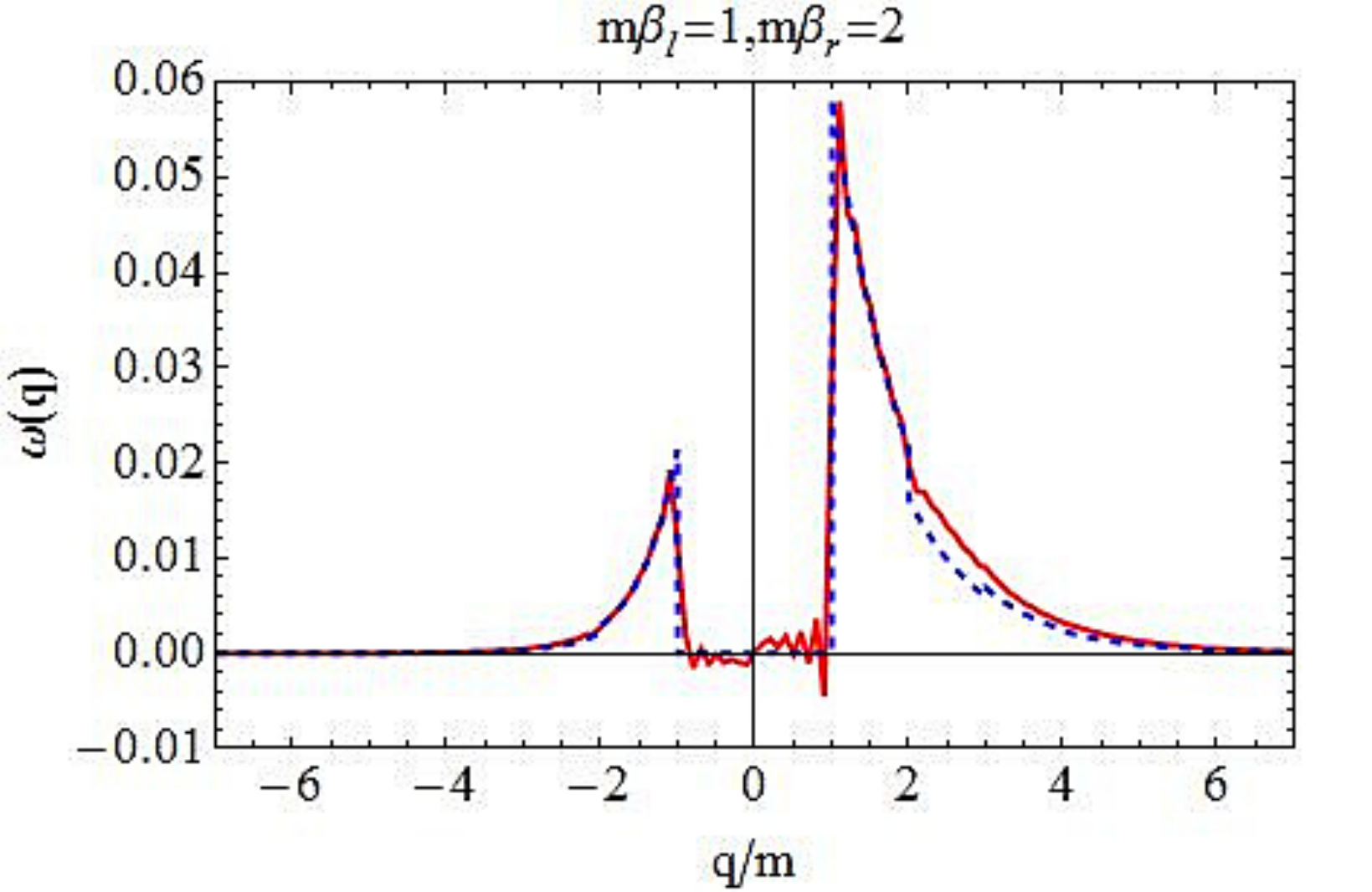}
\caption{The function $\omega(q)$ for the sinh-Gordon (red, solid curve) and Ising (blue, dashed curve) models.}\label{fourier}
\end{center}
\end{figure}

The contributions \eqref{order2} give the exact $\omega_{\rm inter}(q)$ for all $0<|q|<3m$. In order to show that $\omega(q)$ is nonnegative, it is sufficient to show that $\omega_{\rm inter}(q)$ is nonnegative, since $\omega_{\rm Ising}(q)$ is. The contributions \eqref{order2} may be analyzed using the result expressed in the last two lines in \eqref{formula}. Fourier transforming, we find
\begin{align}
	q\,\omega_{\rm inter}(q)
	& \stackrel{0<|q|<3m}= \n
	& \frc{m^2}{4\pi^2} \int_0^\infty d\theta d\gamma\lt(
	e^{-\beta_l(m\cosh\theta + m\cosh\gamma)}
		\delta(m\cosh\theta+m\cosh\gamma-q) \rt. \n & -
		\lt.
	e^{-\beta_r(m\cosh\theta + m\cosh\gamma)}
		\delta(m\cosh\theta+m\cosh\gamma+q) \rt)
	\cosh\theta \,\varphi(\theta-\gamma)\,(\sinh\theta+\sinh\gamma) \n
	& +
	\frc{m^2}{8\pi^2} \int_0^\infty d\theta d\gamma\,
	e^{-\beta_lm\cosh\theta -\beta_r m\cosh\gamma}
		\delta(m\cosh\theta-m\cosh\gamma-q)\n & \times\,
		\varphi(\theta+\gamma)(\sinh(2\theta)-\sinh(2\gamma)+
		2\sinh(\theta-\gamma))\label{corr}
\end{align}
As is apparent, the right-hand side is exactly zero for $0<|q|<2m$, and is positive for $q>2m$ and negative for $q<-2m$. This shows that $\omega_{\rm inter}(q)$ is a nonnegative function of $q$ for all $0<|q|<3m$, hence that there is a Poisson process interpretation in this whole range of energy transfer.

A proof to all orders, however, is more difficult. Instead, we compute $\omega(q)$ by numerically evaluating the Fourier transform in \eqref{omega}. Figure~\ref{fourier} shows the result (solid curve) for the sinh-Gordon model. This is clearly a nonnegative function, up to oscillations that are stronger near sharp changes due to numerical imprecisions.

A comparison between the function $\omega(q)$ in the sinh-Gordon model  and the function $\omega_{\text{Ising}}(q)$ is also presented in Figure \ref{fourier}. The two functions display very similar features. For values of $|q|<2m$ we should find complete agreement between both curves. This indeed appears to be the case up to the limited precision of our numerical procedure. For example, for $|q|<m$ the Fourier transform should be zero (as it is for Ising). Instead, we see oscillations about the value zero. As explained above, for $2m<|q|<3m$ the function $\omega(q)$ of the sinh-Gordon model should start differing from the Ising case, through the correction (\ref{corr}). Although we are not plotting the contribution from this correction here, it is apparent that the two curves start to disagree more for $2m<|q|$.  The two curves become almost indistinguishable for large values of $|q|/m$ (CFT limit) where they both display exponential decay $\omega(q)\propto e^{-\beta_{l,r}|q|}$ according to \eqref{FCFTomega}. Although the coefficient of this exponential decay should be twice as large for the sinh-Gordon model (since $c=1$) this is hard to observe explicitly by comparing the two functions, as the exponential decay is the dominating feature. We have also evaluated numerically $\omega(q)$ for the roaming trajectories model, confirming that it is nonnegative.

The form \eqref{Poisson} of the scaled cumulant generating function means that the complete, large-time scaled statistics of the energy transfer is that of independent, classical packets of energies $|q|$ jumping towards the right ($q>0$) or left ($q<0$) in a Poissonian fashion with intensity $dq\,\omega(q)$ for some positive weight $\omega(q)$. This Poisson process description may be seen as an improvement on the idea behind the Landauer formula for non-equilibrium currents. In the Landauer formula, one assumes that two baths at different temperatures / chemical potentials with independent equilibrium distributions are emitting excitations through available channels. A similar picture holds for all fluctuations in CFT, as the weight function $\omega_{\rm CFT}(q)$ is proportional to the Boltzmann distribution times the flat state density, as observed in \cite{BD1}. However, this is not the case in general: the weight function is not just that determined by the Boltzmann distribution times the state density, but is a more complicated function, encoding nontrivially the full interaction.

\section{(Non-)additivity of the current} \label{sectna}

As recalled above, in \cite{BD1} it was shown that in CFT the current takes the form
\beq\label{jf}
	J_{\rm CFT}(\beta_l,\beta_r) = f(\beta_l)-f(\beta_r)
\eeq
for some function $f$. This form of the current is equivalent to the {\em additivity property}
\beq\label{additivity}
	J_{\rm CFT}(\beta_1,\beta_2) + J_{\rm CFT}(\beta_2,\beta_3) -
	J_{\rm CFT}(\beta_1,\beta_3) = 0.
\eeq
As was observed in \cite{Moore}, additivity, or the existence of the function $f$ as in \eqref{jf}, implies that the non-equilibrium current is completely determined by the linear (equilibrium) conductance
\beq
	G^{\rm eq}_{\rm CFT}(T) = \lt.
	\frc{\p}{\p T_1} J_{\rm CFT}(T_1^{-1},T_2^{-1})\rt|_{T_1=T_2=T},
\eeq
that is
\beq\label{jGeq}
	J_{\rm CFT}(T_1^{-1},T_2^{-1}) =
	\int_{T_2}^{T_1} G^{\rm eq}_{\rm CFT}(T)\, dT.
\eeq
Although this observation is not useful in CFT because the current can be evaluated exactly by other means, if the additivity property were to hold beyond CFT, eq.~\eqref{jGeq} could have deep and important physical consequences. It was indeed suggested in \cite{Moore}, based on a numerical analysis, that additivity  holds in the critical XXZ chain beyond the low-temperature limit, hence beyond the universal region that is expected to be described by CFT. In fact, it was also suggested in \cite{Dhouches} that additivity could hold in massive quantum field theory.

We show that the additivity property {\em does not hold} in general: the non-equilibrium current is not determined by the equilibrium conductance. However, as we observe in the sine-Gordon model, the breaking of additivity may be very small. The breaking is due to a delicate imbalance that arises because, in strongly interacting models, level densities are affected by level occupations at different energies.

\subsection{Proof and numerical observations}

Let us define the additivity deficit
\beq\label{adddef}
	P(\beta,\sigma) := \frc{
	J(\beta,\sigma\beta) + J(\sigma\beta,\sigma^2\beta)
	- J(\beta,\sigma^2\beta)}{J(\beta,\sigma^2\beta)}.
\eeq
We show in Appendix \ref{appnonadd} that
\beq
	P(\beta,\sigma)\neq 0.
\eeq
That is, we show, in the formulation \eqref{ness}, \eqref{W} of the non-equilibrium steady state and in general diagonal-scattering integrable QFT, that for any low enough temperature, $\beta m> \beta_0m$ for some $\beta_0$, and any large enough temperature ratio, $\sigma>\sigma_0$ for some $\sigma_0>1$, the quantity $P(\beta,\sigma)$ is nonzero. For technical reasons, we will assume that the differential scattering phases $\varphi_{ij}(\theta)$ have a fixed sign (i.e.~$\varphi_{ij}(\theta)$ is positive for every $i,j$ and $\theta$, or is negative for every $i,j$ and $\theta$); this condition is satisfied for all models studied in this paper. Our proof will also show that, for instance, $P(\beta,\sigma)$ for $\sigma>\sigma_0$ has a positive sign for the sinh-Gordon and roaming trajectories model, and a negative sign for the reflectionless sine-Gordon model studied above; hence in general, it does not have a fixed, model-independent sign.

\begin{figure}[h!]
\begin{center}
  \includegraphics[width=7.5cm]{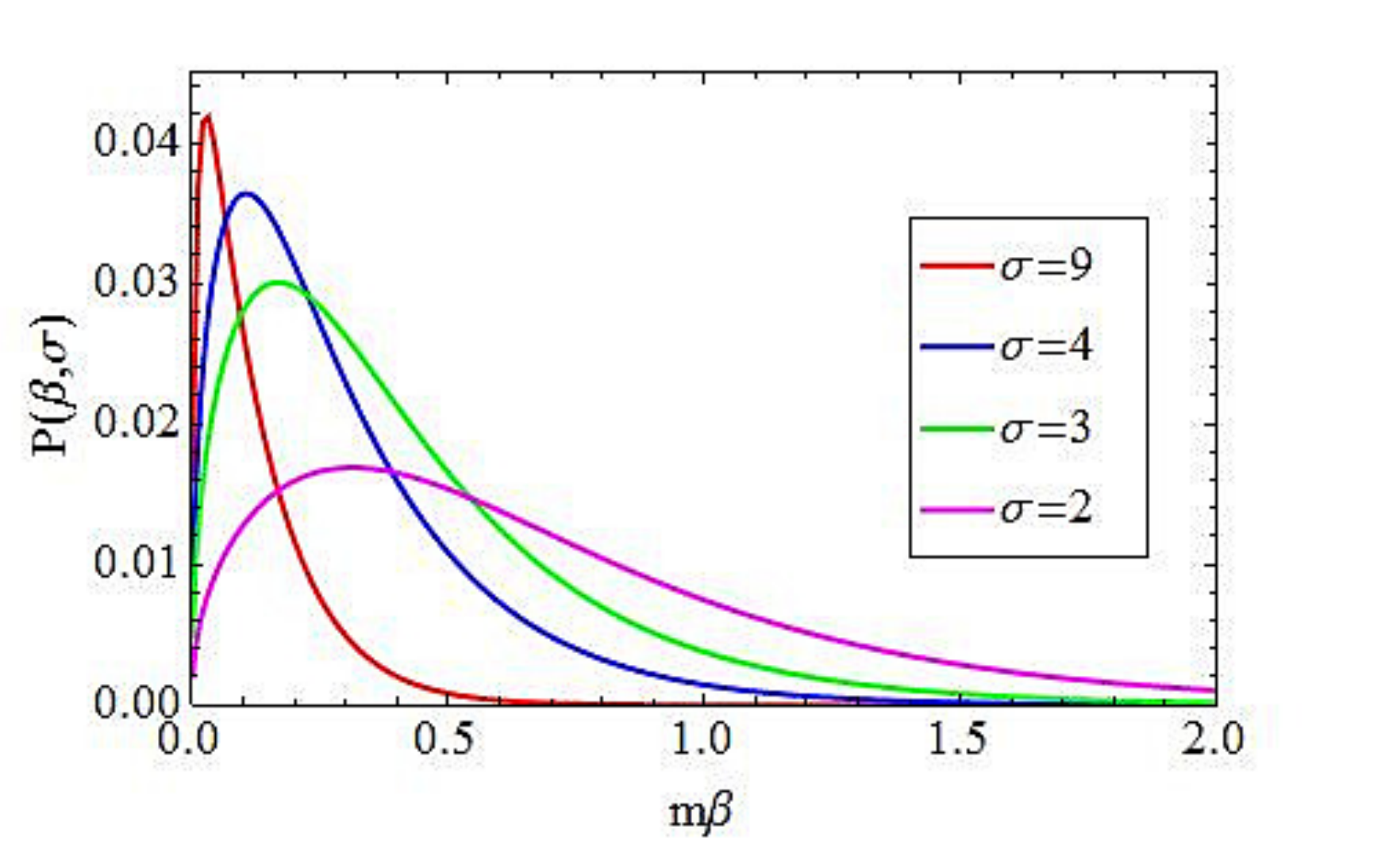}
  \includegraphics[width=7.5cm]{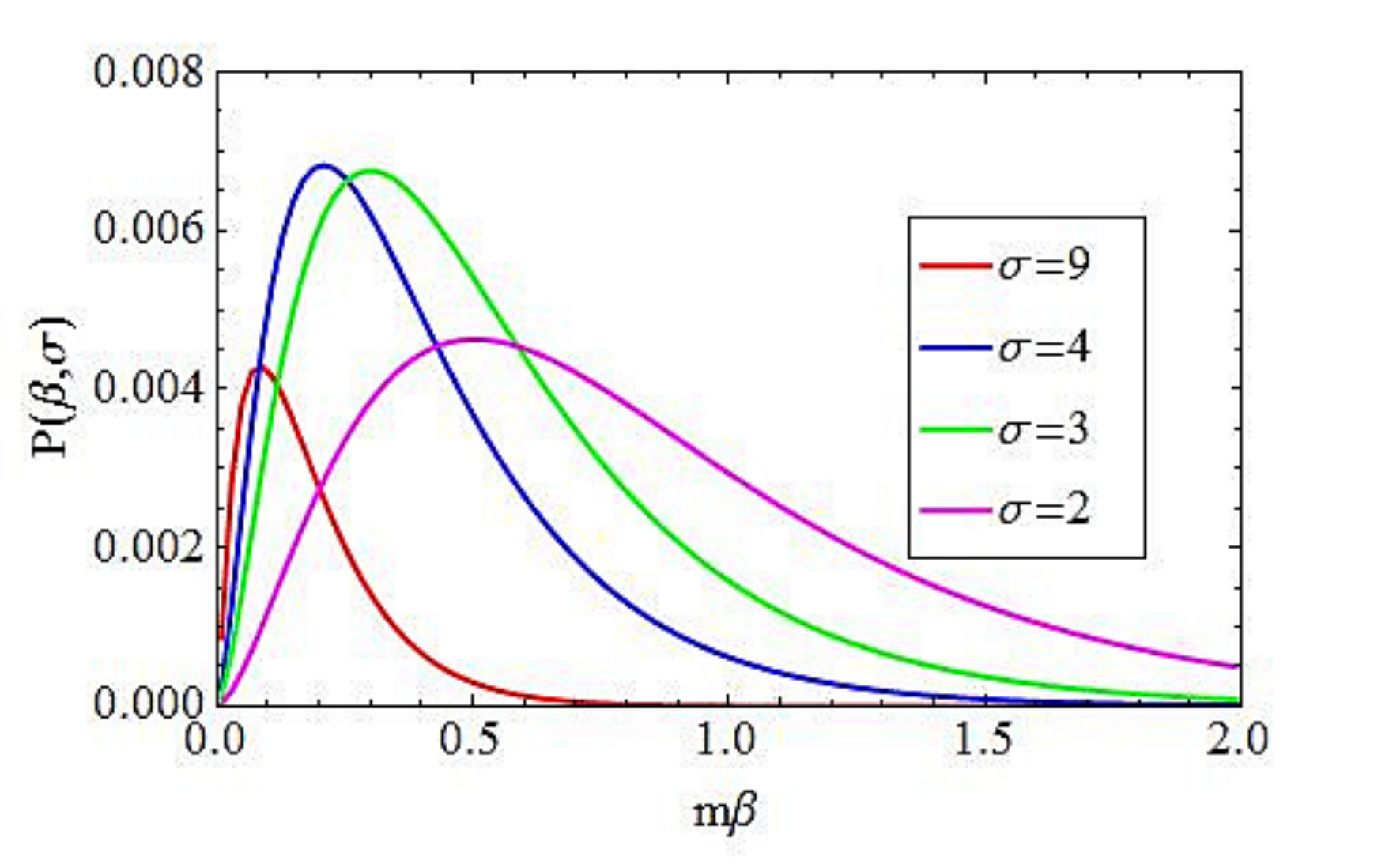}
\caption{The function $P(\beta,\sigma)$ for various values of $\sigma$ and $\theta_0$ in the sinh-Gordon model.}\label{figpsinh}
\end{center}
\end{figure}

Numerical evaluations of the additivity deficit in the sinh-Gordon, roaming trajectories and reflectionless sine-Gordon model also confirm that it is nonzero. The evaluations yield another important insight: the additivity deficit, although nonzero, is actually very small in most cases. The maximum value of the additivity deficit $P(\beta,\sigma)$ as function of $\beta$ depends on the ratio of temperatures $\sigma$, but for ratios below 10 the maximum does not go beyond about 4\% in the sinh-Gordon model. It increases as $\sigma$ is increased, but the region where $P(\beta,\sigma)$ is large becomes smaller. All these features can be observed in the first graph of Figure~\ref{figpsinh} where the values depicted have been obtained by numerically solving the NESSTBA equations. The second graph represents the same quantity as obtained by approximating the current by its low temperature expansion given in appendix \ref{applowt}. We can see that both figures show a similar set of features, however they differ considerably in magnitude, especially for high energies.

\begin{figure}[h!]
\begin{center}
  \includegraphics[width=8cm]{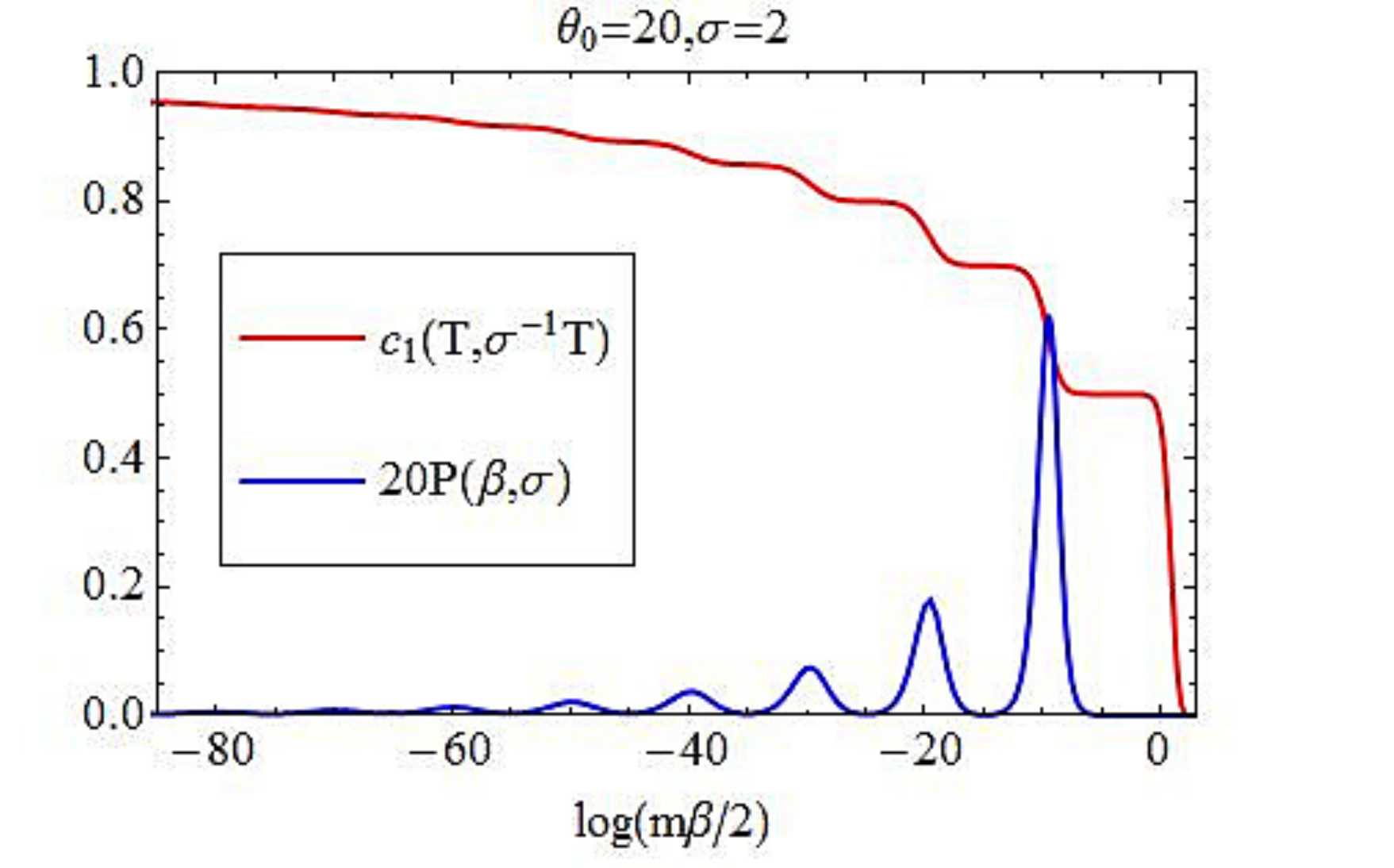}
 \includegraphics[width=7.5cm]{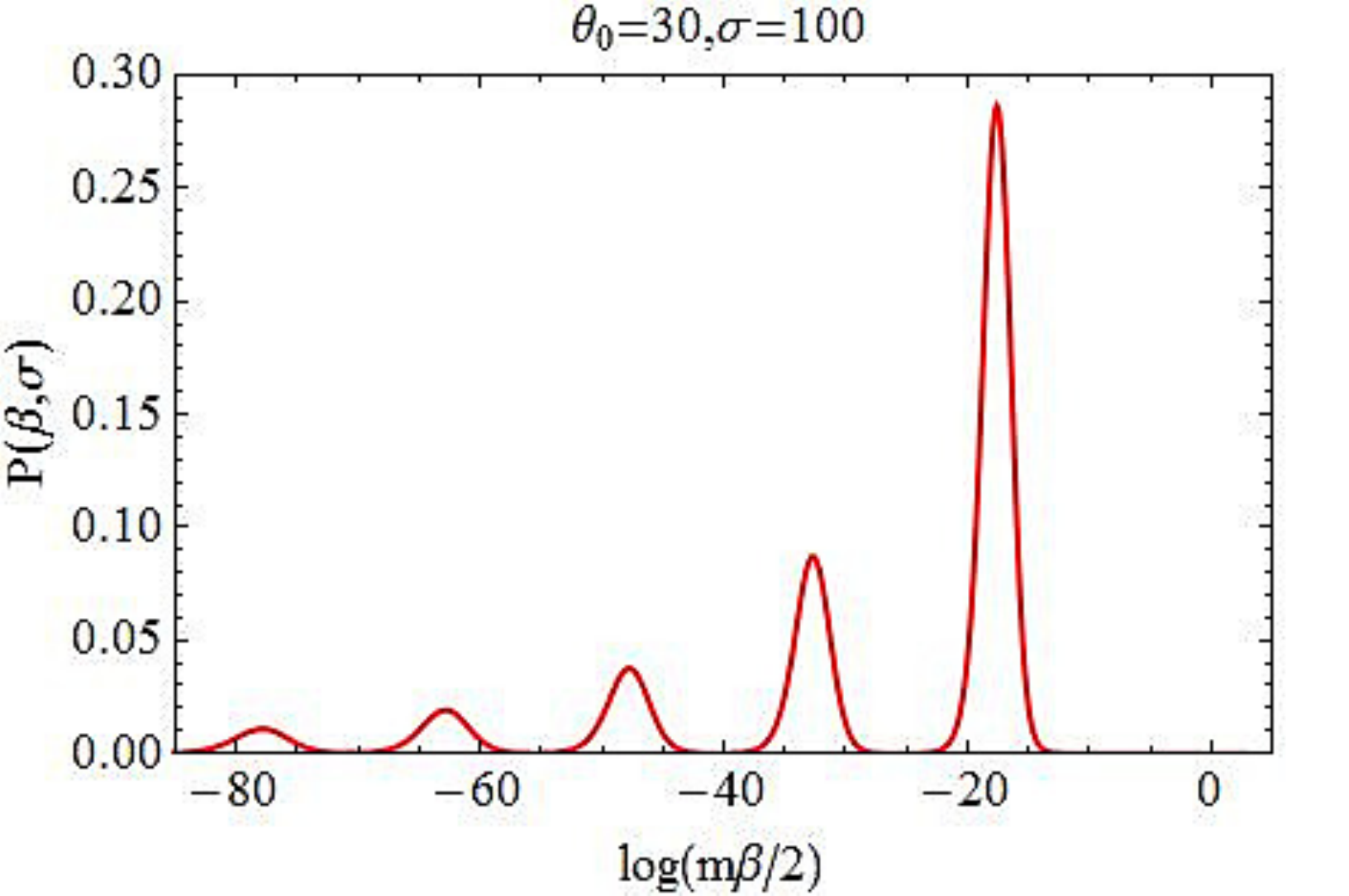}
\caption{The function $P(\beta,\sigma)$ for two values of $\sigma$ in the roaming trajectories model.}\label{addroaming}
\end{center}
\end{figure}

A more involved structure is found when studying the function $P(\beta,\sigma)$ in the roaming trajectories model (see Figure~\ref{addroaming}). We have used a logarithmic scale in this case to shown more clearly the correspondence between regions where the additivity deficit is zero and plateaux in the corresponding $c$-functions. In order to make this correspondence even more explicit we present both the function $c_1$ and the (re-scaled) additivity deficit in the first figure. As we can see, maxima of the additivity deficit are situated at values of $\log(m\beta/2)$ which are exactly multiples of $\frac{\theta_0+\log\sigma}{2}$. These values correspond to the middle point of $c$-function steps which connect two consecutive plateaux. As expected, the additivity deficit vanishes between such maxima exactly in those regions where the $c$-function develops plateaux, where the system is essentially described by a CFT. From this description, it is interesting to note that there is an exception, namely the first step of the $c$-function (connecting the values $c_1=0$ and $c_1=\frac{1}{2}$) does not have an associated maximum of the additivity deficit. In fact this first step and the subsequent plateau at $c_1=\frac{1}{2}$ give us the $c$-function associated to the Ising model. As is easy to observe from \eqref{Ising} the Ising model (both massive and massless) enjoys the additivity property. Therefore the additivity deficit vanishes in this region. The same features are observed as well in the second figure.

It is also worth noting that the maximum value of the additivity deficit increases with $\sigma$, as for the sinh-Gordon model. However, for the roaming trajectories model it seems that the additivity deficit can be rather large for large $\sigma$. For example, in the second figure it sometimes reaches values near $30\%$. We observe that for the flow between the minimal models ${\cal M}_5$ and ${\cal M}_4$ (central charge $7/10$ and $1/2$ respectively), the additivity deficit is the largest.

In contrast, for the reflectionless sine-Gordon model, with $\sigma$ in the range between 3 and 9 the maximum deviation does not go beyond 0.15\%.
\begin{figure}[h!]
\begin{center}
  \includegraphics[width=8cm]{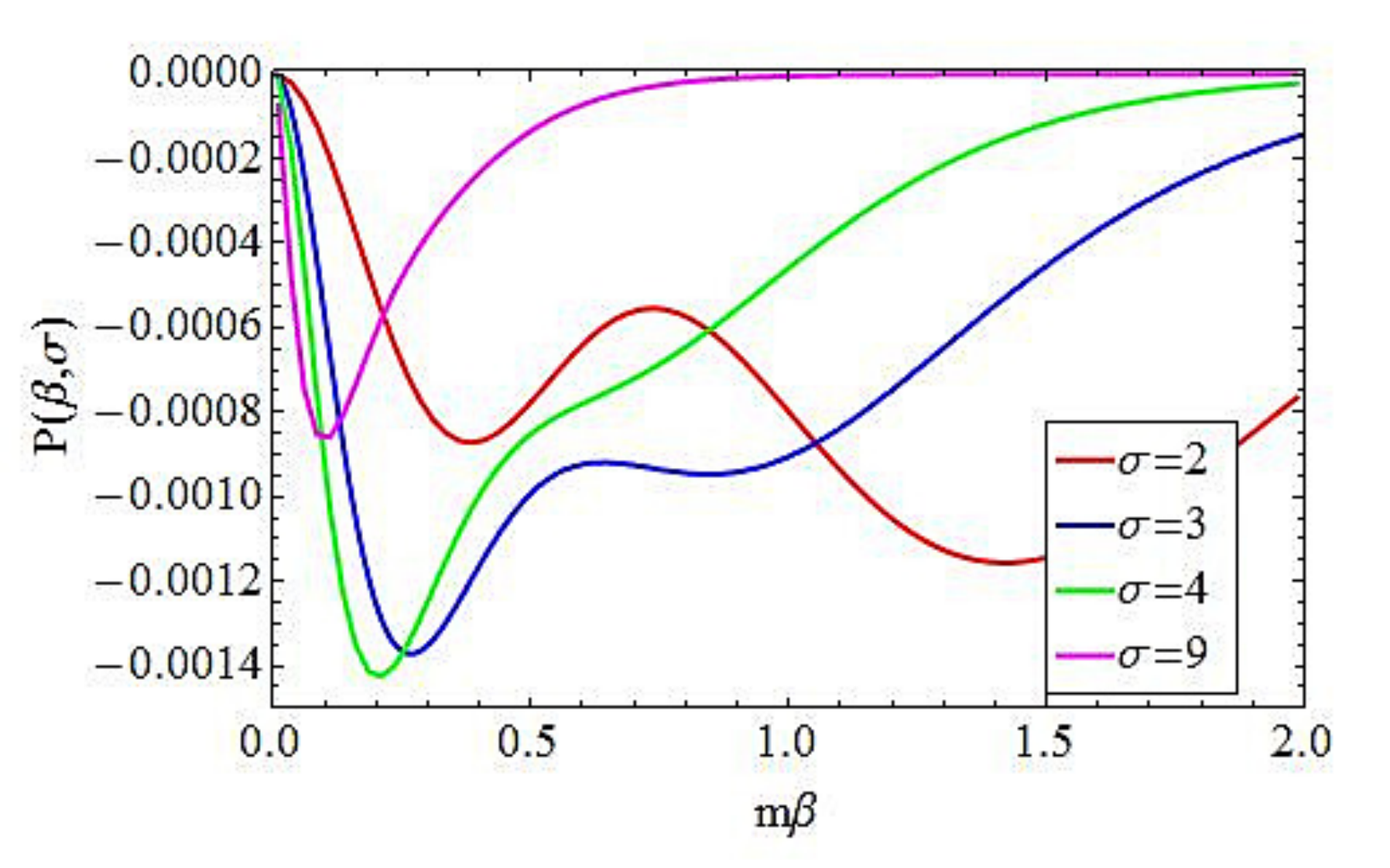}
\caption{The function $P(\beta,\sigma)$ for various values of $\sigma$ in the reflectionless sine-Gordon model.}\label{figpsine}
\end{center}
\end{figure}
This is important, because the numerical analysis of \cite{Moore}, from which it was concluded that additivity exactly held, is not expected to reach the accuracy necessary to detect such small effects. There is of course {\em a priori} no contradiction between our results and the conclusions of \cite{Moore}, because our results are in the context of near-critical, universal low-temperature systems, while the analysis of \cite{Moore} is in the context of the non-universal, higher temperature regime of gapless chains. Yet, our remark that the additivity deficit may be lower than the precision achieved in \cite{Moore} sheds doubt about the precise conclusions reached there. It is however rather interesting that the results of \cite{Moore} and our present results indicate that additivity may be {\em approximately} satisfied in many cases.

\subsection{Physical interpretation}

The additivity property holds exactly in massive quantum field theory in three situations: (i) at high temperatures, where the CFT regime is recovered; (ii) in free-particle models; and (iii) at low temperatures.

In order to understand, at least in the context of integrable QFT, why in the three above-mentioned situations the additivity property should hold, and why it should break, although sometimes weakly, in other situations, let us recall and analyze in a bit more depth the argument for additivity presented in \cite{Dhouches}. We start with the expression \eqref{JL} for the finite-$L$ counterpart of the non-equilibrium current (the latter being obtained by taking the limit $L\to\infty$). As we argued, this makes good sense in integrable QFT, thanks to the description of the finite-$L$ Hilbert space in terms of Bethe ansatz quasi-particles. Labeling quasi-particles in a state $|v\ket$ by $k$, they have well-defined energies $e_k$ momenta $p_k$. Clearly, for every state $|v\ket$ with $n$ quasi-particles, we have $P|v\ket = \lt(\sum_{k=1}^n p_k \rt)|v\ket$. Let us introduce the operators $P_\pm$ and $E_\pm$ that measure the momenta and energies of positive- and negative-momentum quasi-particles respectively:
\[
	P_\pm|v\ket = \lt(\sum_{k:\;p_k{\gtrless} 0} p_k \rt)|v\ket,\quad
	E_\pm|v\ket = \lt(\sum_{k:\;p_k{\gtrless} 0} e_k \rt)|v\ket.
\]
Clearly, we have $P=P_++P_-$, and also, on every state $|v\ket$, the finite-$L$ density matrix $\rho_{\rm stat}^L$ given by \eqref{rhostatL} factorizes, $\rho_{\rm stat}^L |v\ket = e^{-\beta_l E_+ - \beta_r E_-}|v\ket$. Using the fact that $e_{k'} = e_k$ if $p_{k'} = -p_k$, and the fact that by parity invariance, the spectrum of momenta is symmetric under change of signs, it was then concluded in \cite{Dhouches} that the average \eqref{JL} should give, in the large-$L$ limit, the form $f(\beta_l) - f(\beta_r)$ for $J(\beta_l,\beta_r)$, hence the additivity property.

There is, however, one missing argument: in order to reach this conclusion, we also need the trace itself to factorize, $\Tr_L = \Tr_{\{p_k<0\}} \Tr_{\{p_k>0\}}$. This, unfortunately, does not hold in general. Indeed, according to the Bethe ansatz equations, the density of quasi-particles around a certain momentum depends in a nontrivial and nonnegligible way on the density of quasi-particles at other momenta. This non-factorization is true at finite $L$, and in the large-$L$ limit factorization {\em is not} recovered. This is of course what leads to coupled integral equations determining the pseudo-energy in the TBA formulation above. The fact that the relatively small effect of the breaking of the factorization of the trace is at the basis of the breaking of the additivity property may explain why additivity is broken only very slightly for the sinh-Gordon and sine-Gordon models studied here.

Factorization of the trace, hence additivity, does hold, however, in the three situations mentioned above. It is clear that it holds in the situation (ii), in the case of free particles: the defining property of free particles is that the density of states of the quasi-particles is unaffected by the presence or absence of other quasi-particles. In the situation (iii), at low temperatures, it is also clear that the trace factorizes: in this case, one may restrict the trace to excitations containing only one quasi-particle, so that the level density is not affected by the presence of any other quasi-particle. In the situation (i), at high temperatures, factorization of the trace occurs because of the large-energy {\em rapidity-space factorization}: the scattering matrix asymptotically tends to a constant, whence the differential scattering phases $\varphi_{ij}(\theta)$ tends to zero. This implies that the density of particles at momenta largely separated do not influence each other, so that the trace can be factorized into large-energy right-movers and left-movers.

\section{Conclusion}

% exact bounds on all cumulants
% exact stochastic process description generalizing Landauer

In this paper we considered non-equilibrium steady states with energy flows in integrable QFT. The steady states are obtained from the setup where an initial state representing two independently thermalized halves of the system is let to evolve for an infinite time. In this setup, it is the asymptotic regions of the system itself that play the roles of reservoirs providing and absorbing energy (Hamiltonian reservoirs). The exact non-equilibrium density matrix describing averages of local observables in the steady state was proposed, and derived from QFT arguments, in \cite{BD1,Dhouches}. Using this density matrix, we obtained for the fist time the exact energy current in any integrable model of relativisitc QFT with diagonal scattering matrix by generalizing the thermodynamic Bethe ansatz (TBA) method first developed by Al.B.~Zamolodchikov in  \cite{tba1}. Employing the extended fluctuation relation (\ref{EFR}) derived by Bernard and Doyon \cite{BD3}, we further obtain the scaled cumulant generating function of the heat current.

We have verified that our formula for the current (\ref{exactJ}) agrees with known results and techniques in both the high and low temperature limits. For high temperatures, Bernard and Doyon's CFT result \cite{BD1} is exactly recovered by generalizing the high temperature limit analysis of TBA equations in \cite{tba1}. The low temperature expansion of (\ref{exactJ}) also agrees with the result obtained by using finite-volume regularization methods of Pozgay and Takacs in \cite{Pozsgay}.

We have solved the equations (\ref{exactJ}) numerically for three models: the sinh-Gordon model, the roaming trajectories model and the sine-Gordon model at a particular reflectionless point. The numerical results for the currents in these three models agree well with the analytic results both for high and low temperatures. Also the numerics, in particular of the roaming trajectories model, suggest the existence of an infinite family of $c$-functions which are given in terms of the cumulants (these functions are verified to satisfy the required properties: positivity, monotonicity, constant limiting values corresponding to the CFT central charges at RG fixed points). They display a very similar behaviour to that of the standard TBA scaling function, recovering for instance CFT central charges of minimal models in the roaming trajectories model. These allowed us to establish upper bounds for cumulants in any integrable model with diagonal scattering.

We verified that the scaled cumulant generating function can be seen as that of a family of independent Poisson processes, one for every value of the transferred energy. Hence, the complete, large-time scaled statistics of the energy transfer is that of independent, classical packets of energies jumping towards the right or left in a Poissonian fashion with weight $\omega(q)$. We have evaluated $\omega(q)$ numerically for the sinh-Gordon and roaming trajectories models, thus verifying that it is nonnegative. The Poisson process description may be seen as an improvement on the idea behind the Landauer formula.

Finally, we have proved that the additivity of the non-equilibrium current does not hold in a general diagonal theory both analytically and numerically, except at very large and very small temperatures. The additivity deficit has been found to be very small in the numerical analysis of the sinh-Gordon and sine-Gordon models considered here although it can be rather large for large temperature ratios in the roaming trajectories model.

There are many directions to explore from our results. First, an independent verification of our results, perhaps from perturbation theory or from numerical calculations, would be very helpful for confirming both that our techniques work, and that the non-equilibrium density matrix derived in \cite{Dhouches} is the correct one. In particular, DMRG numerics has been performed in \cite{Moore} for the XXZ spin chain in the quench situation that we consider here. It would be possible to adapt this to gapped chains whose massive scaling limit is the reflectionless sine-Gordon model that we studied. It would also be very interesting to extend our analysis of the sine-Gordon model to the non-diagonal case or even to more complex reflectionless points and to check if we still have $c$-functions and a Poisson process interpretation, and if the deviation from additivity is small or not. In the non-diagonal case, TBA becomes much more complicated. The development of nonlinear integral equations for the current, paralleling the equilibrium case, could be helpful. Further, we have only evaluated the Poisson weight by numerically Fourier transforming the current. It would be interesting to have an exact expression in some form (like an integral equation). Additionally, it is natural to think that the $c$-functions we proposed apply much more generally than in integrable QFT; it would be very interesting to have general arguments or a general proof that they are $c$-functions. Finally, an interesting question is to understand if the Poisson process interpretation still holds beyond integrability.

{\bf Acknowledgments.} OCA and BD thank Francesco Ravanini for comments on our non-equilibrium thermodynamic Bethe ansatz and on the non-diagonal generalization. BD thanks Denis Bernard for general comments on the manuscript and fruitful past (and future) collaborations, as well as Fabian Essler and Joel Moore for discussions and email communications. MH thanks Jorn Mossel for useful advice.

\appendix

\section{Cumulants in terms of correlators} \label{appcumul}

The expression \eqref{exprF} for the scaled cumulant generating function can be simplified to its ``naive'' form in the case of pure-transmission systems (and in slightly more general cases as well), as was shown in \cite{BD3}:
\beq
	F(z) = \lim_{t\to\infty} t^{-1} \log\lt( \bra e^{z \Delta Q(t)}\ket_{\rm stat}
	\rt)
\eeq
where we recall that $\Delta Q(t) = Q(t)-Q$. Hence, cumulants have the form
\beq
	C_n = \lim_{t\to\infty} t^{-1}\,\bra (\Delta Q(t))^n\ket_{\rm stat}^{\rm connected}
\eeq
where the superscript ``connected'' has the usual combinatoric meaning, e.g.
\[
	\bra A^2\ket_{\rm stat}^{\rm connected} = \bra A^2\ket_{\rm stat} -
	\lt(\bra A\ket_{\rm stat}\rt)^2,\quad
	\bra A^3\ket_{\rm stat}^{\rm connected} = \bra A^3\ket_{\rm stat} -
	3\bra A\ket_{\rm stat}\,\bra A^2\ket_{\rm stat}
	+2\lt(\bra A\ket_{\rm stat}\rt)^3.
\]
We may now use, by definition of the current operator,
\[
	\Delta Q(t) = \int_0^t ds\,{\cal J}(t)
\]
so that
\beq
	C_n =
	\lim_{t\to\infty} t^{-1}
	\lim_{\ep\to0^+} \int_{0}^t ds_0\cdots ds_{n-1}\,
	\bra {\cal J}(s_{n-1}+(n-1)\ri\ep)\cdots{\cal J}(s_1+\ri\ep){\cal J}(s_0)\ket_{\rm stat}^{\rm connected}.
\eeq
where we have included an imaginary-time shift that agrees with the ordering of the operators in the product $(\Delta Q(t))^n$, and that regularizes the correlation functions. By stationarity of the steady state, we shift all time variables by a common value. Hence, changing variables to $u_j = s_j-s_0$ for $j=1,\ldots,n-1$, we find
\beqa
	C_n &=&
	\lim_{t\to\infty} t^{-1}
	\lim_{\ep\to0^+} \int_0^t ds_0\int_{-t}^t du_1\cdots du_{n-1}\,
	\bra {\cal J}(u_{n-1}+(n-1)\ri\ep)\cdots{\cal J}(u_1+\ri\ep){\cal J}(0)\ket_{\rm stat}^{\rm connected} \n
	&=&
	\lim_{t\to\infty} t^{-1}
	\lim_{\ep\to0^+} \int_{-t}^t du_1\cdots du_{n-1}\,
	\bra {\cal J}(u_{n-1}+(n-1)\ri\ep)\cdots{\cal J}(u_1+\ri\ep){\cal J}(0)\ket_{\rm stat}^{\rm connected} \times  \n
	&& \times \int_{{\rm max}(-u_1,\ldots,-u_{n-1},0)}^{{\rm min}(
	t-u_1,\ldots,t-u_n,t)} ds_0.
\eeqa
Since we expect that connected correlation functions decay exponentially at large time separations, whenever any $u_j$ variable is large the integrand is exponentially suppressed. Hence the last factor can be approximated, for large $t$, by its value with $u_1=\ldots=u_{n-1}=0$, giving a factor of $t$. This shows \eqref{Cn}.

\section{Derivation of the NESSTBA equations} \label{appTBA}

In this section we follow the arguments presented in \cite{tba1} in order to obtain the TBA representation \eqref{exactJ} of the exact non-equilibrium current. For simplicity, we restrict ourselves to models with a single-particle spectrum, but the derivation easily generalizes to many particle types with diagonal scattering.

The idea of \cite{tba1}, in the present context, is to consider the trace representation \eqref{fa} of the free energy $f^a$ on a periodic space of finite length $L$, and to evaluate it in the limit $L\to\infty$ by analyzing the state that provides the leading contribution in this limit. At finite $L$, one assumes that the states are described by the Bethe ansatz associated with the QFT two-particle scattering matrix $S(\theta)$. In particular, there is a discrete set of states, and the distribution of Bethe roots is fixed from Bethe ansatz equations, the periodicity condition of the wave function on the circle of length $L$. One must also make a choice of density matrix $\rho_{\rm stat}^L$ which extrapolates to $\rho_{\rm stat}$ at $L\to\infty$, and as said in the text, here we choose \eqref{rhostatL}.

One describes the Bethe ansatz states using densities of ``levels'' or single-particle states $\rho(\theta)$, and of particles $\nu(\theta)$, as functions of the rapidity (so that $\rho(\theta)\,d\theta$ and $\nu(\theta)\,d\theta$ are the number of available levels and the number of particles, respectively, between rapidities $\theta$ and $\theta+d\theta$). The Bethe ansatz equations relate these two as follows:
\beq\label{rhorho1}
	2\pi \rho(\theta) = L \frc{d p(\theta)}{d\theta} +
	\int d\gamma\,\varphi(\theta-\gamma)\, \nu(\gamma)
\eeq
where $p(\theta) := m\sinh\theta$ is the momentum at rapidity $\theta$ and $\varphi(\theta) = -\ri\p_\theta S(\theta)$. This equation simply means that the non-interacting level density $\frc1{2\pi}\frc{dp(\theta)}{d\theta}$ at rapidity $\theta$ is modified, thanks to the interaction, by the presence of particles at other rapidities. The densities giving rise to the leading contribution of the trace of the operator $\rho_{\rm stat}^L e^{-aP}$ in \eqref{fa} are evaluated by minimizing the free energy functional ${\cal F}^a[\rho,\nu]:=W^a[\nu] -{\cal L}[\rho,\nu]$. The term $W^a[\nu]$ is the functional of the particle density $\nu(\theta)$ that corresponds to the negative of the logarithm of the operator $\rho_{\rm stat}^L e^{-aP}$; this is, thanks to \eqref{rhostatL} and \eqref{W},
\[
	W^a[\nu] = \int d\theta\,(W(\theta) + a\,p(\theta))\,\nu(\theta).
\]
The term ${\cal L}[\rho,\nu]$ is the number of actual Bethe states (entropy) corresponding to $\rho$ and $\nu$ in the ``fermionic case'' (the one that applies in the models considered here), see \cite{tba1},
\[
	{\cal L}[\rho,\nu] = \int d\theta\,\big(\rho(\theta)\log\rho(\theta) -
	\nu(\theta)\log\nu(\theta) -
	(\rho(\theta)-\nu(\theta))\log (\rho(\theta)-\nu(\theta)) \big).
\]
The minimization of ${\cal F}^a[\rho,\nu]$ under the constraint \eqref{rhorho1} leads to the equation
\beq
	\ep(\theta) = W(\theta) + a\,p(\theta) -
	\int \frc{d\gamma}{2\pi} \,\varphi(\theta-\gamma)
	\log(1+e^{-\ep(\gamma)})\label{epeq}
\eeq
for the pseudo-energy
\beq\label{pseudo}
	\ep(\theta) := \log\lt(\frc{\rho(\theta)}{\nu(\theta)}-1\rt).
\eeq
The value of the trace $Tr_L\lt(\rho_{\rm stat}^L\,e^{-aP}\rt)$ in \eqref{fa} as $L\to\infty$ is hence dominated by the choice of the densities satisfying simultaneously \eqref{rhorho1} and \eqref{epeq}. Let us denote these densities by $\rho^a(\theta)$ and $\nu^a(\theta)$, and let us similarly use $\ep^a(\theta)$ for the associated pseudo-energy. The value of the trace is then $e^{-Lf^a + O(1)}$ where $Lf^a = {\cal F}^a[\rho^a,\nu^a]$. Using \eqref{rhorho1} and \eqref{epeq} it turns out \cite{tba1} that this value can be expressed solely in terms of the pseudo-energy,
\beq\label{faapp}
	f^a = -\int \frc{d\theta}{2\pi} \frc{d p(\theta)}{d\theta} \log(
	1+e^{-\ep^a(\theta)}).
\eeq
This is easily generalizable to a spectrum of more than one particle type, as long as the scattering is diagonal; in this case we have many pseudo-energies $\ep_i^a(\theta)$ for $i=1,\ldots,\ell$ the particle types (as well as many level densities and particle densities). Along with the expression of $J$ in \eqref{fa}, we then find \eqref{exactJ}, where we note that $x_i(\theta) = d\ep^a_i(\theta)/da\big|_{a=0}$ and $\ep_i(\theta) = \ep^{a=0}_i(\theta)$.

In order to obtain \eqref{exactJ2}, we observe that the current \eqref{exactJ} can be written as
\beq\label{Jep}
	J(\beta_l,\beta_r) = (E,(1-\varphi)^{-1}\star p).
\eeq
Here we define the inner product
\beq
	(f,g) := \frc1{2\pi} \sum_i
	\int \frc{d\theta}{1+e^{\ep_i(\theta)}} f_i(\theta)\,g_i(\theta)
\eeq
and the linear operation
\beq
	(\Or\star f)_i(\theta) := \frc1{2\pi} \sum_j
	\int \frc{d\gamma}{1+e^{\ep_j(\gamma)}}
	\Or_{ij}(\theta-\gamma)f_j(\gamma),
\eeq
and we use $E_i(\theta) := m_i\cosh(\theta)$ and $p_i(\theta) := m_i\sinh(\theta)$. Indeed \eqref{Jep} holds because, in this notation and according to \eqref{exactJ}, we have $J=(E,x)$, where $x$ satisfies the equation $x = p + \varphi\star x$, whose solution is $x = (1-\varphi)^{-1}\star p = p + \varphi\star p + \varphi\star\varphi\star p + \ldots$

We note that by unitarity, $S_{ij}(\theta)S_{ji}(-\theta) = 1$. Hence, $\varphi_{ij}(\theta) = \varphi_{ji}(-\theta)$. This implies that $\varphi$ is a hermitian operator under our inner product and linear operation:
\beq
	(f,\varphi\star g) = (\varphi\star f,g).
\eeq
Hence also $(1-\varphi)^{-1}$ is hermitian, so that we have
\beq
	J(\beta_l,\beta_r) = (p,(1-\varphi)^{-1}\star E).
\eeq
This is \eqref{exactJ2}. It is simple to see that $\mu_i(\theta) = ((1-\varphi)^{-1}\star E)_i(\theta)$ is in fact equal to $2\pi \rho_i(\theta)/L$, proportional to the level density for particle type $i$ at rapidity $\theta$, thanks to \eqref{rhorho1} and \eqref{pseudo} (generalized to many particle types).

Taking linear combinations of \eqref{exactJ} and \eqref{exactJ2}, one may obtain various expressions for the current. For instance, going back to the case of a single-particle spectrum $\ell=1$ for simplicity, we may define
\beq
	\pm m e^{y_\pm(\theta)} = x(\theta) \pm \mu(\theta)
\eeq
and we find the expression
\beqa
	J(\beta_l,\beta_r) &=& \frc{m^2}{8\pi}\int d\theta\,
	\frc{e^{\theta+y_+(\theta)} - e^{-\theta + y_-(\theta)}}{
	1+e^{\ep(\theta)}} \n
	e^{y_\pm(\theta)} &=& e^{\pm\theta} + \int \frc{d\theta}{2\pi}\,
	\frc{\varphi(\theta-\gamma)\,e^{y_\pm(\gamma)}}{
	1+e^{\ep(\gamma)}}.
\eeqa
The equality of \eqref{exactJ} and \eqref{exactJ2} further gives
\beq\label{special}
	\int d\theta\, \frc{e^{\theta+y_-(\theta)}}{1+e^{\ep(\theta)}} =
	\int d\theta\, \frc{e^{-\theta+y_+(\theta)}}{1+e^{\ep(\theta)}}.
\eeq
We note that in combination with the symmetry of $\ep(\theta)$ under the simultaneous exchange $\theta\leftrightarrow -\theta$ and $\beta_l\leftrightarrow\beta_r$, the above implies that both sides of \eqref{special} are actually invariant under $\beta_l\leftrightarrow\beta_r$.

\section{High-temperature limit} \label{apphight}
In this appendix we will consider the expression for the current presented in (\ref{exactJ}) in terms of thermodynamic quantities and carry out its high temperature limit. We will rigourously show how this limit leads to the known CFT expression (\ref{JCFT}). Our computation follows closely the work \cite{tba1}
where a similar analysis was carried out for the effective central charge $c_{\text{eff}}(r)$.

Recall the relationship between the current and the free energy (\ref{fa}) which, in an $\ell$-particle system, is given by
\beq
f^a=-\sum_{i=1}^{\ell}\int \frac{d\theta}{2\pi}m_i\cosh\theta \log(1+e^{-\ep_i(\theta)}).\label{gse}
\eeq
We can rewrite (\ref{gse}) as
\beqa
f^a=-\sum_{i=1}^{\ell}\int_{-\infty}^{0} \frac{d\theta}{2\pi}\frac{r_{i\scriptscriptstyle r}}{\beta_r} \cosh\theta L_i(\theta)-\sum_{i=1}^{\ell}\int^{\infty}_{0} \frac{d\theta}{2\pi}\frac{r_{i\scriptscriptstyle l}}{\beta_l} \cosh\theta L_i(\theta):=f^a_r+f^a_l.\label{gsesplit}
\eeqa
where $r_{i\scriptscriptstyle r}=m_i\beta_r$ and $r_{i\scriptscriptstyle l}=m_i\beta_l$.
We can obtain the steady-state current in CFT by taking the limits $r_{i\scriptscriptstyle r}\to0$ and $r_{i\scriptscriptstyle l}\to0$. In this high temperature limit, $\ep_i(\theta)$ and consequently $L_i(\theta)$ are constants with limiting values $\ep_i(0), L_i(0)$ in the central region $-\log (2/r_{i\scriptscriptstyle r})\ll \theta \ll \log (2/r_{i\scriptscriptstyle l})$ and it goes to infinity at the two edges. Therefore, $L_i(\theta)$ exhibits a typical plateau behaviour in the central region and has a double exponential falloff outside this region (we have seen many graphical examples of this in previous sections). As $r_{i\scriptscriptstyle l}$ and $r_{i\scriptscriptstyle r}$ go to zero, the plateaux become wider and the form of their two edges tends to some universal pattern.  The limiting form of the left edge is determined by the ``kink" solution $L_{ik-}(\theta)$ which satisfies the equation
\beq \label{kink-}
-\left(1-\frac{a}{\beta_{\scriptscriptstyle r}}\right)e^{-\theta}+\ep_{ik-}(\theta)+\sum_{j=1}^{\ell}\int \frac{d\theta'}{2\pi} \varphi_{ij} (\theta-\theta') L_{jk-}(\theta')=0
\eeq
where $\ep_{ik-}(\theta)\equiv \ep_i[\theta-\log(2/r_{i\scriptscriptstyle r})]$ and $L_{ik-}(\theta) \equiv L_i[\theta-\log(2/r_{i\scriptscriptstyle r})]$. Similarly, the limiting form of the right edge is determined by the function $L_{ak+}(\theta)$ which satisfies the equation
\beq \label{kink+}
-\left(1+\frac{a}{\beta_{\scriptscriptstyle l}}\right)e^{\theta}+\ep_{ik+}(\theta)+\sum_{j=1}^{\ell}\int \frac{d\theta'}{2\pi} \varphi_{ij} (\theta-\theta') L_{jk+}(\theta')=0
\eeq
where $\ep_{ik+}(\theta)\equiv \ep_i[\theta+\log(2/r_{i\scriptscriptstyle l})]$ and $L_{ik+}(\theta) \equiv L_i[\theta+\log(2/r_{i\scriptscriptstyle l})]$. These equations follow from (\ref{exactF}) by performing the indicated rapidity shifts. We can now rewrite the high temperature limit of the ``free energies'' $f^a_{r,l}$ in terms of these new ``kink" functions as
\beq \label{E-}
f^a_r=-\frac{1}{\beta_{\scriptscriptstyle r}}\sum_{i=1}^{\ell}\int \frac{d\theta}{2\pi} e^{-\theta} L_{ik-}(\theta), \qquad
f^a_l=-\frac{1}{\beta_{\scriptscriptstyle l}}\sum_{i=1}^{\ell}\int\frac{d\theta}{2\pi} e^{\theta} L_{ik+}(\theta)
\eeq
Differentiating (\ref{kink-}) and (\ref{kink+}) with respect to $\theta$ we have
\beq
\left(1-\frac{a}{\beta_{\scriptscriptstyle r}}\right)e^{-\theta}+\frac{\partial \ep_{ik-}(\theta)}{\partial \theta}+\sum_{j=1}^{\ell}\int \frac{d L_{jk-}(\theta')}{2\pi }\varphi_{ij}(\theta-\theta')=0
\eeq
and
\beq
-\left(1+\frac{a}{\beta_{\scriptscriptstyle l}}\right)e^{\theta}+\frac{\partial \ep_{ik+}(\theta)}{\partial \theta}+\sum_{j=1}^{\ell}\int \frac{d L_{jk+}(\theta')}{2\pi }\varphi_{ij}(\theta-\theta')=0
\eeq
Solving these equations for $e^{\pm \theta}$ and substituting $e^{-\theta}$ in $f^a_r$ and $e^{\theta}$ in $f^a_l$ we obtain
\begin{equation}\label{Er}
    f^a_r=\frac{1}{\beta_{\scriptscriptstyle r}-a}\sum_{i=1}^{\ell}\int \frac{d\theta}{2\pi}\left(\frac{\partial \ep_{ik-}(\theta)}{\partial \theta}+\sum_{j=1}^{\ell}\int \frac{d L_{jk-}(\theta')}{2\pi }\varphi_{ij}(\theta-\theta') \right) L_{ik-}(\theta)
\end{equation}
and
\begin{equation}\label{El}
    f^a_l=-\frac{1}{\beta_{\scriptscriptstyle l}+a}\sum_{i=1}^{\ell}\int \frac{d\theta}{2\pi}\left(\frac{\partial \ep_{ik+}(\theta)}{\partial \theta}+\sum_{j=1}^{\ell}\int \frac{d L_{jk+}(\theta')}{2\pi }\varphi_{ij}(\theta-\theta') \right) L_{ik+}(\theta)
\end{equation}
Let us consider $f^a_r$ in more detail. When $r_{i\scriptscriptstyle r}, r_{i\scriptscriptstyle l} \rightarrow 0$ we can rewrite it as
\begin{equation}\label{ERR}
    f^a_r=\frac{1}{\beta_{\scriptscriptstyle r}-a}\frac{1}{2\pi}\left[\sum_{i=1}^{\ell}\int_{\epsilon_i(0)}^\infty d\epsilon\log(1+e^{-\epsilon})-\sum_{i,j=1}^{\ell}\int_{\epsilon_i(0)}^\infty \frac{d \epsilon_{ik-}}{1+e^{\epsilon_{ik-}} }\varphi_{ij}*L_{jk-}(\theta)\right],
\end{equation}
where we assumed there is parity invariance $\varphi_{ij}(\theta)=\varphi_{ji}(\theta)$ and used $dL=-{d\epsilon}/(1+e^\epsilon)$. We can now substitute the convolution $\varphi_{ij}*L_{jk-}(\theta)$ by its expression from equation (\ref{kink-}) which leads to
\begin{eqnarray}
 \int_{\epsilon_i(0)}^\infty \frac{d \epsilon_{ik-}}{1+e^{-\epsilon_{ik-}} }\varphi_{ij}*L_{jk-}(\theta)&=&-\int_{\epsilon_i(0)}^\infty \frac{\epsilon_{ik-} d \epsilon_{ik-}}{1+e^{\epsilon_{ik-}} }+ \frac{\beta_r-a}{\beta_r}\int_{L_i(0)}^\infty {e^{-\theta} d L_{ik-}}\nonumber\\
 &=& -\int_{\epsilon_i(0)}^\infty \frac{\epsilon_{ik-} d \epsilon_{ik-}}{1+e^{\epsilon_{ik-}} }+ \frac{\beta_r-a}{\beta_r}
 \int_{-\log\left(\frac{r_{i\scriptscriptstyle r}}{2}\right)}^\infty {e^{-\theta} L_{ik-}d\theta}.
\end{eqnarray}
In the last line we have used integration by parts. We note that the last term is (up to constants and summing up in $i$) nothing but the original function $f^a_r$. We can therefore substitute in (\ref{ERR}) and solve for $f_r^a$. An identical calculation can be performed for $f^a_l$ leading to:
\begin{equation}\label{EE}
    f^a=\frac{1}{4\pi}\left(\frac{1}{\beta_r-a}-\frac{1}{\beta_l+a} \right)\sum_{i=1}^\ell \int_{\epsilon_i(0)}^\infty d\epsilon \left(\log(1+e^{-\epsilon})+\frac{\epsilon}{1+e^\epsilon}\right).
\end{equation}
Differentiating (\ref{EE}) with respect to $a$ and setting $a=0$, finally we have
\beq \label{JUV}
J(\beta_l,\beta_r)=\frac{1}{4\pi}\left[\frac{1}{\beta_{\scriptscriptstyle l}^2}-\frac{1}{\beta_{\scriptscriptstyle r}^2}\right]
\sum_i^\ell \int_{\epsilon_i(0)}^\infty d\epsilon \left(\log(1+e^{-\epsilon})+\frac{\epsilon}{1+e^\epsilon}\right)=\frac{\pi c}{12}(T_l^2-T_r^2).
\eeq
The last equality follows from the identification of the central charge with the integral above (up to the constant $\pi^2/3$). This integral provides a representation of Roger's dilogarithm function. The connection of this function with the central charge has been discussed in detail in the classic literature on the subject \cite{tba1,tba2}. The values $\epsilon_i(0)$ can be obtained by solving the constant TBA equations which were introduced in \cite{tba2}. These can be obtained from the original (both at and out of equilibrium) TBA equations by exploiting the additional fact that the TBA kernel is usually picked about $\theta=0$. This implies that in the high energy (temperature) limit the kernels $\varphi_{ij}*L_j(\theta)$ may be approximated by assuming that $L_j(\theta)$ is constant for $\theta$ near zero
\begin{equation}
\label{kernel}
    \varphi_{ij}*L_j(\theta) \approx N_{ij}\log(1+e^{-\epsilon_i(0)}),
\end{equation}
with $N_{ij}=-\frac{1}{2\pi}\int \varphi_{ij}(\theta) d\theta$ so that the TBA equations become
\beq
\ep_i(0)=\sum_{j=1}^{\ell}N_{ij}\log (1+e^{-\ep_j(0)}).
\eeq
This is a set of coupled algebraic equations which have been solved for many families of models and which exhibit very interesting structures (see e.g.~\cite{Kuniba} for a recent review).

\begin{figure}[h!]
\begin{center}
 \includegraphics[width=7.5cm]{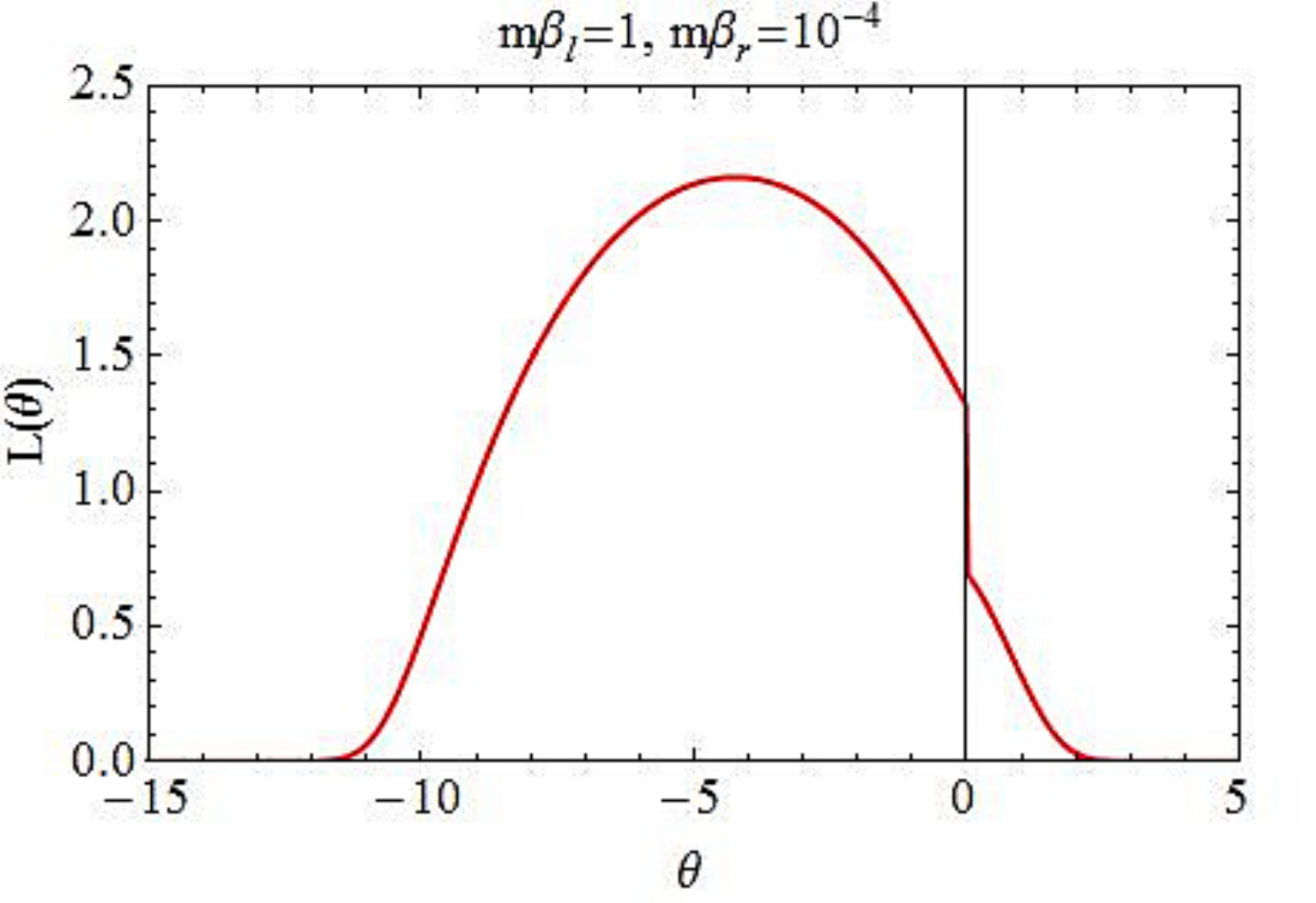}
 \includegraphics[width=7.5cm]{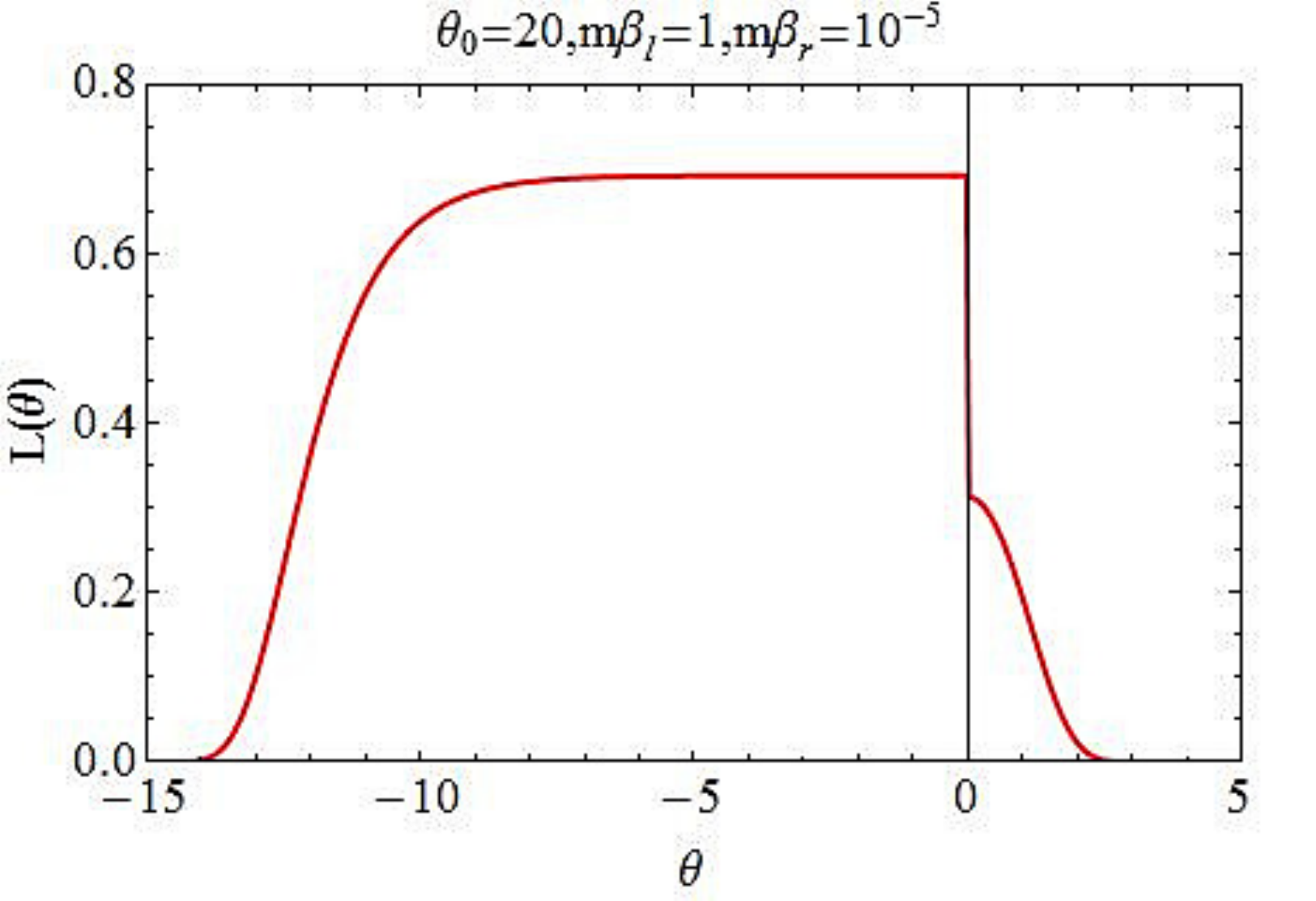}
\caption{$L$-functions for large values of $m\beta_l$ and small values of $m \beta_r$ in the sinh-Gordon and roaming trajectories model.}\label{jump}
\end{center}
\end{figure}

Another interesting physical situation corresponds to the case when one temperature is kept constant and low while the other temperature becomes very large. Thus we have $r_{i\scriptscriptstyle l}=\text{const.}$ and $r_{i\scriptscriptstyle r}\to 0$. In order to study the current in this case we may follow identical steps as above, starting with the separation of the ground state energy into the two contributions (\ref{gsesplit}). In this situation the functions $\ep_i(\theta)$ are not continuous any more at $\theta=0$ due to the presence of $W(\theta)$ in the TBA equations. Thus, they have two different limiting values at $\theta=0$ with the relation:
\beqa
\ep_{i+}(0)=\ep_{i-}(0)-r_{i\scriptscriptstyle l},
\eeqa
where we use the notation $\ep_{i+}(\theta)$ and $\ep_{i-}(\theta)$ to represent $\ep(\theta)$ in the regions $\theta>0$ and  $\theta<0$, respectively. Also we have to note that the left-right asymmetry of the $L$-function has been broken. The left part of the $L$-function remains the same: it has a plateau at $\log(1+e^{-\ep_{i-}(0)})$ in the region $-\log (2/r_{i\scriptscriptstyle r})\ll \theta <0$. But the right part has only a quick exponential falloff, since $\epsilon_{i+}(\theta) \approx  r_{i\scriptscriptstyle l} \cosh\theta$ for $\theta>0$ and $r_{i\scriptscriptstyle l} \geq 1$. Therefore it does not show any plateau behaviour. These features are particularly clear in the second plot of Figure~\ref{jump}, which gives the $L$-function of the roaming trajectories model. A clear discontinuity is also observed in the first figure, albeit the $L$-function does not display a plateau on the left hand side due to the particular features of the sinh-Gordon model already discussed in Section~\ref{Lfunctions}.

Thus $f_l^a$ is simply defined by the second term in (\ref{gsesplit}) and can not be further simplified. $f_r^a$ can be computed exactly as before leading to the same result, that is the first term in (\ref{EE}). By differentiating with respect to $a$ and setting $a=0$ the following result is obtained:
\beqa
J(\beta_l,\beta_r)&=&\sum_{i=1}^{\ell}\left[\int_{0}^{\infty} \frac{d\theta}{2\pi}m_i \cosh\theta \frac{x_{i+}(\theta)}{1+e^{\varep_{i+}(\theta)}}+\frac{1}{4\pi\beta_{r}^2}
 \int_{\epsilon_{i-}(0)}^\infty d\epsilon \left(\log(1+e^{-\epsilon})+\frac{\epsilon}{1+e^\epsilon}\right)\right]\nonumber\\
 &=& \sum_{i=1}^{\ell}\int_{0}^{\infty} \frac{d\theta}{2\pi}m_i \cosh\theta \frac{x_{i+}(\theta)}{1+e^{\varep_{i+}(\theta)}} + \frac{\pi c}{12\beta_r^2}.
\eeqa
where $x_{i+}(\theta)$ can be obtained from (\ref{epeq}) as $x_{i+}(\theta)=\frac{d\ep_{i+}(\theta)}{da}\Big|_{a=0}$.

An interesting conclusion to be drawn from this is that a configuration where one temperature is high and the other is low will still give rise to $c$-functions $c_n(T_l,T_r)$ such as the ones shown in Figure~\ref{roamingc}. This is because when dividing the cumulants by $T_l^{n+1}+(-1)^n T_r^{n+1}$ the coefficient of the singularities of the cumulants is extracted and this is the same as when both right and left temperatures are high. This is the reason why in the last graph of Figure~\ref{roamingc} we can still see the usual plateau structure, even though $m\beta_l=1$ is rather large.

\section{Low-temperature expansion}\label{applowt}
Although the TBA equations generally need to be solved numerically, at low temperatures a perturbative expansion may be used leading to analytic results. In particular these analytic results will allow us to establish rigourously the result $P(\beta,\sigma\beta)\neq 0$ as discussed in appendix \ref{appnonadd}.
  In order to do this, we will consider once more the original TBA equations (\ref{exactJ}). For simplicity, let us consider a theory with a single particle spectrum (such as the sinh-Gordon model). We will start by rewriting the TBA equation as follows:
\begin{equation}\label{yf}
    y(\theta)=\eta(\theta) \exp \left((\varphi*L)(\theta)\right),
\end{equation}
where $y(\theta):=e^{-\epsilon(\theta)}$ and $\eta(\theta):=e^{-W(\theta)}$. For low temperatures, the function $W(\theta)$ takes large values, thus $y(\theta)$ is generally very small. We can therefore expand the exponential on the r.h.s. of (\ref{yf}) up to and including the second order in $\eta(\theta)$ as
\begin{eqnarray}
    y(\theta)= \eta(\theta)\left(1+(\varphi*L)(\theta)+\frac{1}{2}(\varphi*L)^2(\theta)+\cdots \right),
\end{eqnarray}
and approximate the $L$-function as
\begin{equation}
    L(\theta)=\log(1+y(\theta)) \approx y(\theta)-\frac{y(\theta)^2}{2}+\frac{y(\theta)^3}{3!}+\cdots.
\end{equation}
Combining these two expansions and keeping only up to order $\eta(\theta)^2$ terms we find
\begin{eqnarray}
    y(\theta)\approx \eta(\theta)\left(1+(\varphi*\eta)(\theta)\right).
\end{eqnarray}
We now want to use this result to study the non-additivity of the current $J(\beta_l,\beta_r)$. As seen earlier, the current is of the form
\begin{equation}\label{current2}
   J(\beta_l,\beta_r)=\frac{m}{2\pi}\int_{-\infty}^\infty d\theta \cosh\theta \frac{x(\theta)y(\theta)}{1+ y(\theta)}\approx \frac{m}{2\pi}\int_{-\infty}^\infty d\theta \cosh\theta{x(\theta)y(\theta)}(1- y(\theta)),
\end{equation}
The product $x(\theta) y(\theta)$ can be approximated as
\begin{equation}
    -x(\theta) y(\theta) =y'(\theta)=\eta'(\theta)\left(1+(\varphi*\eta)(\theta)\right)+\eta(\theta)(\varphi*\eta')(\theta),
\end{equation}
where $y'(\theta)=\left.\frac{d y(\theta)}{da}\right|_{a=0}$ and $y(\theta,z)$ is the function $e^{\epsilon(\theta)}$ as defined by the equation (\ref{epeq}) and similarly for $\eta'(\theta)$. Thus, at order $\eta^2$ the current is
\begin{equation}\label{Jlowtexp}
   J(\beta_l,\beta_r)\approx \frac{m^2}{2\pi}\int_{-\infty}^\infty d\theta \cosh\theta \eta(\theta) \left(\sinh \theta \left(1-\eta(\theta)+(\varphi*\eta)(\theta)\right)-(\varphi*\eta')(\theta)\right),
\end{equation}
where we used the fact that $\eta'(\theta)=-m \sinh\theta \,\eta(\theta)$. Due to the structure of $W(\theta)$ it is clear that the terms
\begin{equation}
 e(\beta_l,\beta_r):=\frac{m^2}{4\pi}\int_{-\infty}^\infty d\theta \sinh 2\theta\, \eta(\theta)  \left(1-\eta(\theta)\right),\label{order1}
\end{equation}
provide an additive contribution in the sense that they have the structure $f(\beta_l)-f(\beta_r)$. We will now show that this is not the case for the other two terms. We will call this contribution to the current $r(\beta_l, \beta_r)$ and it is given by
\begin{equation}\label{order2}
r(\beta_l, \beta_r):=\frac{m^2}{2\pi}\int_{-\infty}^\infty d\theta \cosh\theta \,\eta(\theta) \left(\sinh \theta (\varphi*\eta)(\theta)-(\varphi*\eta')(\theta)\right).
\end{equation}
Let us write the function explicitly:
\begin{eqnarray}
  r(\beta_l, \beta_r) &=&m^2 \int_{-\infty}^\infty\int_{-\infty}^\infty \frac{d\theta d\gamma}{2(2\pi)^2} \sinh2\theta\, e^{W(\theta)}
  \varphi(\theta-\gamma) e^{W(\gamma)} \nonumber \\
  &+& m^2\int_{-\infty}^\infty\int_{-\infty}^\infty  \frac{d\theta d\gamma}{2\pi} \cosh\theta \sinh \beta\, e^{W(\theta)} \varphi(\theta-\gamma) e^{W(\gamma)}.\label{r2}
\end{eqnarray}
Employing the definition of $W(\theta)$ the integrals above split into four regions, depending on the sign of $\theta$ and $\gamma$:
\begin{eqnarray}
  r(\beta_l, \beta_r) &=&m^2 \int_{0}^\infty\int_{0}^\infty \frac{d\theta d\gamma}{2(2\pi)^2} \sinh(2\theta) e^{-\beta_l m\cosh\theta}
  \varphi(\theta-\gamma) e^{-\beta_l m\cosh \gamma}\nonumber  \\
  &+&m^2 \int_{0}^\infty\int_{-\infty}^0 \frac{d\theta d\gamma}{2(2\pi)^2} \sinh(2\theta) e^{-\beta_lm \cosh\theta}
  \varphi(\theta-\gamma) e^{-\beta_r m\cosh \gamma}\nonumber  \\
 &+&m^2 \int_{-\infty}^0\int_{-\infty}^0 \frac{d\theta d\gamma}{2(2\pi)^2} \sinh(2\theta) e^{-\beta_r m\cosh\theta}
  \varphi(\theta-\gamma) e^{-\beta_r m\cosh \gamma}\nonumber  \\
 &+&m^2 \int_{-\infty}^0\int_{0}^\infty \frac{d\theta d\gamma}{2(2\pi)^2} \sinh(2\theta) e^{-\beta_r m\cosh\theta}
  \varphi(\theta-\gamma) e^{-\beta_l m\cosh \gamma}\nonumber  \\
  &+& m^2\int_{0}^\infty\int_{0}^\infty  \frac{d\theta d\gamma}{2\pi} \cosh\theta \sinh \gamma e^{-\beta_l m \cosh \theta} \varphi(\theta-\gamma) e^{-\beta_l m \cosh \gamma}\nonumber\\
 &+& m^2\int_{0}^\infty\int_{-\infty}^0 \frac{d\theta d\gamma}{2\pi} \cosh\theta \sinh \gamma e^{-\beta_l m\cosh \theta} \varphi(\theta-\gamma) e^{-\beta_r m\cosh \gamma}\nonumber\\
 &+& m^2\int_{-\infty}^0\int_{-\infty}^0 \frac{d\theta d\gamma}{2\pi} \cosh\theta \sinh \gamma e^{-\beta_r m\cosh \theta} \varphi(\theta-\gamma) e^{-\beta_r m\cosh \gamma}\nonumber\\
 &+& m^2\int_{-\infty}^0\int_{0}^\infty \frac{d\theta d\gamma}{2\pi} \cosh\theta \sinh \gamma e^{-\beta_r m\cosh \theta} \varphi(\theta-\gamma) e^{-\beta_l m^2\cosh \gamma}.
\end{eqnarray}
It is now a tedious though simple computation to group these integrals together by changing variables $\theta\rightarrow -\theta$ and $\gamma \rightarrow -\gamma$ where appropriate. When doing this we will use the fact that the kernel $\varphi(\theta)$ is, by definition, an even function. Grouping terms together we finally obtain
\begin{eqnarray}
 && r(\beta_l, \beta_r) =m^2 \int_{0}^\infty\int_{0}^\infty \frac{d\theta d\gamma}{2(2\pi)^2} \left( e^{-\beta_l m( \cosh\theta+\cosh\gamma)}
-e^{-\beta_r m(\cosh\theta+\cosh\gamma)}
  \right)\sinh(2\theta)\varphi(\theta-\gamma)\nonumber \\
&+& m^2\int_{0}^\infty\int_{0}^\infty  \frac{d\theta d\gamma}{2\pi} \left( e^{-\beta_l m( \cosh\theta+\cosh\gamma)}
-e^{-\beta_r m(\cosh\theta+\cosh\gamma)}\right)\varphi(\theta-\gamma) \cosh\theta \sinh \gamma\\
  &+&m^2 \int_{0}^\infty\int_{0}^\infty \frac{d\theta d\gamma}{2(2\pi)^2} e^{-\beta_l m\cosh\theta}
   e^{-\beta_r m\cosh \gamma}\varphi(\theta+\gamma)\left(\sinh(2\theta)-\sinh(2\gamma)+2\sinh(\theta-\gamma)\right).\nonumber
\end{eqnarray}
Examining the function $r(\beta_l,\beta_r)$ we see that the first two lines above provide again additive contributions, that is, they have the structure $f(\beta_l)-f(\beta_r)$. However the last contribution, on the last line, is not additive as it is possible to show both analytically and numerically that it is non-zero. An analytical proof is provided in Appendix \ref{appnonadd}, whereas numerical evidence has been provided in Section \ref{sectna} for the sinh-Gordon and sine-Gordon models. Generalizing to an $\ell$-particle spectrum and using the short-hand notation $l_\theta^i:=\exp(-\beta_l m_i \cosh\theta)$ and $r_\theta^i:=\exp(-\beta_r m_i \cosh\theta)$ we find
the following low temperature approximation
\begin{eqnarray}
 J(\beta_l,\beta_r) &\approx& \sum_{i=1}^\ell m_i^2 \int_{0}^\infty \frac{ d\theta}{2(2\pi)}\left( l_\theta^i
-r_\theta^i- (l_\theta^i)^2
+(r_\theta^i)^2
  \right) \sinh(2\theta)\label{formula}
 \\
 &+&\sum_{i,j=1}^\ell  m_i^2 \int_{0}^\infty\int_{0}^\infty \frac{d\theta d\gamma}{(2\pi)^2} \left( l_\theta^i l_\gamma^j -r_\theta^i r_\gamma^j
  \right)\cosh(\theta)\varphi_{ij}(\theta-\gamma)(\sinh\theta+\sinh\gamma)\nonumber \\
  &+&\sum_{i,j=1}^\ell  m_i^2 \int_{0}^\infty\int_{0}^\infty \frac{d\theta d\gamma}{2(2\pi)^2} l_\theta^i r_\gamma^j
  \varphi_{ij}(\theta+\gamma)\left(\sinh(2\theta)-\sinh(2\gamma)+2\sinh(\theta-\gamma)\right).\nonumber
\end{eqnarray}
This low-temperature expansion can be compared with an expansion obtained directly from the trace expression \eqref{ness}-\eqref{W} (we again assume a single particle spectrum for simplicity). For convention, the finite-volume multi-particle states can be denoted as
\beqa
|\theta_1,...,\theta_n\ket_L \no
\eeqa
The corresponding energy levels are determined by the well-known Bethe ansatz equations
\beq
Q_k(\theta_1,...,\theta_n)=m L \sinh\theta_k+ \sum_{l\neq k}\delta(\theta_k-\theta_l)=2\pi I_k \quad,\quad k=1,...,n
\eeq
where $I_k$ are momentum quantum numbers and $\delta(\theta)=-i\log S(\theta)$  is the two-particle scattering phase-shift.
The density of multi-particle states can be obtained by
\beqa
\rho(\theta_1,...,\theta_n)=\det {\cal J}^{(n)} \quad,\quad {\cal J}^{(n)}_{kl}=\frac{\partial Q_k(\theta_1,...,\theta_n)}{\partial\theta_l}\quad,\quad k,l=1,...,n
\eeqa
Let us expand the traces in (\ref{JL}):
\beqa \label{numerator}
\Tr_L(\rho^{L}_{\text{stat}}P)&=&\sum_{\theta^{(1)}}e^{-W(\theta^{(1)})}m\sinh\theta^{(1)}+\frac{1}{2}\sum^{'}_{\theta^{(2)}_{1}\theta^{(2)}_{2}}
e^{-\sum_{i=1}^2 W(\theta^{(2)}_i)} \sum_{i=1}^2 m\sinh\theta^{(2)}_i\n
&&+\frac{1}{6}\sum^{'}_{\theta^{(3)}_{1}\theta^{(3)}_{2}\theta^{(3)}_{3}}e^{-\sum_{i=1}^3 W(\theta^{(3)}_i)}\sum_{i=1}^3 m\sinh\theta^{(3)}_i+O(e^{-4W})
\eeqa
 and
\beqa
\Tr_L(\rho^{L}_{\text{stat}})&=&1+\sum_{\theta^{(1)}}e^{-W(\theta^{(1)})}+\frac{1}{2}\sum^{'}_{\theta^{(2)}_{1}\theta^{(2)}_{2}}e^{-\sum_{i=1}^2 W(\theta^{(2)}_{i})}\n
&&+\frac{1}{6}\sum^{'}_{\theta^{(3)}_{1}\theta^{(3)}_{2}\theta^{(3)}_{3}}e^{-\sum_{i=1}^3 W(\theta^{(3)}_i)}+O(e^{-4W})
\eeqa
At low temperature, we have the expansion
\beqa \label{denominator}
\frac{1}{\Tr_L(\rho^{L}_{\text{stat}})}&=&1-\sum_{\theta^{(1)}}e^{-W(\theta^{(1)})}+\left( \sum_{\theta^{(1)}}e^{-W(\theta^{(1)})}\right)^2-\frac{1}{2}\sum^{'}_{\theta^{(2)}_{1}\theta^{(2)}_{2}}e^{-\sum_{i=1}^2 W(\theta^{(2)}_{i})}\n
&&-\left( \sum_{\theta^{(1)}}e^{-W(\theta^{(1)})} \right)^3+\left( \sum_{\theta^{(1)}}e^{-W(\theta^{(1)})} \right)\sum^{'}_{\theta^{(2)}_{1}\theta^{(2)}_{2}}e^{-\sum_{i=1}^2 W(\theta^{(2)}_{i})}\n
&&-\frac{1}{6}\sum^{'}_{\theta^{(3)}_{1}\theta^{(3)}_{2}\theta^{(3)}_{3}}e^{-\sum_{i=1}^3 W(\theta^{(3)}_i)}+O(e^{-4W})
\eeqa
The prefactors $1/n!$ for every multi-particle sum take into account overcounted states with different ordering of the same set of rapidities. The prime in the multi-particle sum indicates all quantum numbers (rapidities) for the state are different. The upper indices of the rapidities and $W$ represent the number of particles in the state.

With (\ref{numerator}) and (\ref{denominator}), we can now compute the current up to exponential corrections at finite volume $L$. In the limit $L\to \infty$, the low temperature expansion of the current performed in the first part of this appendix should be recovered. Here, we present the calculation of the current up to the first three orders. Let us discuss the first order contribution $J_1$ in detail. At first order we have the contribution
\beq
J^L_1= \frac{1}{L}\sum_{\theta^{(1)}}e^{-W(\theta^{(1)})}m\sinh\theta^{(1)}.
\eeq
In order to compare with the low temperature expansion we must consider infinite volume. This amounts to taking the limit $L\to \infty$ by replacing the sum over rapidities by an integral over the states in the rapidity space, that is
$
\sum_{\theta^{(1)}} \to \int \frac{d\theta}{2\pi}\rho_1(\theta)
$. The density of one-particle states $\rho_1(\theta)$ is obtained by
\beq
\rho_1(\theta)=L \frac{d \,p(\theta) }{d \theta}=\frac{d \,m \sinh\theta }{d \theta}=m L \cosh\theta.
\eeq
Therefore, the first order term of the current is
\beq
J_1=\lim_{L\to \infty}J^L_1 = m^2\int \frac{d\theta}{2\pi}e^{-W(\theta)}\sinh\theta  \cosh\theta.
\eeq
It can be easily seen that this perfectly agrees with the first contribution (e.g. order $\eta$) in equation (\ref{order1}). A similar computation can be performed for second order terms. We define
\begin{eqnarray*}
J^L_2&=&\frac{1}{L}\left[-\sum_{\theta^{(1)}}e^{- W(\theta^{(1)})}m\sinh\theta^{(1)}\sum_{\theta^{(1)}}e^{- W(\theta^{(1)})}+\frac{1}{2}\sum^{'}_{\theta^{(2)}_{1}\theta^{(2)}_{2}}e^{-\sum_{i=1}^2 W(\theta^{(2)}_i)}\sum_{i=1}^2 m\sinh\theta^{(2)}_i \right]\\
&=&\frac{1}{L}\Bigg[-\sum_{\theta^{(1)}}e^{- W(\theta^{(1)})}m\sinh\theta^{(1)}\sum_{\theta^{(1)}}e^{-W(\theta^{(1)})}+\frac{1}{2}\sum_{\theta^{(2)}_{1}\theta^{(2)}_{2}}e^{-\sum_{i=1}^2 W(\theta^{(2)}_i)}\sum_{i=1}^2 m\sinh\theta^{(2)}_i\\
&&-\frac{1}{2}\sum_{\theta^{(2)}_{1}=\theta^{(2)}_{2}}e^{-2W(\theta^{(2)}_1)}2m\sinh\theta^{(2)}_{1} \Bigg]
\end{eqnarray*}
The last term has $\theta^{(2)}_{1}=\theta^{(2)}_{2}$ and corresponds to a two-particle state with equal quantum numbers of the two particles. In this case, the two-particle Bethe ansatz equations degenerate to a one-particle equation, which means that density of this two-particle state is again $\rho_1$. In the large $L$ limit, we may replace the sums with integrals as $\sum_{\theta^{(1)}} \to \int \frac{d\theta}{2\pi}\rho_1(\theta)$,  $\sum_{\theta_1^{(2)}=\theta_2^{(2)}}\to \int \frac{d\theta}{2\pi}\rho_1(\theta)$ and $ \sum_{\theta_1^{(2)}\theta_2^{(2)}}\to \int \frac{d\theta_1}{2\pi}\frac{d\theta_2}{2\pi}\rho_2(\theta_1,\theta_2)$.
The Bethe ansatz equations for a two-particle state are
\beqa
mL\sinh\theta_1+\delta(\theta_1-\theta_2)=Q_1(\theta_1\theta_2)\n
mL\sinh\theta_2+\delta(\theta_2-\theta_1)=Q_2(\theta_1\theta_2),
\eeqa
so that the relevant density of two-particle states is given by
\[  \rho_2(\theta_1,\theta_2)=\det\left( \begin{array}{cc}
                                    mL\cosh\theta_1+\varphi(\theta_1-\theta_2) &
                                     -\varphi(\theta_1-\theta_2)\\
-\varphi(\theta_1-\theta_2) & mL\sinh \theta_2+\varphi(\theta_1-\theta_2)
\end{array}\right)
\]
Exploiting the fact that $\varphi(\theta)=\varphi(-\theta)$, relabeling integration variables and exchanging the order of integration, we obtain
\beqa
J_2&=&m^2 \int \frac{d\theta_1 d\theta_2}{(2\pi)^2} \cosh\theta_1(\sinh\theta_1+\sinh\theta_2) \varphi(\theta_1-\theta_2) e^{-W(\theta_1)-W(\theta_2)}\n
&&-m^2\int \frac{d\theta}{2\pi}\cosh\theta \sinh\theta e^{-2W(\theta)}.
\eeqa
Once more this is in full agreement with (\ref{r2}) plus the second contribution in (\ref{order1}). Finally, we look at third order contributions
\beqa
J^L_3&=&\frac{1}{L}\Bigg[\sum_{\theta^{(1)}}e^{-W(\theta^{(1)})}m\sinh\theta^{(1)}\bigg(\sum_{\theta^{(1)}}e^{-W(\theta^{(1)})}\bigg)^2
-\frac{1}{2}\sum_{\theta^{(1)}}m\sinh\theta^{(1)}\sum^{'}_{\theta^{(2)}_{1}\theta^{(2)}_{2}}e^{-\sum_{i=1}^2 W(\theta^{(2)}_i)}\n
&&-\frac{1}{2}\sum^{'}_{\theta^{(2)}_{1}\theta^{(2)}_{2}}e^{-\sum_{i=1}^2 W(\theta^{(2)}_i)}\sum_{i=1}^2 m\sinh\theta^{(2)}_i\sum_{\theta^{(1)}}e^{-W(\theta^{(1)})}\n
&&+\frac{1}{6}\sum^{'}_{\theta^{(3)}_{1}\theta^{(3)}_{2}\theta^{(3)}_{3}}e^{-\sum_{i=1}^3 W(\theta^{(3)}_i)}\sum_{i=1}^3 m\sinh\theta^{(3)}_i\Bigg].
\eeqa
By using the relations
\beqa
\sum^{'}_{\theta^{(2)}_1\theta^{(2)}_2}&=&\sum_{\theta^{(2)}_1\theta^{(2)}_2}-\sum_{\theta^{(2)}_1=\theta^{(2)}_2}\n
\sum^{'}_{\theta^{(3)}_1\theta^{(3)}_2\theta^{(3)}_3}&=&\sum_{\theta^{(3)}_1\theta^{(3)}_2\theta^{(3)}_3}- 3\sum_{\theta^{(3)}_1,\theta^{(3)}_2=\theta^{(3)}_3}+2\sum_{\theta^{(3)}_1=\theta^{(3)}_2=\theta^{(3)}_3}\no
\eeqa
we can rewrite
\beqa
J^L_3&=&\frac{1}{L}\Bigg[\sum_{\theta^{(1)}}e^{-W(\theta^{(1)})}m\sinh\theta^{(1)}
\bigg(\sum_{\theta^{(1)}}e^{-W(\theta^{(1)})}\bigg)^2-\frac{1}{2}\sum_{\theta^{(1)}}
m\sinh\theta^{(1)}\sum_{\theta^{(2)}_{1}\theta^{(2)}_{2}}e^{-\sum_{i=1}^2 W(\theta^{(2)}_i)}\n
&&+\frac{1}{2}\sum_{\theta^{(1)}}m\sinh\theta^{(1)}\sum_{\theta^{(2)}_{1}=\theta^{(2)}_{2}}e^{-2 W(\theta^{(2)}_1)} -\frac{1}{2}\sum_{\theta^{(2)}_{1}\theta^{(2)}_{2}}e^{-\sum_{i=1}^2 W(\theta^{(2)}_i)}\sum_{i=1}^2 m\sinh\theta^{(2)}_i\sum_{\theta^{(1)}}e^{-W(\theta^{(1)})}\n
&&+\frac{1}{2}\sum_{\theta^{(2)}_{1}=\theta^{(2)}_{2}}e^{-2 W(\theta^{(2)}_1)}2 m\sinh\theta^{(2)}_1\sum_{\theta^{(1)}}e^{-W(\theta^{(1)})}+\frac{1}{6}\sum_{\theta^{(3)}_{1}\theta^{(3)}_{2}\theta^{(3)}_{3}}e^{-\sum_{i=1}^3 W(\theta^{(3)}_i)}\sum_{i=1}^3 m\sinh\theta^{(3)}_i\n
&&-\frac{1}{2}\sum_{\theta^{(3)}_{1},\theta^{(3)}_{2}=\theta^{(3)}_{3}}e^{- W(\theta^{(3)}_1)-2W(\theta^{(3)}_2)} (m\sinh\theta^{(3)}_1+2m\sinh\theta^{(3)}_2)\n
&&+\frac{1}{3}\sum_{\theta^{(3)}_{1}=\theta^{(3)}_{2}=\theta^{(3)}_{3}}e^{-3 W(\theta^{(3)}_1)} 3m\sinh\theta^{(3)}_1\Bigg].
\eeqa
In the large $L$ limit, we replace the sums with integrals as in previous cases with the addition of
$\sum_{\theta_1^{(2)}\theta_2^{(2)}\theta_1^{(3)}}\to \int \frac{d\theta_1}{2\pi}\frac{d\theta_2}{2\pi}\frac{d\theta_3}{2\pi}\rho_3(\theta_1,\theta_2,\theta_3)$, $
 \sum_{\theta_1^{(2)},\theta_2^{(2)}=\theta_1^{(3)}}\to \int \frac{d\theta_1}{2\pi}\frac{d\theta_2}{2\pi}\rho_3(\theta_1,\theta_2=\theta_3)$ and $
\sum_{\theta_1^{(2)}=\theta_2^{(2)}=\theta_1^{(3)}}\to \int \frac{d\theta}{2\pi}\rho_1(\theta)$. The density $\rho_3(\theta_1,\theta_2=\theta_3)$ can be obtained from the Bethe ansatz equations for a three-particle state with two equal quantum numbers
\beqa
mL\sinh\theta_1+2\delta(\theta_1-\theta_2)&=&Q_1(\theta_1,\theta_2)\n
mL\sinh\theta_2+\delta(\theta_2-\theta_1)&=&Q_2(\theta_1,\theta_2).
\eeqa
The relevant density is given by
\[  \rho_3(\theta_1,\theta_2=\theta_3)=\det\left( \begin{array}{cc}
                                    mL\cosh\theta_1+2\varphi(\theta_1-\theta_2) &                           -2\varphi(\theta_1-\theta_2)\\
-\varphi(\theta_1-\theta_2) & mL\sinh \theta_2+\varphi(\theta_1-\theta_2)
\end{array}\right)
\]
Similarly, we can obtain $\rho_3(\theta_1,\theta_2,\theta_3)$ from the equations
\beqa
mL\sinh\theta_1+\delta(\theta_1-\theta_2)+\delta(\theta_1-\theta_3)&=&Q_1(\theta_1,\theta_2,\theta_3)\n
mL\sinh\theta_2+\delta(\theta_2-\theta_1)+\delta(\theta_2-\theta_3)&=&Q_2(\theta_1,\theta_2,\theta_3)\n
mL\sinh\theta_3+\delta(\theta_3-\theta_1)+\delta(\theta_3-\theta_2)&=&Q_3(\theta_1,\theta_2,\theta_3),
\eeqa
as
\beq\det\left( \begin{array}{ccc}
                                    E_1L+\varphi(\theta_{12})+\varphi_(\theta_{13}) & -\varphi(\theta_{12}) & -\varphi(\theta_{13})\\
-\varphi(\theta_{12}) &E_2L+\varphi(\theta_{12})+\varphi(\theta_{23}) & -\varphi(\theta_{23})\\
-\varphi(\theta_{13}) & -\varphi(\theta_{23}) & E_3L+\varphi(\theta_{13})+\varphi(\theta_{23})
\end{array}\right),
\eeq
where for convenience we used the notation $E_i\equiv m\cosh\theta_i$ and $\varphi(\theta_{ij})\equiv \varphi(\theta_i-\theta_j)$.
Therefore, by doing a similar but more tedious computation, we can obtain the third order term of the current
\beqa
J_3&=&-\frac{1}{2}m^2 \int \frac{d\theta_1 d\theta_2}{(2\pi)^2}\cosh\theta_1(\sinh\theta_1+2\sinh\theta_2)\varphi(\theta_1-\theta_2)e^{-W(\theta_1)-2W(\theta_2)}\no\\
&&+m^2 \int \frac{d\theta_1 d\theta_2 d\theta_3}{(2\pi)^3}\cosh\theta_1 \sum_{i=1}^3 (\sinh\theta_i)\varphi(\theta_1-\theta_2)\varphi(\theta_2-\theta_3)e^{-\sum_{i=1}^3 W(\theta_i)}\no\\
&&+\frac{1}{2}m^2 \int \frac{d\theta_1 d\theta_2 d\theta_3}{(2\pi)^3}\cosh\theta_1 \sum_{i=1}^3 (\sinh\theta_i) \varphi(\theta_1-\theta_2)\varphi(\theta_1-\theta_3)e^{-\sum_{i=1}^3 W(\theta_i)}\no\\
&&-m^2\int \frac{d\theta_1 d\theta_2}{(2\pi)^2} \cosh\theta_1 (2\sinh\theta_1+\sinh\theta_2)\varphi(\theta_1-\theta_2)e^{-2W(\theta_1)-W(\theta_2)}\no\\
&&+m^2\int \frac{d\theta}{2\pi} \cosh\theta \sinh\theta e^{-3W(\theta)}
\eeqa

\section{Proof of non-additivity of the current} \label{appnonadd}
We consider a diagonal-scattering integrable QFT, assuming that all kernels $\varphi_{ij}(\theta)$ have a fixed sign (i.e. $\varphi_{ij}(\theta)$ is positive for every $i,j$ and $\theta$, or is negative for every $i,j$ and $\theta$). This assumption is valid both for the sinh-Gordon model and the reflectionless sine-Gordon model studied above whose kernels only involve cosh functions. We analyze the second leading order in the low-temperature expansion of the additivity deficit \eqref{adddef}, and show that it is nonzero for large enough $\beta m$ and large enough $\sigma$. This is sufficient to prove that additivity does not hold in general. The derivation below is mathematically rigorous.

To second leading order in the low-temperature expansion, the current in a general diagonal-scattering model is
\beq
	J(\beta_l,\beta_r) = \sum_{i=1}^\ell \frc{m_i^2}{2\pi} \int d\theta\,\cosh\theta\,
	\eta_i(\theta)\,\Big(
	\sinh\theta\,(1-\eta_i(\theta)) + (\varphi_{ij} * \eta_j)(\theta)
	-(\varphi_{ij} * \eta_j')(\theta),
	\Big)\label{current3}
\eeq
which generalizes (\ref{Jlowtexp}) to a multi-particle theory.  As seen in section \ref{applowt} the only non additive contribution to (\ref{current3}) takes the form $\sum_{i,j} r_{ij}$ with
\beq\label{rijsym}
	r_{ij}=\frc{m_i^2}{2(2\pi)^2}\int_0^\infty d\theta \int_0^\infty d\gamma\,
	e^{-\beta_l m_i\cosh\theta} e^{-\beta_r m_j\cosh\gamma}\,
	\varphi_{ij}(\theta+\gamma)\,\Big(
	\sinh 2\theta - \sinh 2\gamma + 2\sinh (\theta-\gamma)\Big).
\eeq
that is, the last term in (\ref{formula}). Note that the function
\beq
	U(\theta,\gamma):=\varphi_{ij}(\theta+\gamma)\,\Big(
	\sinh 2\theta - \sinh 2\gamma + 2\sinh (\theta-\gamma)\Big),
\eeq
is anti-symmetric under exchange of the rapidities, has a fixed sign for all $\theta>\gamma$. Let us denote $m_i := am$ and $m_j := bm$ for some mass scale $m$ and positive numbers $a,b$ (one may always take either $a=1$ or $b=1$ by an appropriate choice of scale $m$). Further, let us use the notation
\beq
	e_\theta := e^{-\beta m\cosh\theta}.
\eeq
Then the contribution of $r_{ij}$ to the ``normalized" additivity deficit $\tilde{P}(\beta,\sigma) = \sum_{i,j=1}^\ell \tilde{P}_{ij}(\beta,\sigma)$ \eqref{adddef} is
\beq
	\tilde{P}_{ij}(\beta,\sigma) = \frc{m_i^2}{2(2\pi)^2} \int_0^\infty d\theta \int_0^\infty
	d\gamma\,U(\theta,\gamma)\,\lt(
	e_\theta^a e_\gamma^{b\sigma}
	+
	e_\theta^{a\sigma} e_\gamma^{b\sigma^2}
	-
	e_\theta^a e_\gamma^{b\sigma^2}
	\rt),
\eeq
where $\tilde{P}(\beta,\sigma)=J(\beta, \sigma^2 \beta)\sum_{i,j=1}^\ell P_{ij}(\beta,\sigma)$.
Symmetrizing under $\theta\leftrightarrow \gamma$ we find
\beq\label{Psymm}
	\tilde{P}_{ij}(\beta,\sigma) = \frc{m_i^2}{2(2\pi)^2}
	\int_0^\infty d\theta \int_0^\theta
	d\gamma\,U(\theta,\gamma)\,\lt(
	e_\theta^a e_\gamma^{b\sigma} - e_\gamma^a e_\theta^{b\sigma}
	+
	e_\theta^{a\sigma} e_\gamma^{b\sigma^2}
		- e_\gamma^{a\sigma} e_\theta^{b\sigma^2}
	-
	e_\theta^a e_\gamma^{b\sigma^2}
		+ e_\gamma^a e_\theta^{b\sigma^2}
	\rt).
\eeq

We now show that for every $\sigma>{\rm max}(a/b,1)$, every $\beta m$ large enough, and every $\theta>\gamma$, the sum of the terms in the parenthesis on the right-hand side is positive. This then implies that the integrand has a fixed sign, and since it is obviously nonzero, it shows that the additivity deficit is nonzero.

First we have, for the middle two terms,
\beq
	e_\theta^{a\sigma} e_\gamma^{b\sigma^2}
		- e_\gamma^{a\sigma} e_\theta^{b\sigma^2}
	 = e_\theta^{a\sigma} e_\gamma^{a\sigma} \lt(
	 e_\gamma^{b\sigma^2 - a\sigma} - e_\theta^{b\sigma^2-a\sigma}\rt).
\eeq
If $\sigma>a/b$, then thanks to $\theta>\gamma$ the factor in the parenthesis on the right-hand side is positive. The first factor is also obviously positive, hence the left-hand side is positive.

Second, for the remaining four terms, we find
\beq
	e_\theta^a e_\gamma^{b\sigma} - e_\gamma^a e_\theta^{b\sigma}
	-
	e_\theta^a e_\gamma^{b\sigma^2}
		+ e_\gamma^a e_\theta^{b\sigma^2}
	= e_\theta^a e_\gamma^a \lt(
	e_\gamma^{b\sigma-a}(1-e_\gamma^{b\sigma(\sigma-1)})
	-e_\theta^{b\sigma-a}(1-e_\theta^{b\sigma(\sigma-1)})
	\rt)\label{secondpart}
\eeq
Let us consider the two terms in the outermost parenthesis on the right-hand side
\beq
	K:=e_\gamma^{b\sigma-a}(1-e_\gamma^{b\sigma(\sigma-1)})
	-e_\theta^{b\sigma-a}(1-e_\theta^{b\sigma(\sigma-1)}).
\eeq
The ratio of these two terms is
\beq
	\lt(\frc{e_\gamma}{e_\theta}\rt)^{b\sigma-a}
	\frc{1-e_\gamma^{b\sigma(\sigma-1)}}{
	1-e_\theta^{b\sigma(\sigma-1)}}>
	\lt(\frc{e_\gamma}{e_\theta}\rt)^{b\sigma-a}
	\lt(1-e_\gamma^{b\sigma(\sigma-1)}\rt)>
	\lt(\frc{e_\gamma}{e_\theta}\rt)^{b\sigma-a}
	\lt(1-e^{-b\sigma(\sigma-1)\beta m}\rt)
\eeq
where the two inequalities hold if $\sigma>1$. Then if $\sigma>a/b$ and if $\cosh\theta-\cosh\gamma$ is greater than
\[
	\frac{\log\lt(1-e^{-b\sigma(\sigma-1)\beta m}\rt)}{(b\sigma-a)\beta m},
\]
which can be made as small as desired by taking $\beta m$ large enough,
the right-hand side of the last inequality above is greater than one, whence $K$ is positive. Further, it is clear that $K$ is zero at $\gamma=\theta$ and continuous and differentiable for $\gamma\in[0,\theta]$. We now show that it has no minimum in that range. Denoting $x=e_\gamma$ and $y=e_\theta$, we look for a minimum in $x$ in the range $x\in[y,1]$, for $y>0$. Differentiating with respect to $x$, a minimum may occur if
\[
	(b\sigma-a) x^{b\sigma -a-1} - (b\sigma^2-a)x^{b\sigma^2-a-1} = 0
	\quad\Rightarrow\quad x = \frc1{b\sigma(\sigma-1)} \log \lt(
	\frc{b\sigma-a}{b\sigma^2-a}\rt).
\]
The right-hand side is smaller than zero if $\sigma>1$, which puts the minimum outside of the range for any $y>0$.

\small
%\bibliography{Ref}

\end{document}